\documentclass[a4paper,11pt]{article}
\pdfoutput=1 

\usepackage{jheppub} 

\usepackage[T1]{fontenc} 
\usepackage{caption,subcaption,multirow}

\usepackage{wrapfig, float}
\usepackage{indentfirst}
\usepackage{enumitem}

\usepackage{amsmath,amsfonts,amssymb,mathtools,bbm,bm,slashed}

\numberwithin{equation}{section}

\title{Quantum entanglement and top spin correlations in SMEFT at higher orders}

\author[a]{Claudio Severi,}
\author[a]{Eleni Vryonidou}
\affiliation[a]{Department of Physics and Astronomy, University of Manchester, Oxford Road, Manchester M13 9PL, United Kingdom}

 \abstract{
    We present the first analysis of top spin polarizations, $t \bar t$ spin correlations, and $t \bar t$ spin entanglement at the LHC in the context of the Standard Model Effective Field Theory, that goes beyond Leading Order QCD accuracy. The complete set of independent dimension-6 operators entering $t \bar t$ production is identified, and their effects on all top spin observables are extracted at linear and quadratic order in $c/\Lambda^2$. By comparing results at LO and NLO, we note that the inclusion of higher orders, while not dramatically changing the picture, often amounts to notable numerical differences, that are not fully captured by LO scale variation. We also find that the expected deviations from the SM have an intricate phase space structure, and show up predominantly at large top $p_T$. For this reason, we advocate for the measurement of spin observables differentially or doubly-differentially in the $t \bar t$ phase-space. We show how the inclusion of present and future top spin measurements will improve global fits to top LHC data, by also addressing the issue of flat directions.
    }

\begin{document}

\maketitle

\clearpage

\section{Introduction}

	The Standard Model Effective Field Theory (SMEFT) is an extension of the Standard Model (SM), that includes a stack of higher--dimensional operators $\mathcal O_k$,
	\begin{equation}
	\mathcal L_\text{SMEFT} = \mathcal L_\text{SM} + \sum_{d = 5}^\infty \sum_k \frac{c_k}{\Lambda^{d-4}} \mathcal O_k, \label{smeftL}
	\end{equation}
	parameterized by a set of Wilson coefficients $c_k$ and a global scale $\Lambda$. This framework offers the significant advantage that only minimal assumptions are put on the new physics (NP): $\mathcal L_{\text{SMEFT}}$ is constructed to be the most general Lagrangian that reproduces the SM at low energy, and is Lorentz and $SU(3)_\text C \times SU(2)_\text L \times U(1)_\text Y$ gauge invariant. The SMEFT provides a promising framework to identify and parameterize deviations from the SM predictions, and a huge effort is underway to combine a wide range of experimental measurements with theoretical predictions in global fits, aimed at constraining the Wilson coefficients of SMEFT operators. The correlating nature of the SMEFT between different observables, different scales, and sometimes different sectors, mandates the inclusion of the largest possible dataset in the fits. Over the last few years, global combinations of LHC data in the top, Higgs and EW sectors have been achieved \cite{Buckley:2015lku, Aguilar-Saavedra:2018ksv, Hartland:2019bjb, Brivio:2019ius, Ellis:2020unq, Ethier:2021bye}.  While there has been enormous progress in the SMEFT interpretation of LHC measurements, several operators exist for which the current sensitivity is still limited. Further, global fits are often plagued by the presence of flat directions in their parameter space. One such example arises in the top sector, where several 2-quark and 4-quark operators can modify top interactions. The sheer number of these operators, and the consequent need to break the degeneracies that arise, demands the consideration of the largest possible set of observables. 

We are particularly fortunate, in that respect, that almost 30 years after the discovery of the top quark, the remarkable performance of the LHC has moved the top sector of the SM in the realm of precision measurements. A broad campaign is underway involving the study of processes observed at the LHC for the first time: the production of tops in association with a vector boson, with a Higgs, and the simultaneous production of four tops. At the same time, the luminosity and energy reach of the LHC allows precise differential measurements of the main top production channels, both $t \bar t$ and and single tops, in several final state configurations.

A particularly interesting class of measurements is those focusing on the top spin. Top spin measurements are experimentally very challenging. The first measurement of top spin, establishing the presence of spin correlation in top pairs, was completed in 2012 \cite{ATLAS:2012ao}. Subsequent studies at $\sqrt{s} = 7 \, \text{TeV}$ \cite{ATLAS:2014aus}, $8 \, \text{TeV}$ \cite{CMS:2015cal, CMS:2016piu, ATLAS:2016bac}, and $13 \, \text{TeV}$ \cite{CMS:2019nrx, ATLAS:2019hau, ATLAS:2019zrq} considered more refined observables, and included several different top decay topologies. However, to this date, only one experimental measurement of the full $t \bar t$ spin density matrix, that completely characterizes top spin polarization and $t \bar t$ spin correlations, has been published \cite{CMS:2019nrx}. Furthermore, all the existing experimental measurements were performed inclusively in  the $t \bar t$ phase space; as we will argue in the following, this is a major limitation, as all top spin degrees of freedom exhibit a complex structure in phase space, that is largely washed out if no binning is applied.

Top spin correlation measurements can also be studied in the framework of pure Quantum Mechanics (QM). As first noted in \cite{Afik:2020onf}, sufficiently strong top spin correlations can signal the presence of quantum entanglement between top quarks. Entanglement, arguably the most essential and counter-intuitive prediction of QM, has been detected experimentally in a variety of physical systems, but never at the TeV scale. As argued in \cite{Afik:2020onf}, a simple selection based on the $t \bar t$ invariant mass will be enough to observe entanglement at the LHC. The structure of spin entanglement in $t \bar t$ pairs has since been studied multiple times, both in the SM and in the SMEFT \cite{Afik:2022kwm, Aguilar-Saavedra:2022uye, Aoude:2022imd}. When entanglement between two particles is sufficiently strong, measurements can be set up so as to violate Bell Inequalities (BI), the ultimate signal of non-classical behavior. A violation of BIs signals correlations so strong that no classical theory, even with arbitrary and hidden degrees of freedom, can explain. Top spin correlations grow with the transverse momentum of tops so that, at extreme $p_T$, BIs are violated at parton level. A first analysis \cite{Fabbrichesi:2021npl} suggested a violation of BIs may be observed at the LHC. Further studies \cite{Severi:2021cnj} suggested that the experimental verification of BIs will likely be very challenging, and if no new physics is present to enhance spin correlations, the detection of a violation may just escape the reach of the LHC. The investigation of possible BI measurements with top spin in colliders is ongoing, with progress having been reported recently \cite{Aguilar-Saavedra:2022uye}. Other collider processes have also been investigated \cite{Barr:2021zcp,Barr:2022wyq,Aguilar-Saavedra:2022wam,Aguilar-Saavedra:2022mpg,Ashby-Pickering:2022umy,Fabbrichesi:2022ovb}.

    Standard Model predictions for the observables with which experiments can probe all the $t \bar t$ spin degrees of freedom are available at NLO QCD + EW \cite{Bernreuther:2015yna} and NNLO QCD \cite{Czakon:2020qbd} accuracy. Before this work, SMEFT corrections to these observables were available at LO in QCD and LO in $c/\Lambda^2$ \cite{Bernreuther:2015yna}. Quadratic SMEFT corrections for a selection of dimension-6 operators were computed point-by-point in $t \bar t$ phase space in \cite{Aoude:2022imd}, for the two LO QCD channels $gg \to t \bar t$ and $q \bar q \to t \bar t$.

    In this work we will systematically consider the effect of dimension-six operators on $t \bar t$ entanglement, top spin polarization and $t \bar t$ spin correlation observables, at the $c/\Lambda^2$ and $c^2/\Lambda^4$ level, while also including higher-order QCD corrections.

    This paper is organized as follows. In Section \ref{sec:EFT} we describe the SMEFT operators that enter $t \bar t$ production at the LHC. In Section \ref{sec:spin_obs} we introduce the relevant top spin observables and show how these can be extracted from collider data.  Section \ref{sec:calc} describes in detail the numerical computation we carried. Results are presented in Section \ref{sec:results} at the inclusive and differential level. In Section \ref{sec:constraints} we estimate the improvement to global fits caused by the inclusion of spin observables, once more experimental results will become available in the near future.  We conclude in Section \ref{sec:conclusion}. Finally Appendices \ref{app:inclusive} and \ref{app:binned} contain additional numerical results and plots.

\section{SMEFT operators in top pair production} \label{sec:EFT}

In this section we present the relevant operators entering top pair production at dimension-6.  Within this work, we will assume conservation of baryon number and CP invariance. To keep under control the otherwise unmanageable number of four-quark operators, a flavor assumption must be put in place. Several have been proposed, and are studied in detail in \cite{Aguilar-Saavedra:2018ksv}. We will assume the flavor symmetry as in \cite{Degrande:2020evl}:
\begin{equation}
	U(2)_Q \times U(2)_u \times U(3)_d, \label{symm}
\end{equation}
	that implies new physics couples predominantly to top quarks. This is rooted on the unusually large value of the SM top Yukawa coupling $y_t \approx 0.993$, signaling a special relationship between the two heaviest SM particles, and hinting at the presence of new physics close to the electroweak scale.
	
We focus on operators containing at least one top quark field, assuming other operators are well enough constrained from measurements not involving tops. All such operators are considered, in a systematic approach, with the exception of those that enter top production only in virtual corrections, and those only entering in highly suppressed contributions (e.g. $b \bar b \to t \bar t$). 

	\subsection{Two-fermion operators}

	One can write down three dimension-six operators modifying the top-gluon interaction and thus $t \bar t$ production. To highlight the chiral properties of the induced $t \bar t g$ vertex these can be written as:
	\begin{align}
		\mathcal O_{gt} &= \overline t \, T_A \gamma^\mu  D^\nu t \, G_{\mu \nu}^A, \\
		\mathcal O_{gQ} &= \overline Q T_A \gamma^\mu  D^\nu Q G_{\mu \nu}^A, \\
		 \mathcal O_{tG} &= g_S \, \overline Q T_A \tilde \varphi \sigma^{\mu \nu} t \, G_{\mu \nu}^A = g_S \, \mathcal Q_{uG, 33} .\label{otg}
	\end{align}
	The second equality is in terms of the Warsaw basis \cite{Grzadkowski:2010es}. Once the additional symmetry \eqref{symm} and the EOM are enforced, only $\mathcal O_{tG}$ and $\mathcal O_{tG}^\dagger$ survive, as standard manipulations yield:
	\begin{align}
		\mathcal O_{gt} + \mathcal O_{gt}^\dagger  &= g_s \Big( \text{four-quarks} \Big) \\
	\mathcal O_{gQ} + \mathcal O_{gQ}^\dagger  &= g_s \Big( \text{four-quarks} \Big) \\
		\mathcal O_{gt} - \mathcal O_{gt}^\dagger &= \mathcal O_{gQ}^\dagger - \mathcal O_{gQ}  = \frac{y_t}{g_S} (\mathcal O_{tG} - \mathcal O_{tG}^\dagger)
	\end{align}
   This shows that $\mathcal O_{gt}$ and $\mathcal O_{gQ}$ reduce to four-fermion operators; more specifically, they reduce to operators composed of two light quarks and two tops, that will be considered in the following, and to operators with four top quark fields, that we will neglect due to their large suppression in the process at hand. The only surviving operator, $\mathcal O_{tG}$, has the effect of deforming the top quark color charge distribution, producing a chromo-magnetic and chromo-electric dipole moment proportional to $\Re(c_{tG})$ and $\Im(c_{tG})$ respectively.

    On top of the two-fermion operators listed above, there are five operators involving two fermion fields and electroweak gauge bosons that are in principle relevant for top physics,
    \begin{align}
        \mathcal Q_{\varphi u, \, 33} &= i (\varphi^\dagger \stackrel{\leftrightarrow}{D}_\mu \varphi)(\overline t \gamma^\mu t), \\
	    \mathcal Q_{\varphi q, \, 33}^{(1)} &= i (\varphi^\dagger \stackrel{\leftrightarrow}{D}_\mu \varphi)(\overline Q \gamma^\mu Q),\\	 \mathcal Q_{\varphi q, \, 33}^{(3)} &= i (\varphi^\dagger \stackrel{\leftrightarrow}{D}_\mu \sigma_I \varphi)(\overline Q \gamma^\mu \sigma^I Q), \label{ew1} \\
		\mathcal Q_{uB, \, 33} &= (\overline Q \sigma^{\mu \nu} t) \tilde \varphi B_{\mu \nu},  \\
		\mathcal Q_{uW, \, 33} &= (\overline Q \sigma^{\mu \nu} \sigma_I t) \tilde \varphi W_{\mu \nu}^I \label{ew2}.
	\end{align}
        
    Both at $\mathcal O(c/\Lambda^2)$ and $\mathcal O(c^2/\Lambda^4)$, these operators only enter electroweak top production, and since this process is heavily suppressed with respect to QCD production, their effect will be neglected. An NLO computation including both QCD and EW corrections would be required to fully take into account these contributions. 
    
    We note that $\mathcal Q_{\varphi q, \, 33}^{(3)}$ and $\mathcal Q_{uW, \, 33}$ also induce a modified $t W b$ vertex. Since spin measurements involve both the production and the decay of the particle at hand, operators that modify the decay of tops enter our study. In the following we will show that their actual effect on the spin observables we consider is negligible.

	\subsection{Four-fermion operators}

	There are seven independent dimension-6 operators mediating $t \overline t$ production by $q \overline q$ annihilation that result in a color-octet final state. These operators are constructed with a combination of four quark fields, two light ($u, d, c, s$) and two tops. Following the LHC Top WG notation and convention \cite{Aguilar-Saavedra:2018ksv}, we express SMEFT operators in a basis that highlights their chiral structure:
	\begin{align}
		 \mathcal O_{tu}^8 &= \textstyle \sum_{\text{f} = 1}^2 (\overline t \gamma_\mu T^A t)(\overline u_{\text{f}} \gamma^\mu T_A u_{\text{f}}) = \textstyle \sum_{\text{f} = 1}^2 - \frac 1 6 \mathcal Q_{uu, \text{\text{ff33}}} + \frac 1 2 \mathcal Q_{uu, \text{3ff3}},  \label{our4q_1} \\
		\mathcal O_{td}^8  &=\textstyle \sum_{\text{f} = 1}^3 (\overline t \gamma_\mu T_A t)(\overline d_{\text{f}} \gamma^\mu T^A d_{\text{f}}) = \textstyle \sum_{\text{f} = 1}^3 \mathcal Q_{ud, \text{33ff}}^{(8)}, \\ 
		\mathcal O_{tq}^8  &=\textstyle \sum_{\text{f} = 1}^2 (\overline q_{\text{f}} \gamma_\mu T_A q_{\text{f}})(\overline t \gamma^\mu T^A t) = \textstyle \sum_{\text{f} = 1}^2 \mathcal Q_{qu, \text{ff33}}^{(8)}, \\
		\mathcal O_{Qu}^8  &=\textstyle \sum_{\text{f} = 1}^2 (\overline Q \gamma_\mu T_A Q)(\overline u_{\text{f}} \gamma^\mu T^A u_{\text{f}}) = \textstyle \sum_{\text{f} = 1}^2 \mathcal Q_{qu, \text{33ff}}^{(8)} , \\ 
		\mathcal O_{Qd}^8  &= \textstyle \sum_{\text{f} = 1}^3 (\overline Q \gamma_\mu T_A Q)(\overline d_{\text{f}} \gamma^\mu T^A d_{\text{f}}) = \textstyle \sum_{\text{f} = 1}^3 \mathcal Q_{qd, \text{33ff}}^{(8)} , \\
		 \mathcal O_{Qq}^{1,8}  &=\textstyle \sum_{\text{f} = 1}^2 (\overline Q \gamma_\mu T^A Q)(\overline q_{\text{f}} \gamma^\mu T_A q_{\text{f}}) =\textstyle \sum_{\text{f} = 1}^2 \frac 1 4 \mathcal Q_{qq, \text{3ff3}}^{(1)} - \frac 1 6 \mathcal Q_{qq, \text{33ff}}^{(1)} + \frac 1 4 \mathcal Q_{qq, \text{3ff3}}^{(3)}, \\
		 \mathcal O_{Qq}^{3,8} &=\textstyle \sum_{\text{f} = 1}^2 (\overline Q \gamma_\mu T^A \sigma_I Q)(\overline q_{\text{f}} \gamma^\mu T_A \sigma^I q_{\text{f}}) =\textstyle \sum_{\text{f} = 1}^2 \frac 3 4 \mathcal Q_{qq, \text{3ff3}}^{(1)}  - \frac 1 6 \mathcal Q_{qq, \text{33ff}}^{(3)} - \frac 1 4 \mathcal Q_{qq, \text{3ff3}}^{(3)} . \label{our4q_7}
	\end{align}
	The second equality in \eqref{our4q_1} -- \eqref{our4q_7} again refers to the expression in terms of the Warsaw basis \cite{Grzadkowski:2010es}. The operators \eqref{our4q_1} -- \eqref{our4q_7} can also be considered in their color-singlet version:
	\begin{align}
		 \mathcal O_{tu}^1 &= \textstyle  \sum_{\text{f} = 1}^2 (\overline t \gamma_\mu t)(\overline u_{\text{f}} \gamma^\mu u_{\text{f}}) =\textstyle \sum_{\text{f} = 1}^2 \mathcal Q_{uu, \text{ff33}},  \label{our4q1_1} \\
		\mathcal O_{td}^1  &=\textstyle \sum_{\text{f} = 1}^3 (\overline t \gamma_\mu t)(\overline d_{\text{f}} \gamma^\mu d_{\text{f}}) =\textstyle \sum_{\text{f} = 1}^3 \mathcal Q_{ud, \text{33ff}}^{(1)} , \\ 
		\mathcal O_{tq}^1  &= \textstyle \sum_{\text{f} = 1}^2 (\overline q_{\text{f}} \gamma_\mu q_{\text{f}})(\overline t \gamma^\mu t) =\textstyle \sum_{\text{f} = 1}^2 \mathcal Q_{qu, \text{ff33}}^{(1)} , \\
		\mathcal O_{Qu}^1  &=\textstyle \sum_{\text{f} = 1}^2 (\overline Q \gamma_\mu Q)(\overline u_{\text{f}} \gamma^\mu u_{\text{f}}) =\textstyle \sum_{\text{f} = 1}^2 \mathcal Q_{qu, \text{33ff}}^{(1)}, \\ 
		\mathcal O_{Qd}^1  &=\textstyle \sum_{\text{f} = 1}^3 (\overline Q \gamma_\mu Q)(\overline d_{\text{f}} \gamma^\mu d_{\text{f}}) =\textstyle \sum_{\text{f} = 1}^3 \mathcal Q_{qd, \text{33ff}}^{(1)} , \\
		 \mathcal O_{Qq}^{1,1}  &=\textstyle \sum_{\text{f} = 1}^2 (\overline Q \gamma_\mu Q)(\overline q_{\text{f}} \gamma^\mu q_{\text{f}}) = \textstyle\sum_{\text{f} = 1}^2 \mathcal Q_{qq, \text{33ff}}^{(1)} , \\
		 \mathcal O_{Qq}^{3,1} &=\textstyle \sum_{\text{f} = 1}^2 (\overline Q \gamma_\mu \sigma_I Q)(\overline q_{\text{f}} \gamma^\mu \sigma^I q_{\text{f}}) =\textstyle \sum_{\text{f} = 1}^2 \mathcal Q_{qq, \text{ff33}}^{(3)} . \label{our4q1_7}
	\end{align}
	Again the second equality in \eqref{our4q1_1} -- \eqref{our4q1_7} refers to the expression in terms of the Warsaw basis \cite{Grzadkowski:2010es}. When considering only the leading-order SM QCD amplitudes for the Standard Model, these operators effectively enter top production at the $1/\Lambda^4$ level, since the $\mathcal O(1/\Lambda^2)$ interference arises only with the suppressed electroweak SM production $q \bar q \, \to \, Z/\gamma \, \to \, t \bar t$, or with QCD production at NLO.

\section{Spin of top quarks} \label{sec:spin_obs}

    Top quark pairs are unique candidates for spin polarization and spin correlation measurements. Top production at hadron colliders involves top quarks with negligible individual spin polarization, but large pair spin correlation with an elaborate phase-space structure. In fact, with appropriate cuts, it has been argued that spin correlations become so strong that quantum entanglement can be detected between top pairs \cite{Afik:2022kwm} and, in principle, the violation of Bell inequalities can also be observed \cite{Fabbrichesi:2021npl, Severi:2021cnj, Aguilar-Saavedra:2022uye}.	Furthermore, the spin de-correlation time from non-perturbative strong effects, of order $m_t \, \Lambda^{-2}_{\text{\tiny QCD}}$, is much larger than the top lifetime, so top quarks decay before their initial spin state is lost in the environment. In fact, since the hadronization timescale, $\Lambda_{\text {\tiny{QCD}}}^{-1}$, is also about one order of magnitude larger than the top lifetime, tops behave like free, spinning particles.  
    
	The spin state of a $t \bar t$ pair is described by the density matrix:
	\begin{equation}
		\rho = \frac 1 4 \Big( \mathbbm 1 \otimes \mathbbm 1 + \sum_{i = 1}^3 B_i \, \sigma_i \otimes \mathbbm 1 + \sum_{i = j}^3 \bar B_j \, \mathbbm 1 \otimes \sigma_j +  \sum_{i = 1}^3 \sum_{j = 1}^3 C_{ij} \, \sigma_i \otimes \sigma_j \Big), \label{rho}
	\end{equation}
	where the first term in each tensor product refers to the top and the second term to the anti-top. The parameters entering \eqref{rho} have the physical interpretation of being the expectation values of individual spins and spin correlations,
	\begin{equation}
		\langle S_i \rangle = B_i, \quad \langle \bar S_i \rangle = \bar B_j, \quad \langle S_i \bar S_j \rangle = C_{ij}.
	\end{equation}
	
    The parameters entering \eqref{rho} have well defined C, P, and CP transformation properties \cite{Bernreuther:2015yna}, listed in Table \ref{tab:obs}. In particular, the linear combinations $B_i - \bar B_i$ and $C_{ij} - C_{ji}$ are CP violating, and will not be considered further in this work.

\paragraph{Spin analyzing power.}$ \ $ \smallskip

The measurement of the $t \, \bar t$ spin state may be considered to be very challenging, as measuring a particle's spin traditionally requires careful measurements of its trajectory in a rapidly changing magnetic field. 
However, provided that the particle one is interested in decays electroweakly, and that its decay products are fully recovered, the reconstruction of the spin state becomes experimentally possible even in the difficult environment of a hadron collider. In fact, thanks to the fully chiral nature of weak interactions, the momenta of daughters $X = b, W, \ell, q, \nu$ emerging from the decay of tops are correlated with the spin of the initial top, with the decay width given at LO by:
	\begin{equation}
		\frac{1}{\Gamma} \frac{d\Gamma}{d \cos \theta_X} = \frac{1 + \alpha_X \cos \theta_X}{2}\,, \label{gamma}
	\end{equation}
	where $\alpha_X$ is a parameter known as {\it spin analyzing power} of particle $X$, and $\theta_X$ is the angle between the original top spin and the direction of the emitted $X$ in the top rest frame. 

 Assuming for concreteness that $\alpha_X > 0$, the direction of flight of particle $X$ then follows a cosine distribution around the initial top's spin, with the most likely trajectory being aligned to the spin itself, and the least likely being opposite to it.  As a result of this effect, individual decay products can be considered as proxies for the spin of the corresponding top quarks, and correlations between different decay products as proxies for those between the top quark spins.

	At leading order in the SM, the spin analyzing power of prompt $W$ bosons and charged leptons emerging from the $W$ decay is given by:
	\begin{align}
		&\alpha_W = \frac{m_t^2 - 2  m_W^2 - m_b^2}{(m_b^2 - m_t^2)^2 + (m_b^2 + m_t^2)  m_W^2 - 2  m_W^4} \ \times \nonumber \\
		    &\hspace{5 mm} \times \sqrt{(m_b - m_t -  m_W)(m_b + m_t -  m_W)(m_b - m_t +  m_W)(m_b + m_t +  m_W)} \ \approx \ 0.394,\\
		&\alpha_\ell = 1.
	\end{align}

    It is curious that the charged lepton has a larger spin-analyzing power than its mother, the $W$ boson. This is due to the constructive and destructive interference between amplitudes with intermediate $W$ bosons of different helicities; this information is lost when considering the direction of flight of the prompt $Wb$ pair. 

	The presence of new physics in the top electroweak couplings, \eqref{ew1} and \eqref{ew2}, can affect the value of $\alpha$ \cite{Zhang:2010dr}. The effect of $\mathcal Q_{\varphi q, \, 33}^{(3)}$ is to rescale the SM value of the $tWb$ vertex, with no effect on angular distributions, and thus on $\alpha$. On the other hand, $\mathcal Q_{uW, \, 33}$ can alter spin analyzing powers. The spin analyzing power of the $Wb$ pair is modified at the $c_{uW, \, 33} v^2 / \Lambda^2$ level, as expected from dimensional analysis. However, interestingly, when considering the full decay $t \to b \, \ell \, \nu_\ell$ the on-shell conditions for final-state leptons remove the linear (and cubic) contribution, and we find:
	\begin{align}
		\alpha_\ell(c_{uW, \, 33}) &= 1 - \frac{c_{uW, \, 33}^2 v^4}{\Lambda^4} \frac{4 (2 m_t^6 + 3 m_t^4 m_W^2 - 6 m_t^2 m_W^4 + m_W^6 + 12 m_t^4 m_W^2  \log m_W/m_t)}{(m_W^2 - m_t^2)^2 (m_t^2 + 2 m_W^2)}, \label{alpha_ell_ctw}
	\end{align}	
    up to corrections of order $c_{uW, \, 33}^4 v^8 / \Lambda^8$ and $\Gamma_W^2/m_W^2$.     
    The absence of a linear term in \eqref{alpha_ell_ctw} is the result of delicate cancellations, resulting from the interplay of the equations of motion of the external fermions and the chiral structure of the interaction induced by $\mathcal O_{uW, 33}$. The cancellation is independent of $m_W$ and $m_b$, while it is spoiled by a finite $m_\ell$. This has been noted repeatedly in the literature \cite{Grzadkowski:2002gt, Aguilar-Saavedra:2006qvv}, published results are in agreement with our \eqref{alpha_ell_ctw}, when considering the different conventions.
    This result applies to both the positively and the negatively charged lepton. We note that the small pre-factor makes the $\mathcal Q_{uW, \, 33}$ contribution to the lepton spin analyzing power of order $10^{-3}$ or less, assuming recent bounds for its Wilson coefficient. This suggests the dilepton final state is an exceptionally clean channel to investigate new physics in top production, with little contamination from electroweak effects in the top decay.
	
	Higher order QCD corrections to $\alpha_W$ and $\alpha_\ell$ are known in the SM \cite{Czarnecki:1990pe, Brandenburg:2002xr}. NLO effects in the top decay can be reduced to a shift in $\alpha$, that we consider, and to very limited corrections to the functional relation \eqref{gamma}, that are neglected.

	\paragraph{Measurement of the spin state.} $ \ $ \smallskip

	The differential cross section for the process $p p \to t \bar t \to a b$, where $a$ is a decay product of the top quark and $b$ is a decay product of the anti-top quark can be expressed as \cite{Bernreuther:2015yna}:
	\begin{align}
		&\frac{1}{\sigma} \frac{d \sigma}{d\cos \theta_{i} } = \frac{1 + \alpha_a B_i \cos \theta_{i} }{2}, \quad \frac{1}{\sigma} \frac{d \sigma}{d\cos \bar \theta_{i} } = \frac{1 + \alpha_b \bar B_i \cos  \bar \theta_{i} }{2}  \label{dcos} \\
		&\frac{1}{\sigma} \frac{d \sigma}{d (\cos \theta_{i} \cos \bar \theta_{j})} = -\frac{1 + C_{ij} \alpha_a \alpha_b \, \cos \theta_{i} \cos \bar \theta_{j}}{2} \log \big|\cos \theta_{i} \cos \bar \theta_{j} \, \big|, \label{dcoscos}
	\end{align}
	where $\theta_i$ is the angle between $\vec{p}_a$ and the $i$-th axis in the top rest frame, and $\bar \theta_j$ the angle between $\vec{p}_b$ and the $j$-th axis in the anti-top rest frame. 

	A suitable reference frame must be defined in order to evaluate the angles $\theta_i$, $\bar \theta_j$. An advantageous choice is the {\it helicity basis} $\lbrace \hat k, \hat r, \hat n \rbrace$, defined in the $t \bar t$ pair reference frame as: 
	\begin{equation}
		\hat k = \text{top direction}, \quad	\hat r = \frac{ \hat p - \hat k \, \cos \theta}{\sin \theta}, \quad \hat n = \frac{\hat p \times \hat k}{\sin \theta}	 \label{helbasis}
	\end{equation}
	 where $\hat p$ is the beam axis and $\theta$ is the top/anti-top scattering angle in the $t \bar t$ center of mass frame. There is a sign ambiguity in choosing the beam direction $\hat p$; for every event, $\hat p$ is chosen so that:
	\begin{equation}
		0 \leq \theta \leq \frac{\pi}{2}.
	\end{equation}	
	
    The integration of \eqref{dcos} and \eqref{dcoscos} gives explicit relations to measure the spin state:
	\begin{equation}
		B_i =  \big \langle \frac{3}{\alpha_a} \cos \theta_{i} \big \rangle, \quad \bar B_i = \big  \langle  \frac{3}{\alpha_b} \cos \bar \theta_{i} \big \rangle, \quad C_{ij} =  \big \langle \frac{9}{\alpha_a \alpha_b} \cos \theta_{i} \cos \bar \theta_{j} \big \rangle, \label{dcos_integrated}
	\end{equation}	
	where angle brackets denote the expectation value:
	\begin{equation}
	  S = \langle s \rangle = \frac{\int s \, |\mathcal M|^2 \, d\Pi}{\int |\mathcal M|^2 \, d\Pi} \equiv \frac{\sigma_S}{\sigma}.\label{xsec_ratio}
	\end{equation}

	The above relations show that an arbitrary spin observable $S$ can be extracted as a ratio of two cross sections: the numerator $\sigma_S$ is the phase-space integral of $s$, some dimensionless combination of angles and spin analysing powers, weighted with the matrix element squared, and the denominator is the total cross section. Further, as the various examples proposed in Ref.~\cite{Aguilar-Saavedra:2022uye} show, it is possible to construct combination of angles $s$ so that $\langle s \rangle$ is directly sensitive to several quantities of interest, e.g. $B_i \pm B_j$, $C_{ij} \pm C_{ji}$, $C_{ii} + |C_{jj} + C_{kk}|$.
	
    The observables used in this work, that are those commonly considered in spin measurements, are listed in Tab.~\ref{tab:obs}, together with their transformation properties under P and CP.  Standard Model predictions for the spin observables of Tab.~\ref{tab:obs} for $t \bar t$ pairs produced at the LHC are already available at NNLO QCD \cite{Czakon:2020qbd} and NLO QCD + EW \cite{Bernreuther:2015yna}. 
    
    \begin{table}[]
    \centering
    \begin{tabular}{|c|c||c|c|} \hline \hline
    \multicolumn{2}{c}{CP even} &  \multicolumn{2}{c}{CP odd} \\ \hline
    $B_k + \overline B_k$ & P-odd  & $B_k - \overline B_k$ & P-odd  \\
    $B_r + \overline B_r$ & P-odd  & $B_r - \overline B_r$ & P-odd  \\
    $B_n + \overline B_n$ & P-even  & $B_n - \overline B_n$ & P-even  \\
    $C_{kk}$ & P-even & &  \\
    $C_{rr}$ & P-even & & \\
    $C_{nn}$ & P-even & & \\
    $C_{kr} + C_{rk}$ & P-even & $C_{kr} - C_{rk}$ & P-even  \\
    $C_{kn} + C_{rn}$ & P-odd & $C_{kn} - C_{rn}$ & P-odd  \\
    $C_{rn} + C_{rn}$ & P-odd  & $C_{rn} - C_{rn}$ & P-odd  \\ \hline \hline
    \end{tabular}
    \captionsetup{width=\linewidth}
    \caption{Top spin polarization and $t \bar t$ spin correlation observables with their C and CP transformation properties. The CP-even observables are those considered in this work, the CP-even observables are needed in principle for a full reconstruction of the density matrix $\rho$, but they are not considered here.}
    \label{tab:obs}   
    \end{table}
  
  In the SMEFT, allowing at most one NP insertion in each Feynman diagram, cross sections are quadratic functions of the Wilson coefficients so that:
	\begin{equation}
		S = \frac{\sigma_{S}^{(0)} + \hat c \, \sigma_{S}^{(1)} + \hat c^2 \, \sigma_{S}^{(2)}}{\sigma^{(0)} + \hat c \, \sigma^{(1)} + \hat c^2 \, \sigma^{(2)}}, \qquad \hat c \equiv \frac{c}{\Lambda^2\,/\,(1 \, \text{TeV})^2}.  \label{ratio}
	\end{equation}	
	The expression in \eqref{ratio} can be series expanded around the SM value,
	\begin{equation}
		S = \frac{\sigma_{S}^{(0)}}{\sigma^{0}} + \hat c \> \frac{\sigma_{S}^{(1)} \, \sigma^{(0)} - \sigma_{S}^{(0)} \, \sigma^{(1)}}{(\sigma^{0})^2} + \cdots, \label{ratio_expanded}
	\end{equation}	
	so that a linear, quadratic, and so on, shift can be extracted. However, to maintain generality, we extract the six cross sections in \eqref{ratio} separately, so that the full dependence of $S$ on $c$ can be evaluated.
  
  Accurate SM predictions can be implemented in several ways, such as:
  \begin{itemize}[itemsep=0pt]
    \item[--] the SM cross sections $\sigma_{S}^{(0)}$ and $\sigma^{(0)}$ in \eqref{ratio} are replaced by their most accurate values, EFT terms are not modified, and then the ratio $S$ is taken; 
    \item[--]  numerator and denominator of \eqref{ratio} are globally rescaled by the SM K-factors $K_S = \sigma_{S, \text{accurate}}^{(0)} / \sigma_{S}^{(0)}$ and $K = \sigma_{\text{accurate}}^{(0)} / \sigma^{(0)}$, then and the ratio $S$ is taken; 
    \item[--] the ratio \eqref{ratio} is series expanded, the first term $\sigma_{S}^{(0)} / \sigma^{(0)}$ is replaced with the accurate value, other terms in the series are not modified -- even if they contain $\sigma_{S}^{(0)}$ or $\sigma^{(0)}$.
  \end{itemize}
  
    It is not clear which one, if any, of the prescriptions outlined above entails the best physical meaning, and the spread in results obtained by implementing accurate SM predictions in different ways may be considered to be a theoretical uncertainty. In this work we will report all six cross sections that enter \eqref{ratio}, leaving the possibility open for arbitrary implementations of accurate SM calculations. 

\paragraph{Quantum entanglement.} $ \ $ \smallskip

    As discussed in the introduction, and in more detail in Refs.~\cite{Afik:2020onf, Severi:2021cnj, Aoude:2022imd, Afik:2022kwm}, top spin correlations can signal the presence of quantum entanglement between tops. A calculation shows that, in the case of $t \bar t$, entanglement is signaled by the condition:
\begin{equation}
    \Delta = |C_{kk} + C_{rr}| - C_{nn} - 1 > 0. \label{entanglement}
\end{equation}
    Inequality \eqref{entanglement} is general, and independent of the possible presence of new physics.  We also note that the concurrence of the $t \bar t$ spin quantum state \eqref{rho} is given by:
\begin{equation}
    \mathcal C[\rho] = \begin{dcases} \Delta/2, & \Delta > 0 \\ 0 & \Delta \leq 0 \end{dcases}
\end{equation}
    The absolute value in the definition of $\Delta$ is non-trivial, as, already in the SM, at $t \bar t$ threshold $C_{kk} + C_{rr} < 0$, so $\Delta = -\text{Tr} \, C - 1$, while at large top $p_T$, $C_{kk} + C_{rr} > 0$, so $\Delta$ becomes $C_{kk} + C_{rr} - C_{nn} - 1$. The use of $\text{Tr} \, C$ to detect entanglement at $t \bar t$ threshold has been the topic of extended studies \cite{Afik:2020onf, Afik:2022kwm}; in fact, as shown in Ref.~\cite{Aguilar-Saavedra:2022uye}, $\Delta$ can be measured directly using dedicated angular observables, regardless of the sign of the absolute value. To appreciate directly the effect of SMEFT operators on the amount of spin entanglement, in the following we will consider:
\begin{equation}
    \Delta^\pm = \pm(C_{kk} + C_{rr}) - C_{nn} - 1 = \frac{\pm(\sigma_{C_{kk}} + \sigma_{C_{rr}}) - \sigma_{C_{nn}} - \sigma}{\sigma}  \label{quantum_obs}
\end{equation}
    as additional observables, as an extension to those appearing in \eqref{rho}. We note that $\Delta^-$ is a well known top spin observable, usually called $D \equiv (\Delta^- + 1)/3$.

\section{Computation Setup} \label{sec:calc}

	Since the reconstruction of a spin quantum state requires the analysis of decay products, the analysis of top spin polarization and correlations at NLO would require the generation of events of the form:
	\begin{equation}
		p \ p \to b \, \bar b \, \ell^+ \, \ell^- \, \nu \, \bar \nu \label{processdilep}
	\end{equation}
	at order $\mathcal O(\alpha_{\text{s}}^3 \, \alpha_{\text{EW}}^4)$. This would fully include virtual corrections, and all off-shell effects for both tops and $W$ bosons.  This computation is extremely challenging, and as of now can only be handled by purpose-made tools, e.g. \cite{Campbell:2012uf,Campbell:2014kua},  that lack the versatility needed to include arbitrary SMEFT operators in the matrix element. 

	A useful proxy for the full NLO computation could be given by the process:
	\begin{equation}
		p \ p \to W^+ \, W^- \, b \, \bar{b} \label{processwbwb}
	\end{equation}
	at order $\mathcal O(\alpha_{\text{s}}^3 \, \alpha_{\text{EW}}^2)$, since the prompt $W \, b$ pair emitted by the electroweak decay of tops also has spin analyzing power. For such a computation, suitable phase-space cuts help isolate the doubly resonant contribution $p \ p \to t \, \bar t \to W^+ \, W^- \, b \, \bar{b}$.  Generally speaking, one should cluster parton-level QCD radiation in jets, impose a selection of standard jet cuts, and ask that the invariant mass of the two $W$ + $b$-jet pairs is close to $m_t$. The full SM amplitude for \eqref{processwbwb} has been evaluated at NLO in \cite{Denner:2010jp, Bevilacqua:2010qb}, but the significant complexity of SMEFT amplitudes prevented us from extracting precise enough results for this process. 

	To obtain our results we combine a NLO fixed-order calculation for the process $p \ p \to t \, \bar t$ and the modeling of the electroweak decays $t \to b \, \ell^+ \, \nu$ and $\bar t \to \bar b \, \ell^- \, \bar \nu$ with {\tt MadSpin} \cite{Artoisenet:2012st}. This yields predictions whose accuracy is summarized in Table \ref{tab:madspin}. Accuracy is NLO QCD, up to virtual corrections to the spin state, that can not be handled by {\tt MadSpin}. However, as the kinematic state is constructed before the spin state, and the top quarks' momentum is correlated with their spin\footnote{See for instance the argument given in \cite{Severi:2021cnj} for top pairs emitted at large $p_T$.}, the full-NLO accuracy of the $t \bar t$ kinematics improves the partial-NLO accuracy of the spin state. It is also true that in the SM virtual corrections to the $t \bar t$ spin state amount to a percent level shift for most of the observables we consider \cite{Bernreuther:2015yna, Frederix:2021zsh}. The only exception is $C_{rr}$, for which, given its small value in the SM, the relative correction is $\sim \, 25\%$. We thus consider it reasonable to assume that the impact of virtual corrections remains under control also in the SMEFT.
	
	\hspace{5mm}
	
	\begin{table}[h]
	\centering
    \begin{tabular}{cccc} \hline
    & \hspace{2mm} LO QCD \hspace{2mm} & \multicolumn{2}{c}{NLO QCD} \\
    & & {\small Real emission} & {\small Virtual corrections} \\ \hline
    $t \, \bar t$ kinematics & \checkmark & \checkmark & \checkmark  \\ \hline
    $t \, \bar t$ spin state & \checkmark & \checkmark & {\small Approximate} \\ \hline
    \end{tabular}
    \captionsetup{width=\textwidth}
    \caption{Accuracy of the computation resulting from {\tt Madgraph5\_aMC@NLO} and {\tt MadSpin}.}
    \label{tab:madspin}
    \end{table}

	Hard events featuring a top pair production are generated with {\tt Madgraph5\_aMC@NLO} v3.4.0 at fixed-order NLO QCD accuracy, $\mathcal O(\alpha_s^3 \alpha_{\text{EW}}^0)$.   Electroweak corrections, known to amount to a percent-level correction to the total cross section \cite{Bernreuther:2006vg}, and to very small shifts in the spin observables \cite{Bernreuther:2015yna}, are not considered. The hadronic center of mass energy is set to $\sqrt s = 13.0$ TeV, corresponding to the second run of the LHC.  We use the {\tt NNPDF3.1} parton distribution function \cite{NNPDF:2017mvq} at NLO QCD with $\alpha_s(m_Z) = 0.118$.  The dataset used to extract this PDF does not include any top spin data. The SM renormalization and factorization scales $\mu_\text R$ and $\mu_\text F$ are set on an event-by-event basis to one fourth of the total transverse energy $H_T$ of the process, and the SMEFT renormalization scale $\mu_\text{EFT}$ is set to the top mass. Events are generated within the {\tt SMEFT@NLO} model \cite{Degrande:2020evl}, whose flavor symmetry is \eqref{symm}. Note that the flavor symmetry implies that light SM particles, including the $b$ quark, are massless, and the CKM matrix is the identity. Other SM input parameters are set to their current experimental values. The {\tt SMEFT@NLO} model is restricted to only contain the operators of interest, $\mathcal O_{tG}$ and the four-quark operators \eqref{our4q_1} -- \eqref{our4q_7} and  \eqref{our4q1_1} -- \eqref{our4q1_7}, one at a time. Top quarks are kept on shell.  
	
	The leptonic decay of top quarks $t \to b \, \ell^+ \, \nu$, $\bar t \to \bar b \, \ell^- \, \bar \nu$ is modeled by {\tt MadSpin} \cite{Artoisenet:2012st}, adapted to handle fixed-order calculations with minor improvements to the technique described in \cite{Frederix:2021zsh}. Spin observables are extracted at parton level using charged leptons emerging from the electroweak decay, integrating the expressions that enter in the numerator and denominator of \eqref{xsec_ratio} separately. The individual contributions entering \eqref{ratio} are then extracted by varying the Wilson coefficients and fitting a quadratic dependence. 

    The sources of theoretical uncertainty we consider are MC statistical uncertainty and scale variation. Statistical uncertainty is estimated with standard procedures, scale variation is evaluated from the envelope of the nine points $\mu_R = (H_T/8, \, H_T/4, \, H_T/2)$, $\mu_{\text{f}} = (H_T/8, \, H_T/4, \, H_T/2)$, $\mu_\text{EFT} = (m_t)$.

\paragraph{Validations.}$ \ $ \smallskip

    Before moving to NLO accuracy, we validate our setup at LO. We compare the SM $t \bar t$ spin density matrix we obtain with {\tt MadSpin} with the one obtained from the benchmark process \eqref{processdilep}, that at LO is feasible. In addition, we compare results with and without the narrow width approximation for tops and W bosons. All our checks show deviations at the $10^{-3}$ level, well below current (and presumably future) experimental accuracy, and other sources of theoretical uncertainty. The SMEFT effects on the $t \bar t$ spin density matrix of the two-fermion operator \eqref{otg} and of the four-fermion color-octet operators \eqref{our4q_1} - \eqref{our4q_7} were computed at LO QCD and LO in $c/\Lambda^2$ in \cite{Bernreuther:2015yna}. After accounting for the different choice of operators, we are able to replicate the results found in \cite{Bernreuther:2015yna}, up to relative deviations of order $10 \%$. We note that computational setup in \cite{Bernreuther:2015yna} features a different PDF, different choices of factorization and renormalization scales, and different numerical input parameters with respect to the one presented in this work, so perfect numerical agreement is not expected. However, the pattern of deviations from the SM, and their relative size, is found to be in good agreement. To further check compatibility, we reproduced the exact setup of \cite{Bernreuther:2015yna} in a LO run with the operator $\mathcal O_{tG}$, and obtained agreement at the $10^{-4}$ level.

\section{Results and discussion} \label{sec:results}

	Our numerical integration is carried:
	\begin{itemize}[itemsep=0pt]
		\item Inclusive in $t \bar t$ phase space.
		\item Binned in $t \bar t$ phase space, with a $3 \times 3$ binning in the $t \bar t$ invariant mass $m_{t \bar t}$ and the center of mass scattering angle $\theta$. This binning will be updated to match the actual resolution of experimental measurements, once available.
	\end{itemize}
	    In any case, we do not impose any cuts or  detector acceptance effects on the final state $b \, \ell^+ \, \nu \, \bar b \, \ell^- \, \bar \nu$. The plots in this Section collect results for the fifteen SMEFT operators inclusive in phase space, and for selected operators binned in phase space, using the sample $m_{t \bar t} \, - \, \cos \theta$ binning we considered. Appendix \ref{app:inclusive} contains tables with all cross-sections needed to replicate our plots inclusive in phase space. Plots with binned results for operators not included in this Section are in Appendix \ref{app:binned}. 

	    To highlight the improvement offered by our NLO calculation over the LO, in Appendix \ref{app:inclusive} the cross-sections we obtained are provided separately at LO and NLO accuracy. The large $K$-factor of the total cross-section $\sigma$ makes the numerical comparison between LO and NLO not obvious. However, after taking the appropriate ratios, we find that, frequently, NLO results are appreciably different from LO ones. As expected, NLO results also tend to show a smaller scale variation.
	    
	    \subsection{Results inclusive in $t \bar t$ phase space}
	    
	    The following Figures \ref{fig:DIM62F_24} - \ref{fig:DIM64F_7_8} show the change with respect to the SM value, defined as:
    	\begin{equation}
    		\Delta S = \frac{\sigma_{S}^{(0)} + \hat c \, \sigma_{S}^{(1)} + \hat c^2 \, \sigma_{S}^{(2)}}{\sigma^{(0)} + \hat c \, \sigma^{(1)} + \hat c^2 \, \sigma^{(2)}} - \frac{\sigma_{S}^{(0)}}{\sigma^{(0)}}, \label{smdiff}
    	\end{equation}	    
	    for the top spin polarizations $B_i + \bar B_i$, the $t \bar t$ spin correlations $C_{ii}$ and $C_{ij} + C_{ji}$, and the entanglement markers $\Delta^\pm$, as a function of the Wilson coefficient of the operators we consider, inclusive in $t \bar t$ phase space. Similarly to \eqref{ratio}, in \eqref{smdiff} we defined $\hat c \equiv c \ (1 \, \text{TeV})^2 / \Lambda^2$, which is equivalent numerically to setting $\Lambda = 1 \, \text{TeV}$. For the purposes of plotting results, we employ our own NLO QCD SM predictions $\sigma_{S}^{(0)}$ and $\sigma^{(0)}$ in \eqref{smdiff}. Accurate SM predictions, as discussed above, may also be used in \eqref{smdiff}. However, as the leading SM contribution is subtracted, the actual numerical difference is negligible in the range we consider. While, as we will see in the following, an inclusive measurement hides most of the physical features present, we can still highlight several interesting points, presented in the rest of this Section.
    
    We begin by noting that quadratic terms $\mathcal O(c^2/\Lambda^4)$, that induce a non-linear behaviour of spin observables, are almost always important, and often leading with respect to the SM-SMEFT interference $\mathcal O(c/\Lambda^2)$. This is particularly pronounced for the color-singlet operators for which the deviations from the SM are nearly identical for positive and negative values of the coefficients. This is expected as this class of operators only interferes with the SM at NLO in QCD. For the color-octet operators both linear and quadratic terms are important for the range of coefficients that we consider.  We also notice that the effect of color-singlet operators is more prominent than the effect of color-octet operators, due to a relative color factor of $9/2$ entering the respective squared amplitudes.

\begin{figure}[H]
\centering

\begin{subfigure}{.5\textwidth}
  \centering
  \includegraphics[width=\linewidth]{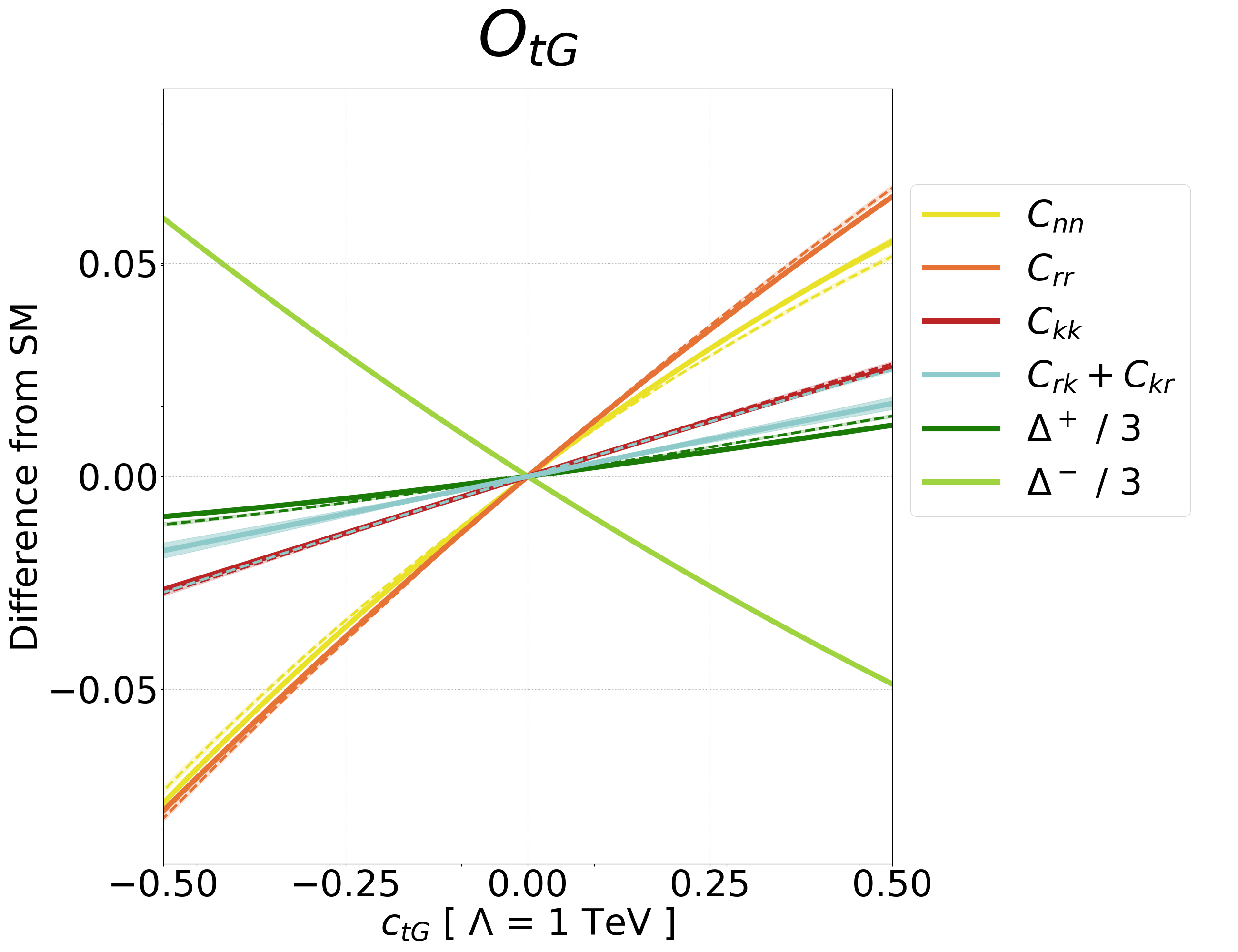}
  \subcaption{}
  \label{fig:DIM62F_1}
\end{subfigure}
\begin{subfigure}{.5\textwidth}
  \centering
  \includegraphics[width=\linewidth]{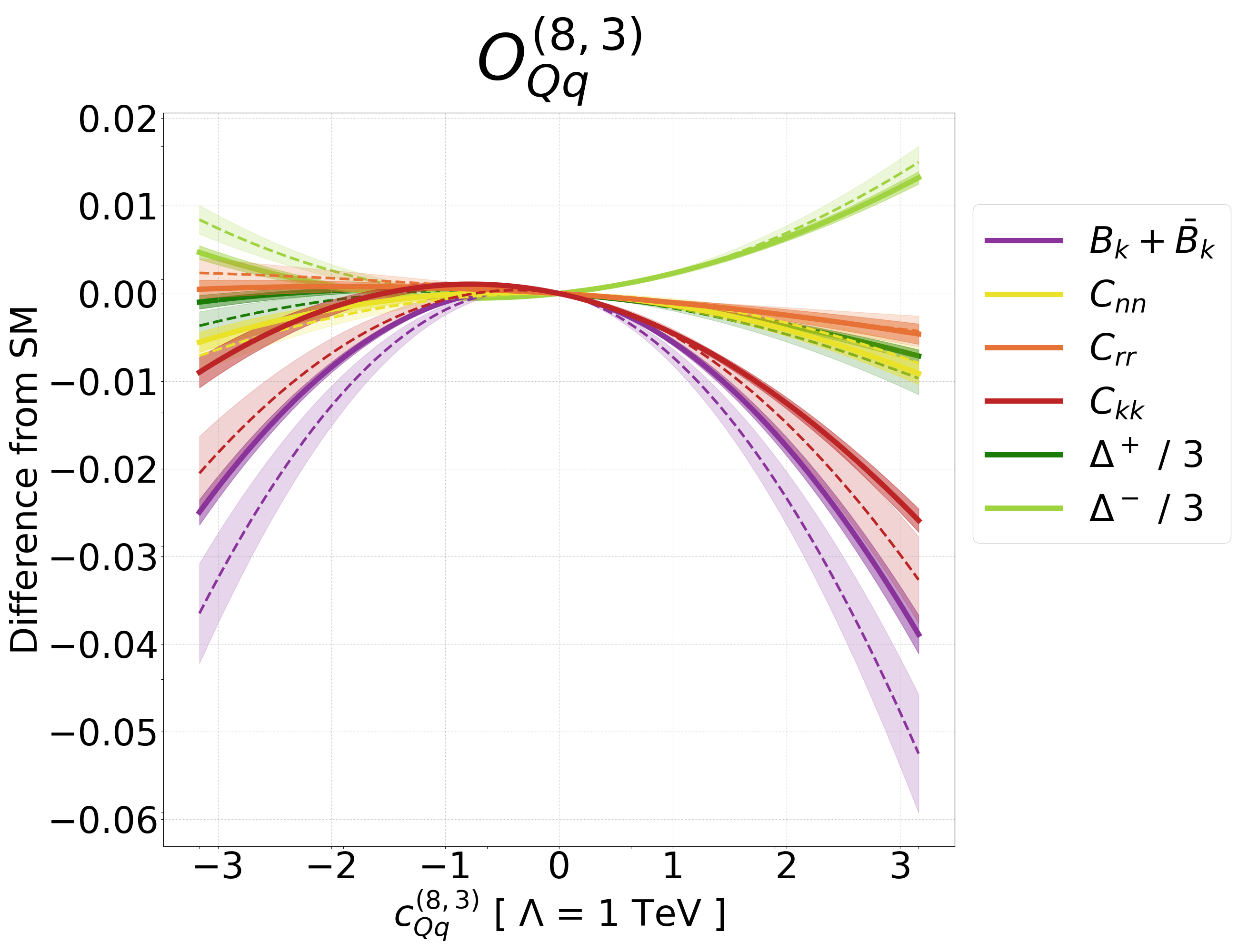}
  \subcaption{}
  \label{fig:DIM64F_1a}
\end{subfigure}%
\begin{subfigure}{.5\textwidth}
  \centering
  \includegraphics[width=\linewidth]{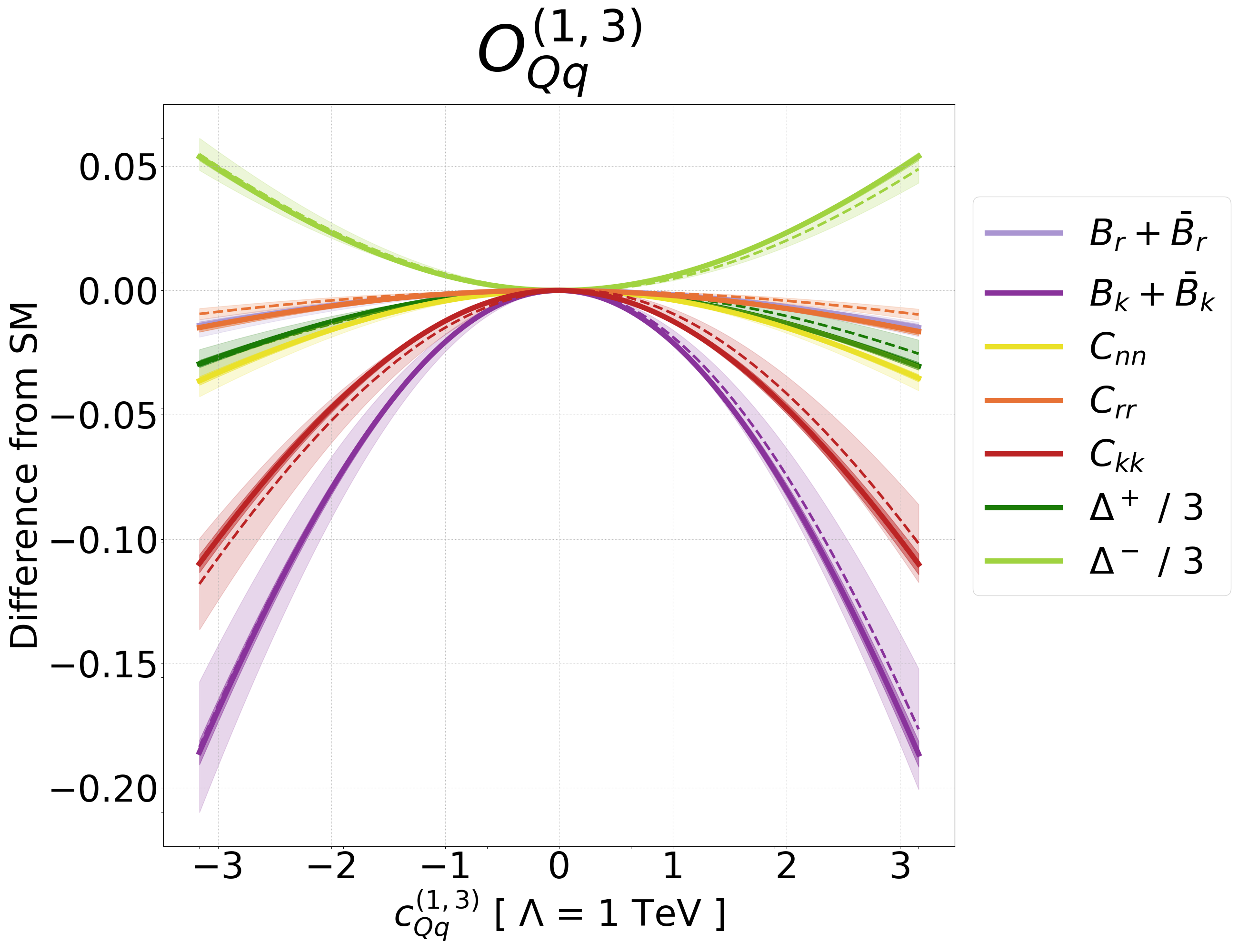}
  \subcaption{}
  \label{fig:DIM64F_1b}
\end{subfigure}
\begin{subfigure}{.5\textwidth}
  \centering
  \includegraphics[width=\linewidth]{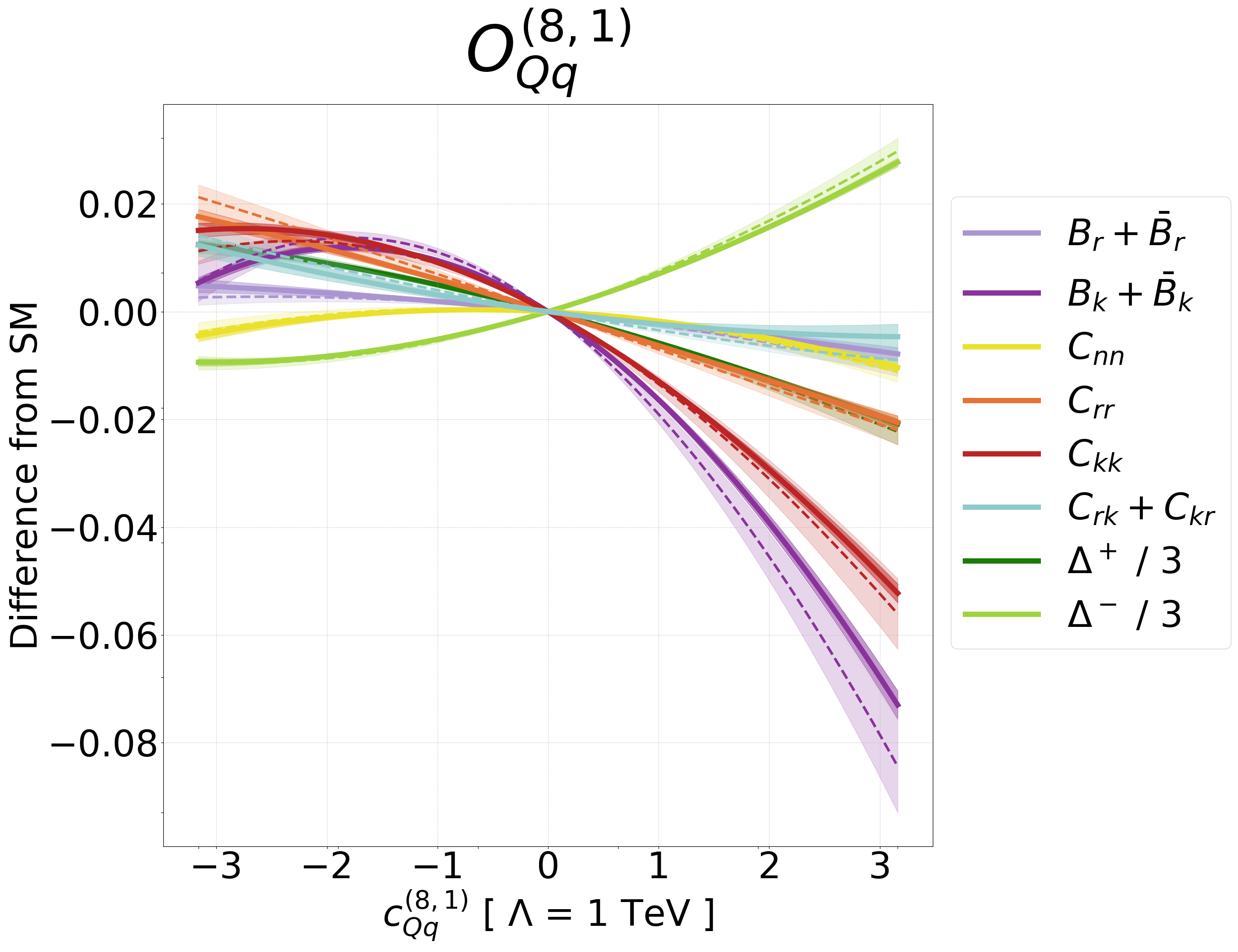}
  \subcaption{}
  \label{fig:DIM64F_2a}
\end{subfigure}%
\begin{subfigure}{.5\textwidth}
  \centering
  \includegraphics[width=\linewidth]{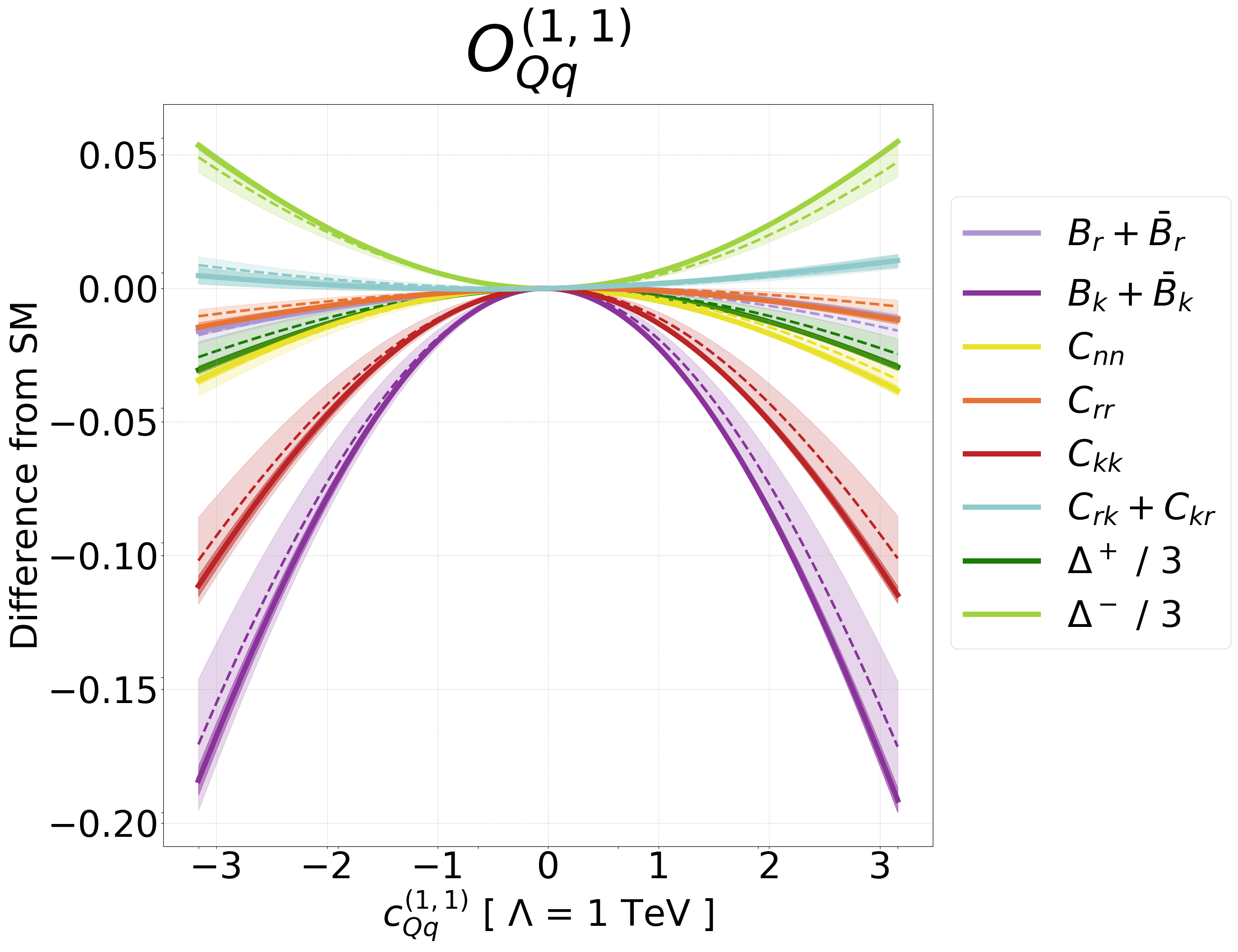}
  \subcaption{}
  \label{fig:DIM64F_2b}
\end{subfigure}

\captionsetup{width=\linewidth}
\caption{Change from the SM value for spin observables for the operators $\mathcal O_{tG}$ (\ref{fig:DIM62F_1}), $\mathcal O_{Qq}^{(8,3)}$ (\ref{fig:DIM64F_1a}), $\mathcal O_{Qq}^{(1,3)}$ (\ref{fig:DIM64F_1b}), $\mathcal O_{Qq}^{(8,1)}$ (\ref{fig:DIM64F_2a}), $\mathcal O_{Qq}^{(1,1)}$ (\ref{fig:DIM64F_2b}),  inclusive in $t \bar t$ phase space. Dashed lines indicate results at LO, continuous lines indicate NLO. The shaded region around each curve represents the combination of scale and MC uncertainty. The MC uncertainty is  always sub-leading compared to scale variation. Only curves that deviate appreciably from zero are shown.}
\label{fig:DIM62F_24}
\end{figure}

\begin{figure}[H]
\centering

\begin{subfigure}{.5\textwidth}
  \centering
  \includegraphics[width=\linewidth]{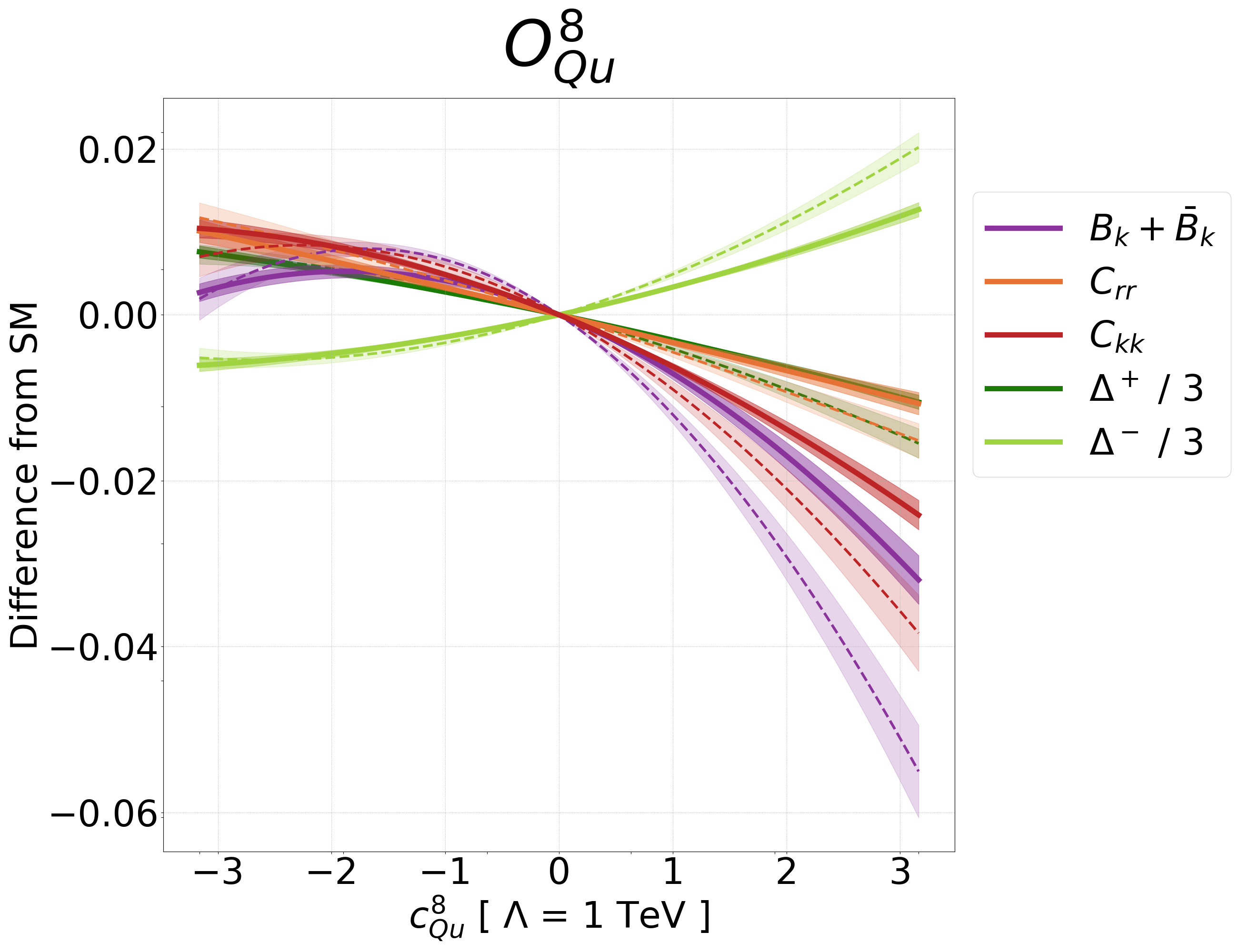}
  \subcaption{}
  \label{fig:DIM64F_3a}
\end{subfigure}%
\begin{subfigure}{.5\textwidth}
  \centering
  \includegraphics[width=\linewidth]{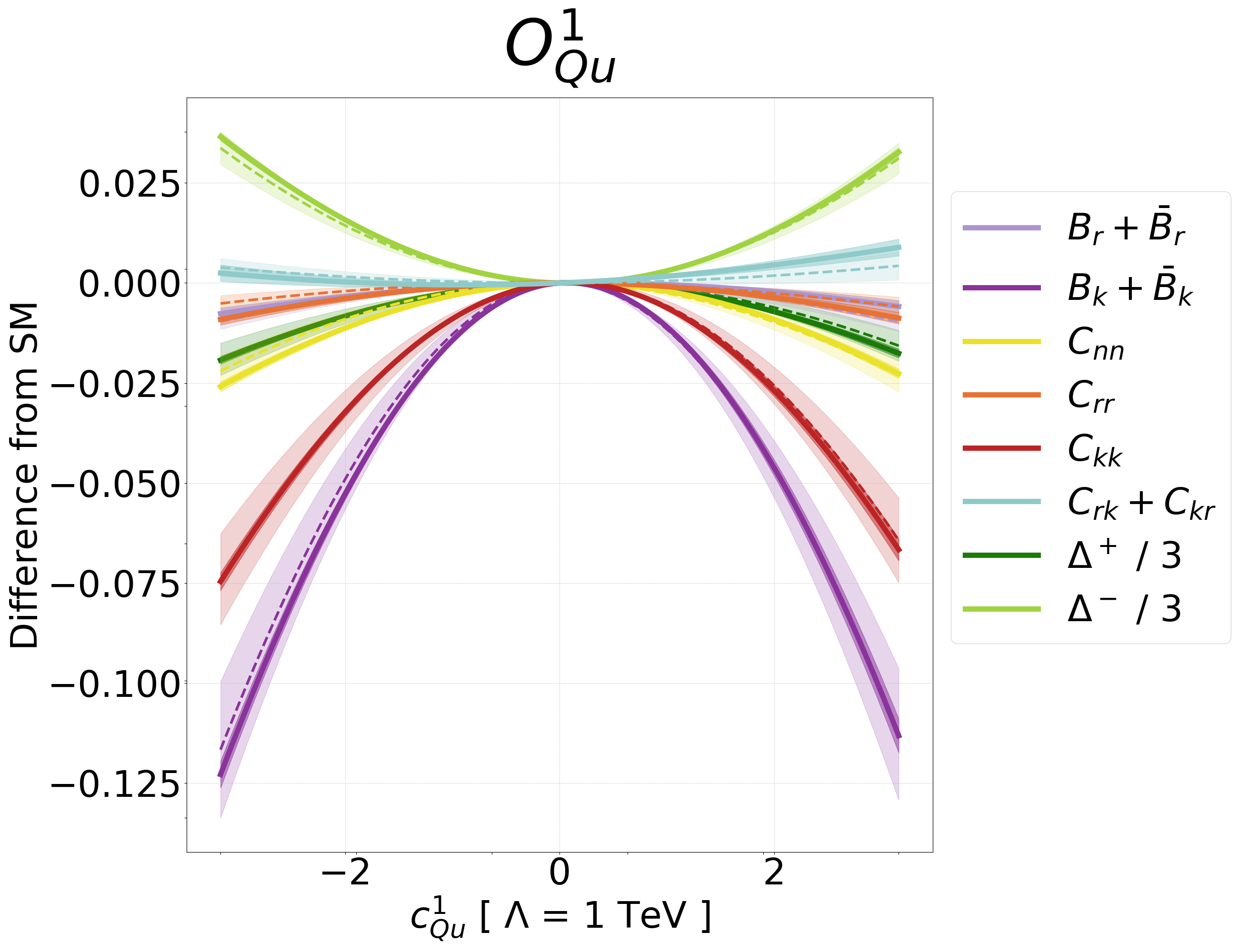}
  \subcaption{}
  \label{fig:DIM64F_3b}
\end{subfigure}

\begin{subfigure}{.5\textwidth}
  \centering
  \includegraphics[width=\linewidth]{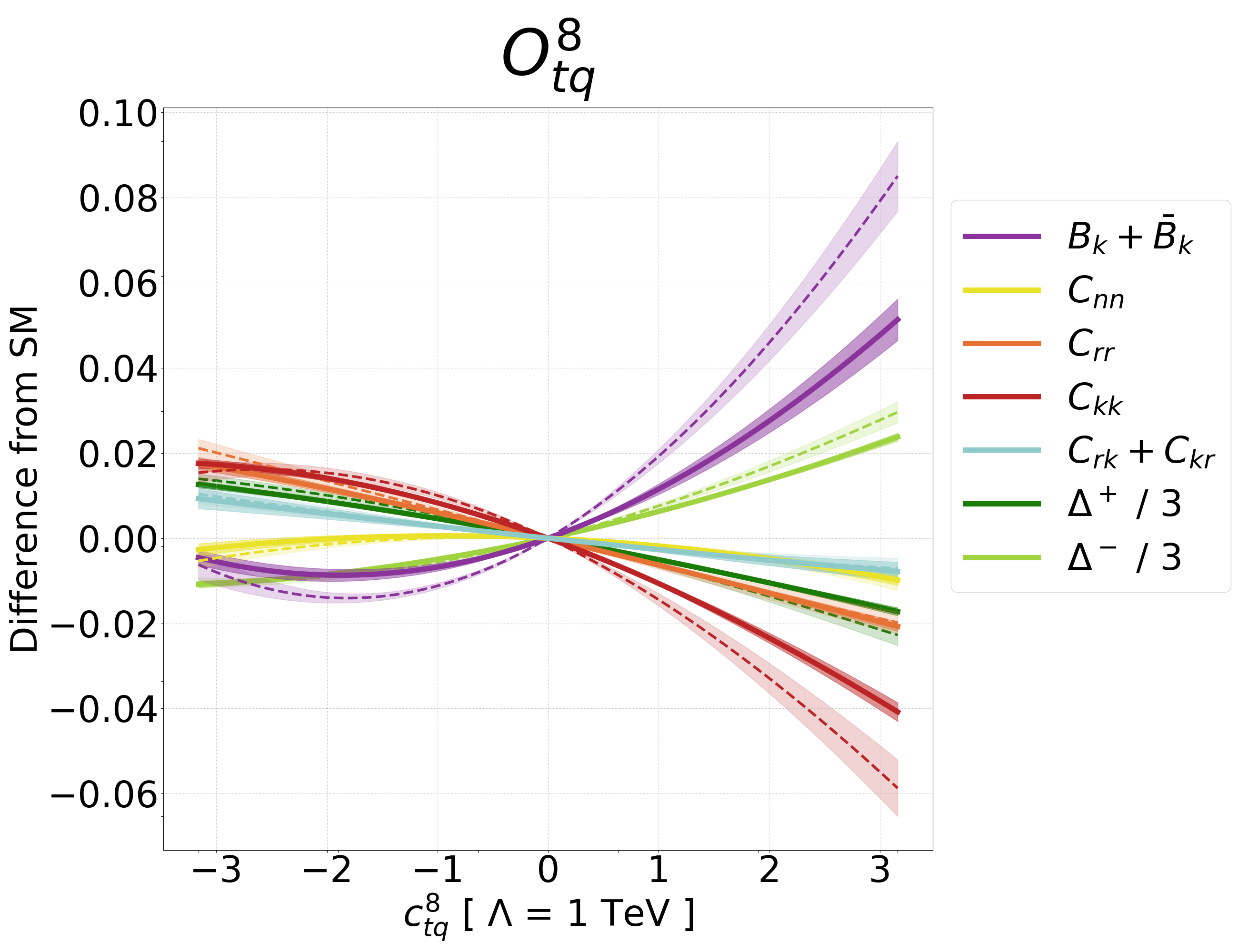}
  \subcaption{}
  \label{fig:DIM64F_4a}
\end{subfigure}%
\begin{subfigure}{.5\textwidth}
  \centering
  \includegraphics[width=\linewidth]{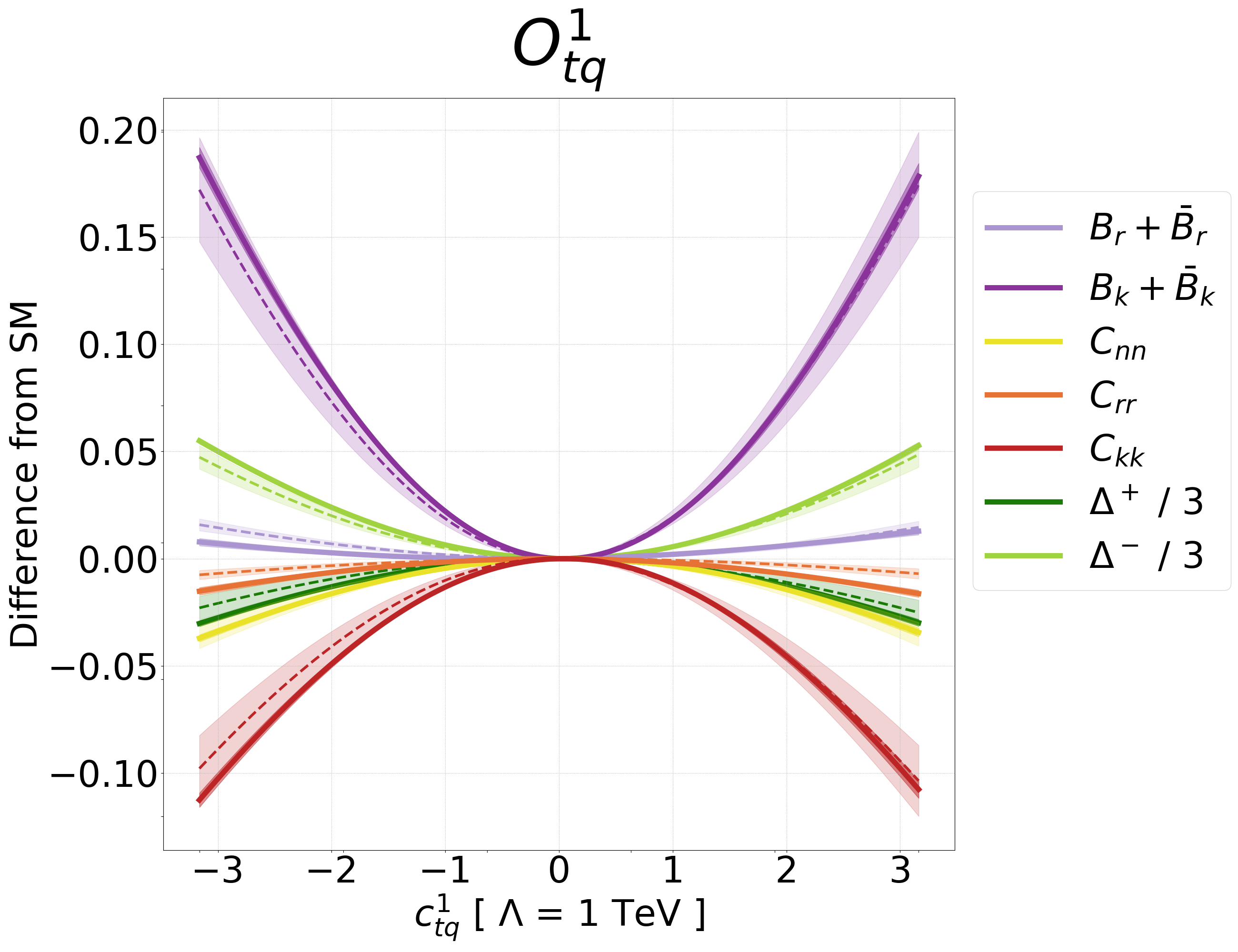}
  \subcaption{}
  \label{fig:DIM64F_4b}
\end{subfigure}

\begin{subfigure}{.5\textwidth}
  \centering
  \includegraphics[width=\linewidth]{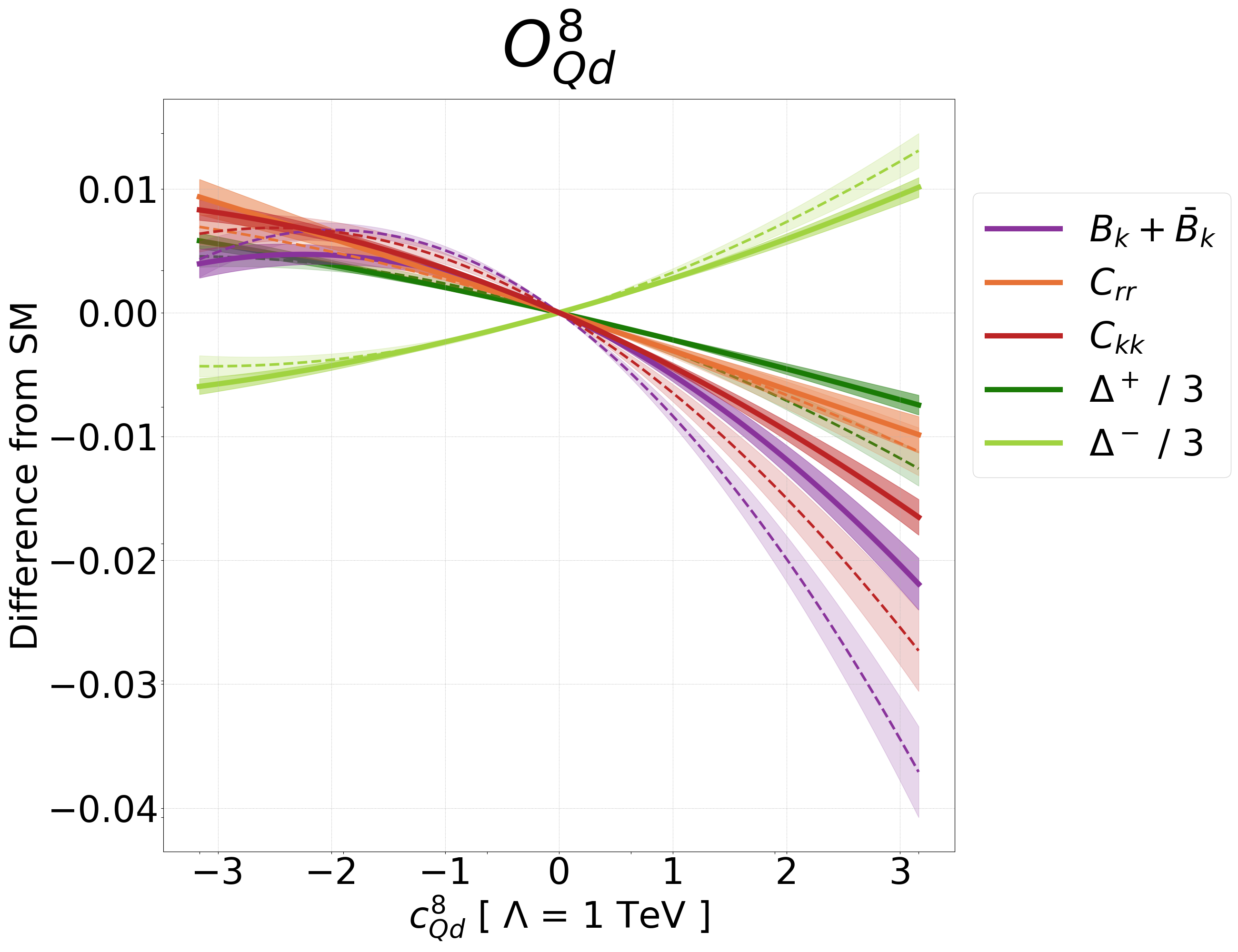}
  \subcaption{}
  \label{fig:DIM64F_6a}
\end{subfigure}%
\begin{subfigure}{.5\textwidth}
  \centering
  \includegraphics[width=\linewidth]{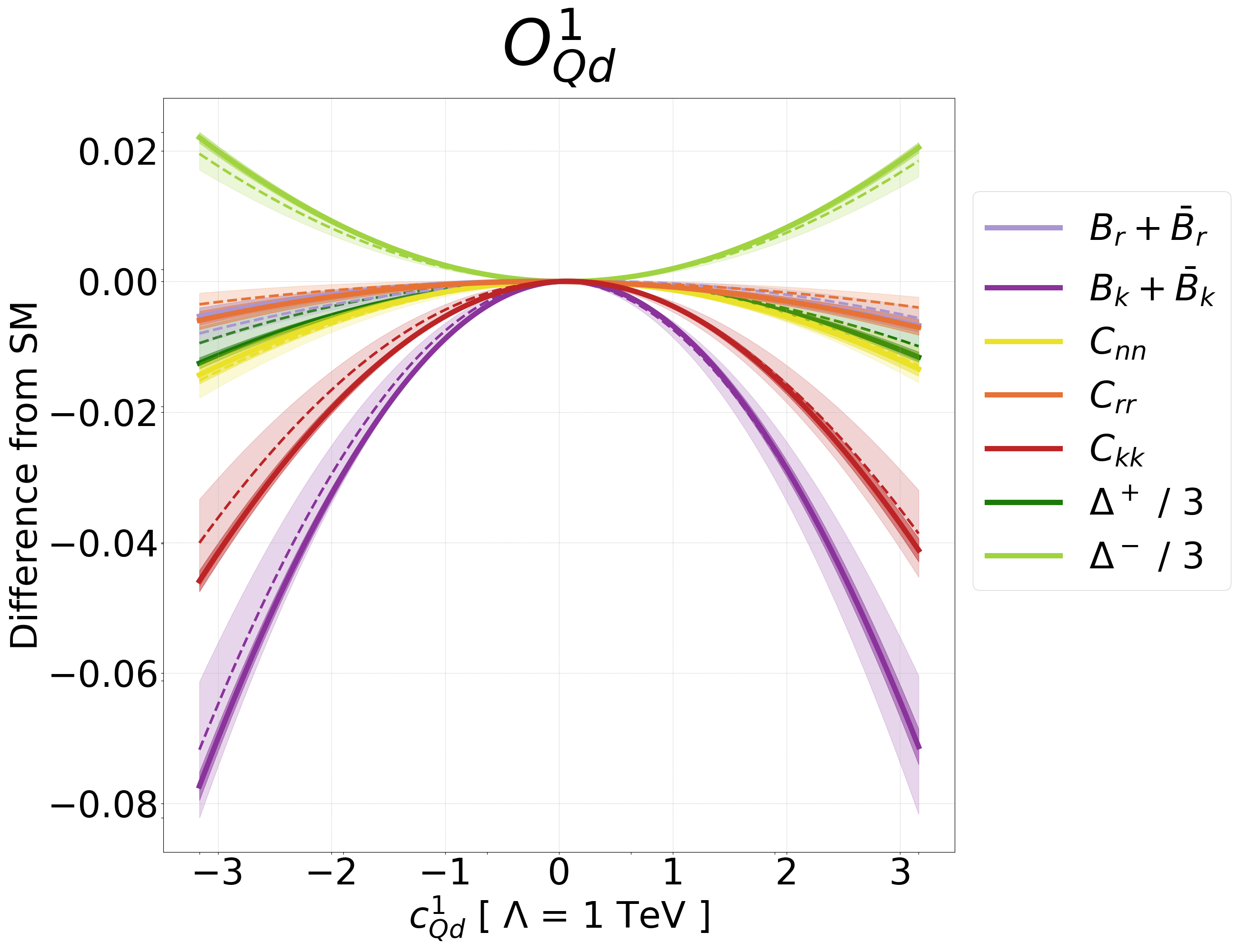}
  \subcaption{}
  \label{fig:DIM64F_6b}
\end{subfigure}

\captionsetup{width=\linewidth}
\caption{Results for the operators $\mathcal O_{Qu}^{8}$ (\ref{fig:DIM64F_3a}), $\mathcal O_{Qu}^1$ (\ref{fig:DIM64F_3b}), $\mathcal O_{tq}^{8}$ (\ref{fig:DIM64F_4a}), $\mathcal O_{tq}^1$ (\ref{fig:DIM64F_4b}), $\mathcal O_{Qd}^{8}$ (\ref{fig:DIM64F_6a}) and $\mathcal O_{Qd}^1$ (\ref{fig:DIM64F_6b}), similar to Figure \ref{fig:DIM62F_24}.}
\label{fig:DIM64F_2}
\end{figure}

\begin{figure}[H]
\centering

\begin{subfigure}{.5\textwidth}
  \centering
  \includegraphics[width=\linewidth]{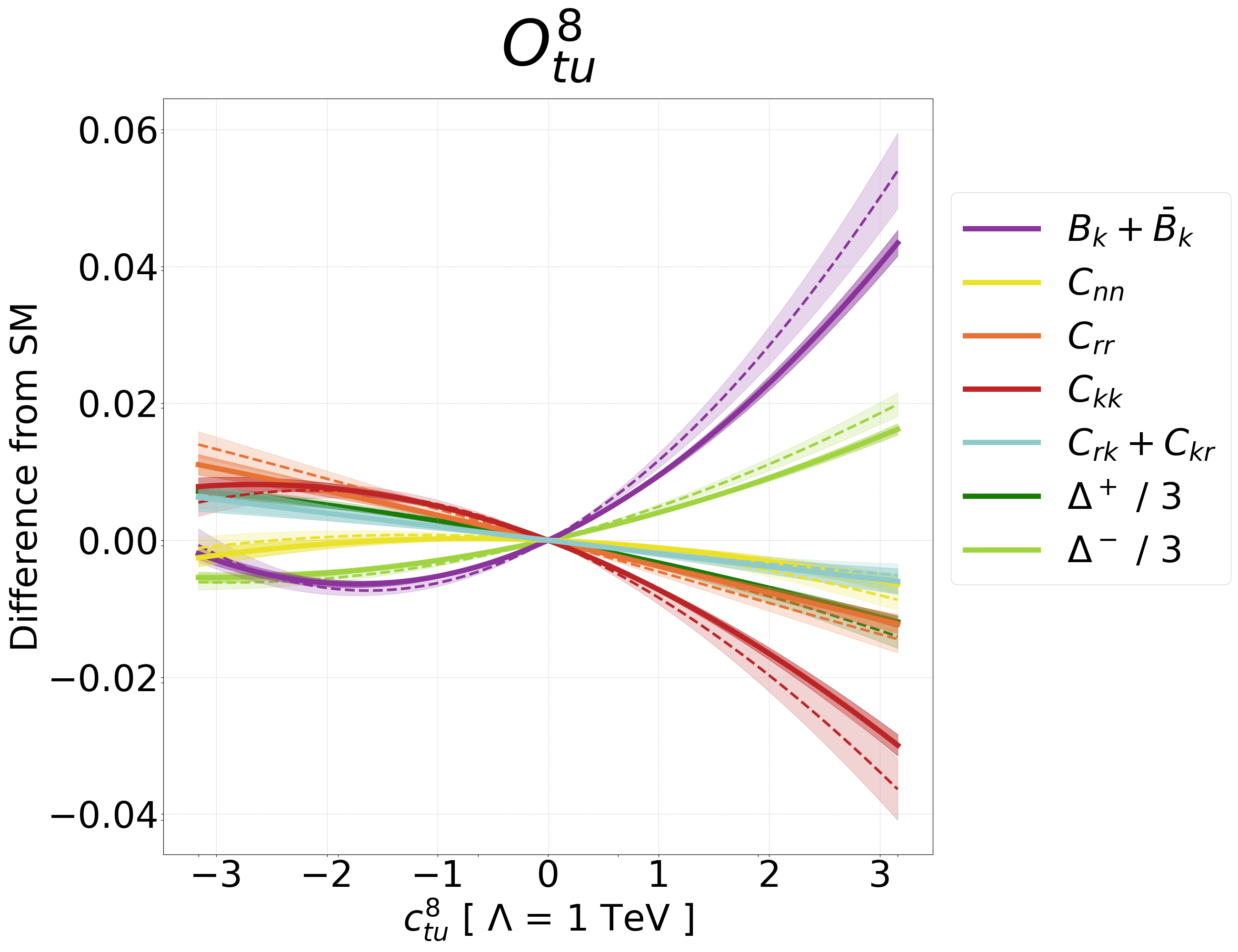}
  \subcaption{}
  \label{fig:DIM64F_7a}
\end{subfigure}%
\begin{subfigure}{.5\textwidth}
  \centering
  \includegraphics[width=\linewidth]{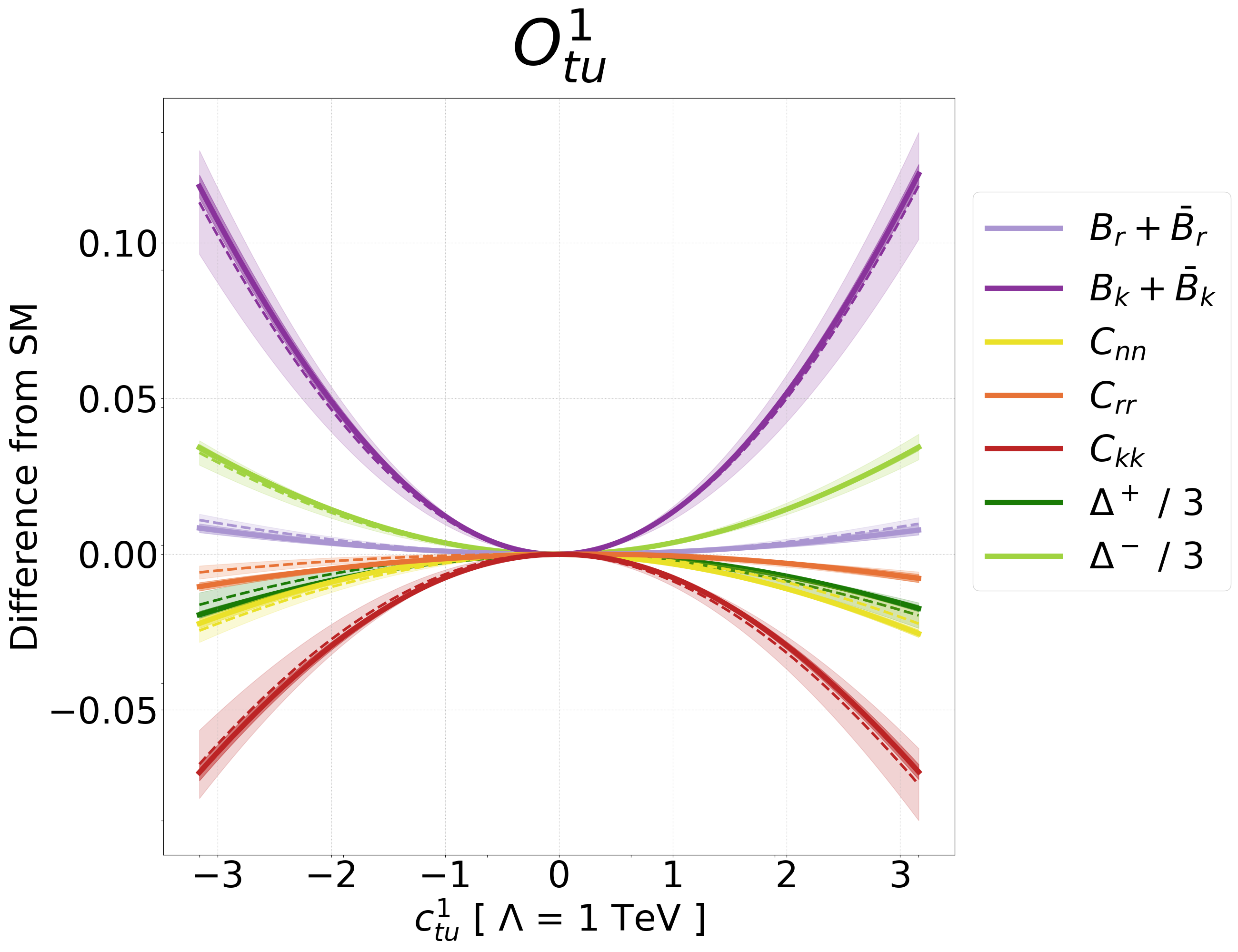}
  \subcaption{}
  \label{fig:DIM64F_7b}
\end{subfigure}

\begin{subfigure}{.5\textwidth}
  \centering
  \includegraphics[width=\linewidth]{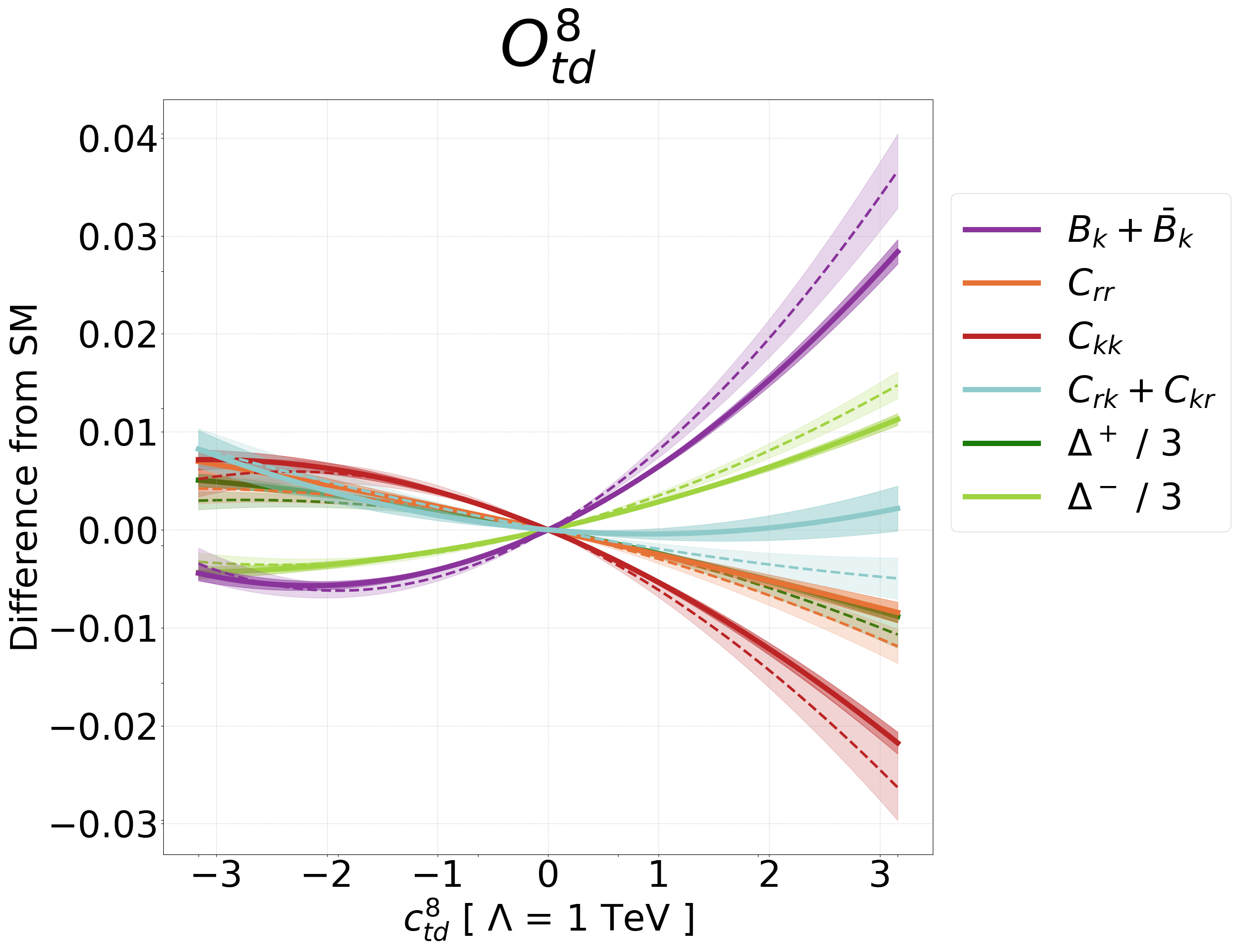}
  \subcaption{}
  \label{fig:DIM64F_8a}
\end{subfigure}%
\begin{subfigure}{.5\textwidth}
  \centering
  \includegraphics[width=\linewidth]{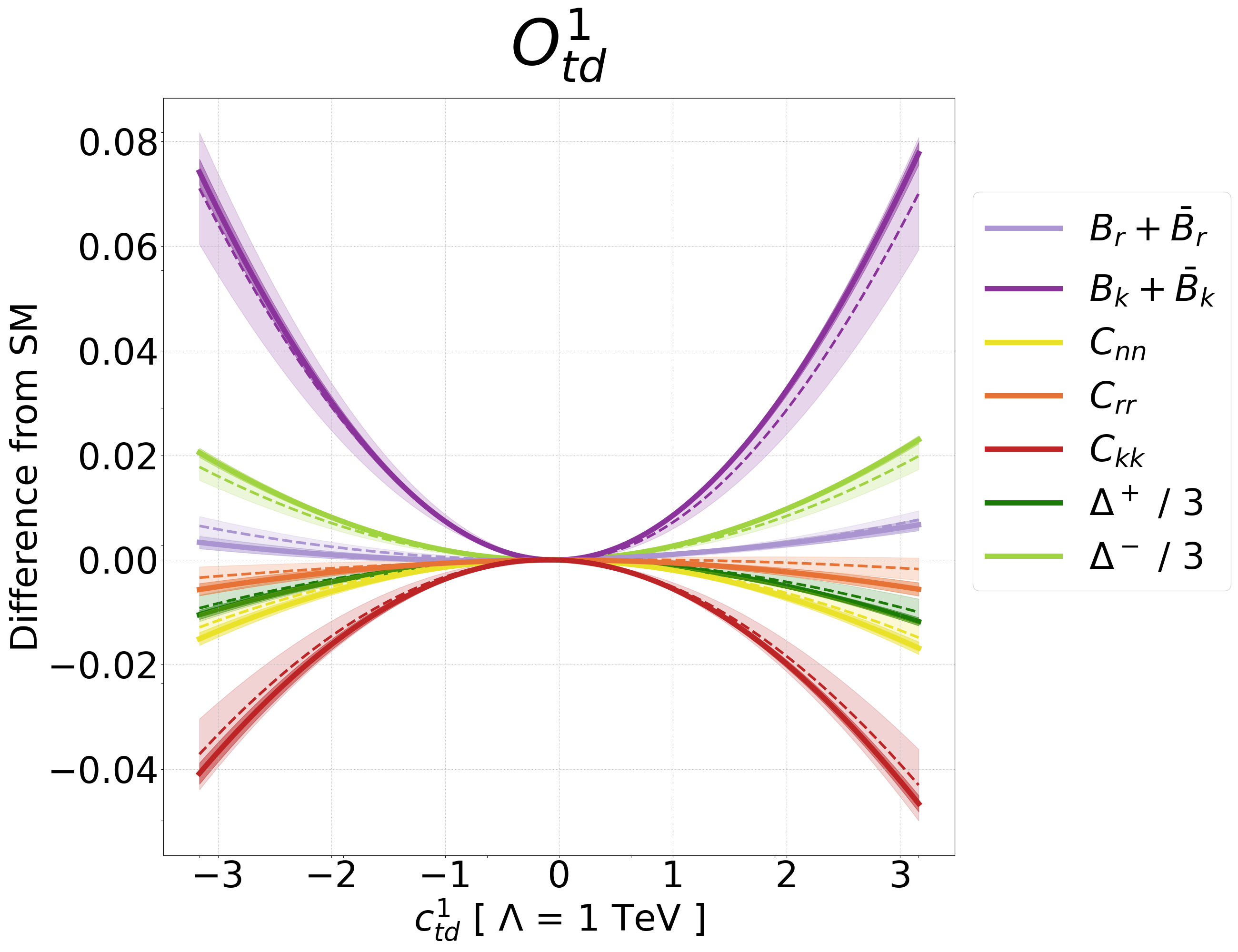}
  \subcaption{}
  \label{fig:DIM64F_8b}
\end{subfigure}

\captionsetup{width=\linewidth}
\caption{Results for the operators $\mathcal O_{tu}^{8}$ (\ref{fig:DIM64F_7a}), $\mathcal O_{tu}^1$ (\ref{fig:DIM64F_7b}), $\mathcal O_{td}^{8}$ (\ref{fig:DIM64F_8a}), $\mathcal O_{td}^1$ (\ref{fig:DIM64F_8b}), similar to Figure \ref{fig:DIM62F_24}.}
\label{fig:DIM64F_7_8}
\end{figure}

        Our results are in qualitative agreement with the findings at LO QCD of \cite{Bernreuther:2015yna, Aoude:2022imd}, signaling that NLO QCD contributions are under control, and our curves appear stable under the inclusion of the NLO with $\mathcal O(10 \%) \sim \mathcal O(\alpha_s)$ deviations. This is particularly true for spin correlation observables, as they arise from ratios of (weighted) cross-sections. It is frequently the case, however, that LO scale variation is not a good estimate for missing higher orders. This is especially true for color-octet operators, where the shift between LO QCD and NLO QCD often exceeds $3 \times$ the LO scale variation. This is despite the fact that our scale variation is conservative, as it includes all nine points obtained from the $1/2, 1, 2$ rescaling of $\mu_R$ and $\mu_{\text{f}}$, and not the often used seven.

Looking at individual spin observables, we note that several are not affected by dimension-6 operators, even in the regime where the NP-squared terms are dominant. The $\hat n$ polarization $B_n + \overline B_n$ never deviates appreciably from its SM value for any of the operators we considered. This is because, as already noted in \cite{Aoude:2022imd}, the only contributions to $B_n + \overline B_n$ come from the absorptive parts of one-loop amplitudes, and are thus heavily suppressed. Amongst those in Table~\ref{tab:obs}, $B_n + \overline B_n$ is the only spin observable for which this happens.

    The quantum entanglement markers $\Delta^\pm$ exhibit a similar pattern of deviations from the SM to the other spin observables. However, they are not included in this work for pure academic interest: their experimental measurement may be significantly simpler than the one of other observables. The enhanced precision is due to the fact that $\Delta^-$ is accessible directly from the opening angle between the leptons, with no need for \eqref{dcos_integrated}. The CMS Collaboration recently published a projection for top spin correlation measurements at the HL-LHC \cite{CMS:2022cqm}, where it is argued that $\Delta^-$ will be the spin observable more easily accessible, with a projected relative accuracy at the end of HL-LHC of $3\%$ \footnote{The notation used in \cite{CMS:2022cqm} and in other papers from experimental collaborations is $D \equiv (\Delta^- + 1)/3$.}.  Given the prospects of precise measurements, the entanglement markers are expected to provide good sensitivity to the dimension-6 coefficients. 

\paragraph{The top chromomagnetic moment operator $\mathcal O_{tG}$.} $ \ $ \smallskip

    Proceeding operator by operator, $\mathcal O_{tG}$ stands out as exhibiting a special behaviour. We immediately note that the effect of $c_{tG}$ is approximately ten times larger than the effect of all other four-fermion Wilson coefficients. Note the $x$-axis scale of Fig.~\ref{fig:DIM62F_1} is different from the one in all other plots. In fact, as noted below, top spin correlation measurements yield a remarkable sensitivity in constraining the value of $c_{tG}$.

    The large contribution of $c_{tG}$ to the spin observables we consider is a consequence of two factors. First, $\mathcal O_{tG}$ is the only SMEFT operator at dimension 6 that enters gluon-fusion top production $g \, g \to t \, \bar t$, and since $\sigma(g \, g \to t \, \bar t)/\sigma(q \, \bar q \to t \, \bar t) \sim 10$
     at the LHC, any BSM effect capable of entering gluon-fusion is heuristically expected to lead over those only appearing in $q \bar q$ annihilation by a similar factor. Second, due to the nature of $\mathcal O_{tG}$, the contributions proportional to $c_{tG}$ exhibit a structure that closely resembles the SM. In fact, the SM--$\mathcal O_{tG}$ interference amplitude is numerically substantial, and addionally, as it can be seen from Table \ref{tab:SM_DIM2F_24}, the scale variation of $c_{tG}$ and SM cross-sections are highly correlated, and in fact cancel in the ratio when computing spin observables.

    We also note that $\mathcal O_{tG}$ is affecting $t \bar t$ spin correlations, that is entries of the $C_{ij}$ matrix, without modifying the SM values of each top's polarization, $B_i$ and $\bar B_j$. This is understood analytically, as both the linear and squared contribution of $c_{tG}$ to all polarization observables vanish at LO QCD. 
    
     \paragraph{A case study: $B_k + \overline B_k$.} $ \ $ \smallskip
    
    In the case of four-fermion operators, $B_k + \overline B_k$ is the spin observable that consistently deviates the most from its SM value. We therefore take it as a case study, and analyse its properties in detail. While an NLO-accurate analytical calculation of the SMEFT shifts of $B_k + \overline B_k$ is laborious, and beyond the scope of this work, interesting patterns emerge from our numerical results, shown in App.~\ref{app:inclusive} for all observables, and reported in Table \ref{tab:bk} for the reader's convenience.
    
    As noted above, $\mathcal O_{tG}$ does not contribute significantly to $B_{k}+\bar{B}_{k}$. The four-fermion operators, that all contribute, can be grouped in two classes, $\mathcal O_{Qq}^{(8,3)}, \mathcal O_{Qq}^{(8,1)}, \mathcal O_{Qu}^{8}, \mathcal O_{Qd}^{8}$, for which the SM shift is negative, and $\mathcal O_{tq}^{8}, \mathcal O_{tu}^{8}, \mathcal O_{td}^{8}$, for which the shift is positive. In the following we will focus on color-octet operators, but our exact arguments apply to color-singlet operators.
    
    At order $c/\Lambda^2$, the trend of positive/negative shifts with respect to the SM is consistent with the phase space integration of the results presented in \cite{Aoude:2022imd}, where it is noted that at LO QCD the coupling of new physics to $B_k + \overline B_k$, at linear order, is only through the Axial-Vector (AV) combinations:
    \begin{align}
        - c^{(8,3)}_{Qq} - c_{Qq}^{(8,1)} - c_{Qu}^{8} + c_{tq}^{8} + c_{tu}^{8}, \quad &\text{for } u \bar u \to t \bar t \\
        c^{(8,3)}_{Qq} - c_{Qq}^{(8,1)} - c_{Qd}^{8} + c_{tq}^{8} + c_{td}^{8}, \quad &\text{for } d \bar d \to t \bar t.
    \end{align}
    The coefficient $c^{(8,3)}_{Qq}$ enters with opposite signs the $u \bar u$ and $d \bar d$ processes; the fact that the $u \bar u$ contribution is observed to dominate in our numerical calculation, as shown in Figure \ref{fig:DIM64F_1a}, is of course understood in terms of PDF luminosity in $p \, p$ collisions. 
    
\begin{table}[h]
\centering
\scalebox{0.82}
{
\begin{tabular}{cc|rr|rr}
 & & \multicolumn{2}{c}{Color-octet} & \multicolumn{2}{c}{Color-singlet} \\
 Operators & Order & $\sigma^{(1)}$ [pb] & $\sigma^{(2)}$ [pb] & $\sigma^{(1)}$ [pb] &  $\sigma^{(2)}$ [pb] \\ \hline
\multirow{ 2}{*}{$\mathcal O_{Qq}^{(8,3)}$ and $\mathcal O_{Qq}^{(1,3)}$ }& \small{LO} & $ -0.079 \scriptstyle{(10)} \, \displaystyle{^{+0.008}_{-0.011}}  $ & $ -0.133 \scriptstyle{(3)} \, \displaystyle{^{+0.011}_{-0.013}}  $  &   $ 0.036 \scriptstyle{(23)} \, \displaystyle{^{+0.007}_{-0.008}}  $ & $ -0.584 \scriptstyle{(5)} \, \displaystyle{^{+0.038}_{-0.043}}  $  \\ 
 & \small{NLO} & $ -0.083 \scriptstyle{(7)} \, \displaystyle{^{+0.004}_{-0.001}}  $ & $ -0.118 \scriptstyle{(2)} \, \displaystyle{^{+0.002}_{-0.001}}  $  &   $ -0.004 \scriptstyle{(16)} \, \displaystyle{^{+0.007}_{-0.004}}  $ & $ -0.761 \scriptstyle{(4)} \, \displaystyle{^{+0.031}_{-0.034}}  $  \\ \hline
 
\multirow{ 2}{*}{  $\mathcal O_{Qq}^{(8,1)}$ and $\mathcal O_{Qq}^{(1,1)}$ }& \small{LO} & $ -0.441 \scriptstyle{(10)} \, \displaystyle{^{+0.054}_{-0.068}}  $ & $ -0.123 \scriptstyle{(3)} \, \displaystyle{^{+0.009}_{-0.011}}  $  &   $ -0.004 \scriptstyle{(28)} \, \displaystyle{^{+0.000}_{-0.001}}  $ & $ -0.557 \scriptstyle{(7)} \, \displaystyle{^{+0.035}_{-0.040}}  $  \\ 
 & \small{NLO} & $ -0.469 \scriptstyle{(7)} \, \displaystyle{^{+0.014}_{-0.003}}  $ & $ -0.129 \scriptstyle{(2)} \, \displaystyle{^{+0.003}_{-0.003}}  $  &   $ -0.044 \scriptstyle{(14)} \, \displaystyle{^{+0.011}_{-0.015}}  $ & $ -0.767 \scriptstyle{(4)} \, \displaystyle{^{+0.032}_{-0.035}}  $  \\ \hline 
 
\multirow{ 2}{*}{$\mathcal O_{Qu}^{8}$ and $\mathcal O_{Qu}^1$ }& \small{LO} & $ -0.272 \scriptstyle{(8)} \, \displaystyle{^{+0.034}_{-0.043}}  $ & $ -0.080 \scriptstyle{(2)} \, \displaystyle{^{+0.006}_{-0.008}}  $  &   $ 0.019 \scriptstyle{(16)} \, \displaystyle{^{+0.002}_{-0.002}}  $ & $ -0.357 \scriptstyle{(4)} \, \displaystyle{^{+0.023}_{-0.026}}  $  \\ 
 & \small{NLO} & $ -0.203 \scriptstyle{(7)} \, \displaystyle{^{+0.021}_{-0.009}}  $ & $ -0.055 \scriptstyle{(2)} \, \displaystyle{^{+0.004}_{-0.003}}  $  &   $ 0.062 \scriptstyle{(10)} \, \displaystyle{^{+0.017}_{-0.012}}  $ & $ -0.462 \scriptstyle{(2)} \, \displaystyle{^{+0.018}_{-0.020}}  $  \\ \hline 
 
\multirow{ 2}{*}{ $\mathcal O_{tq}^{8}$ and $\mathcal O_{tq}^1$ }& \small{LO} & $ 0.446 \scriptstyle{(10)} \, \displaystyle{^{+0.073}_{-0.057}}  $ & $ 0.123 \scriptstyle{(2)} \, \displaystyle{^{+0.012}_{-0.010}}  $  &   $ 0.012 \scriptstyle{(28)} \, \displaystyle{^{+0.002}_{-0.002}}  $ & $ 0.562 \scriptstyle{(6)} \, \displaystyle{^{+0.042}_{-0.036}}  $  \\ 
 & \small{NLO} & $ 0.332 \scriptstyle{(7)} \, \displaystyle{^{+0.016}_{-0.043}}  $ & $ 0.089 \scriptstyle{(2)} \, \displaystyle{^{+0.004}_{-0.006}}  $  &   $ -0.062 \scriptstyle{(15)} \, \displaystyle{^{+0.014}_{-0.019}}  $ & $ 0.747 \scriptstyle{(3)} \, \displaystyle{^{+0.031}_{-0.029}}  $  \\ \hline 
 
\multirow{ 2}{*}{  $\mathcal O_{Qd}^{8}$ and $\mathcal O_{Qd}^1$  }& \small{LO} & $ -0.195 \scriptstyle{(6)} \, \displaystyle{^{+0.026}_{-0.034}}  $ & $ -0.049 \scriptstyle{(2)} \, \displaystyle{^{+0.004}_{-0.005}}  $  &   $ 0.003 \scriptstyle{(11)} \, \displaystyle{^{+0.002}_{-0.002}}  $ & $ -0.217 \scriptstyle{(3)} \, \displaystyle{^{+0.015}_{-0.017}}  $  \\ 
 & \small{NLO} & $ -0.150 \scriptstyle{(7)} \, \displaystyle{^{+0.017}_{-0.008}}  $ & $ -0.033 \scriptstyle{(2)} \, \displaystyle{^{+0.003}_{-0.002}}  $  &   $ 0.036 \scriptstyle{(8)} \, \displaystyle{^{+0.008}_{-0.007}}  $ & $ -0.282 \scriptstyle{(2)} \, \displaystyle{^{+0.011}_{-0.012}}  $  \\ \hline 
 
\multirow{ 2}{*}{ $\mathcal O_{tu}^{8}$ and $\mathcal O_{tu}^1$ }& \small{LO} & $ 0.262 \scriptstyle{(9)} \, \displaystyle{^{+0.040}_{-0.032}}  $ & $ 0.081 \scriptstyle{(2)} \, \displaystyle{^{+0.008}_{-0.007}}  $  &   $ 0.026 \scriptstyle{(17)} \, \displaystyle{^{+0.004}_{-0.003}}  $ & $ 0.360 \scriptstyle{(4)} \, \displaystyle{^{+0.026}_{-0.023}}  $  \\ 
 & \small{NLO} & $ 0.268 \scriptstyle{(7)} \, \displaystyle{^{+0.002}_{-0.008}}  $ & $ 0.078 \scriptstyle{(2)} \, \displaystyle{^{+0.002}_{-0.001}}  $  &   $ 0.024 \scriptstyle{(11)} \, \displaystyle{^{+0.011}_{-0.007}}  $ & $ 0.471 \scriptstyle{(3)} \, \displaystyle{^{+0.023}_{-0.020}}  $  \\ \hline 
 
\multirow{ 2}{*}{ $\mathcal O_{td}^{8}$ and $\mathcal O_{td}^1$ }& \small{LO} & $ 0.188 \scriptstyle{(9)} \, \displaystyle{^{+0.029}_{-0.023}}  $ & $ 0.050 \scriptstyle{(2)} \, \displaystyle{^{+0.005}_{-0.004}}  $  &   $ -0.005 \scriptstyle{(10)} \, \displaystyle{^{+0.001}_{-0.002}}  $ & $ 0.215 \scriptstyle{(3)} \, \displaystyle{^{+0.016}_{-0.014}}  $  \\ 
 & \small{NLO} & $ 0.191 \scriptstyle{(7)} \, \displaystyle{^{+0.005}_{-0.007}}  $ & $ 0.044 \scriptstyle{(2)} \, \displaystyle{^{+0.001}_{-0.000}}  $  &   $ 0.020 \scriptstyle{(9)} \, \displaystyle{^{+0.005}_{-0.004}}  $ & $ 0.288 \scriptstyle{(2)} \, \displaystyle{^{+0.015}_{-0.013}}  $  \\ \hline 
\end{tabular}
}
\captionsetup{width=\linewidth}
\caption{Cross sections appearing in the numerator of \eqref{ratio} for $B_{k}+\bar{B}_{k}$ for the four-fermion operators we consider in this work, extracted from Appendix \ref{app:inclusive} and with the same notation, $\text{central(stat)}^{+\text{scale}}_{-\text{scale}}$}
\label{tab:bk}
\end{table}

    At order $\mathcal O(c^2/\Lambda^4)$ and LO QCD, a calculation based on {\tt FeynCalc} \cite{Shtabovenko:2020gxv} and the {\tt SMEFTatNLO} model \cite{Degrande:2020evl} shows that new physics enters in $B_k + \overline B_k$ through the combinations:
    \begin{align}
        - (c^{(8,3)}_{Qq} + c_{Qq}^{(8,1)})^2 - (c_{Qu}^{8})^2 + (c_{tq}^{8})^2 + (c_{tu}^{8})^2, \quad &\text{for } u \bar u \to t \bar t, \label{bk_uu}\\
       - (- c^{(8,3)}_{Qq} + c_{Qq}^{(8,1)})^2 - (c_{Qd}^{8})^2 + (c_{tq}^{8})^2 + (c_{td}^{8})^2 , \quad &\text{for } d \bar d \to t \bar t, \label{bk_dd}
    \end{align} 
    when no phase-space cuts are applied on the $t \bar t$ pair. This is confirmed by our numerical results in Table \ref{tab:bk}. The relative signs between four-fermion operators that enter the contributions to $B_k + \overline B_k$ at linear and quadratic order in $c/\Lambda^2$ are {\it not} the same as those that enter in the total $t \bar t$ cross-section. For instance, at LO QCD and linearly in $c/\Lambda^2$, the inclusive cross section is only sensitive to the Vector-Vector (VV) combination \cite{Bernreuther:2015yna, Aoude:2022imd}:
     \begin{align}
        c^{(8,3)}_{Qq} + c_{Qq}^{(8,1)} + c_{Qu}^{8} + c_{tq}^{8} + c_{tu}^{8}.
    \end{align}   
    At quadratic SMEFT order $\mathcal O(c^2/\Lambda^4)$ more combinations of Wilson coefficients enter the inclusive cross section, but, again, not the specific ones in \eqref{bk_uu} and \eqref{bk_dd}. This simple observation already highlights the importance of spin observables in breaking flat directions otherwise present in the 4-fermion sector.

    We also note that several operators exhibit a relatively large $K$-factor between the LO and NLO. Most notably, $\mathcal O_{Qu}^8$, $\mathcal O_{tq}^8$, and $\mathcal O_{Qd}^8$, show a $\sim 50 \%$ decrease in the $\sigma^{(2)}$ cross-section when going to NLO. This in turn translates to a significant mismatch between the LO and NLO curves in Figures \ref{fig:DIM64F_3a}, \ref{fig:DIM64F_4a}, and \ref{fig:DIM64F_6a}, that is not predicted by the LO scale variation.
  
    \medskip
    
    We conclude this Section with a remark: the difference from the SM values for all spin observables we considered is of order $0.1$ over the range of Wilson coefficients still allowed by the 2021 global fit \cite{Ethier:2021bye}. The measurement of spin polarizations and spin correlations in a hadron collider is notoriously difficult, and a $\%$ level precision, needed to be competitive on the global fit, may be out of reach. However, as we shall see in the next section, the elaborate structure of spin observables in $t \bar t$ phase space suggests a different route.
 
\subsection{Results differential in $t \bar t$ phase space} \label{sec:differential}
	    
	The following Figures \ref{fig:DIM62F_24_binned} - \ref{fig:DIM64F_2_binned_B}, similarly to Figures \ref{fig:DIM62F_24} - \ref{fig:DIM64F_7_8}, show, for selected SMEFT operators, the change at NLO with respect to the SM value for the top spin polarizations $B_i + \bar B_i$, the $t \bar t$ spin correlations $C_{ii}$ and $C_{ij} + C_{ji}$, and the entanglement markers $\Delta^\pm$, as a function of Wilson coefficients, with $t \bar t$ phase space binned in the pair invariant mass $m_{t \bar t}$ and in the c.m.f.\@ scattering angle $\theta$. Figure \ref{fig:binning} shows our binning, and the fraction of the total $t \bar t$ cross-section included in each bin. Even after accounting for detector effects, already after Run 2 of the LHC, all bins contain several million $t \bar t$ events decaying leptonically.
	
	\begin{figure}[H]
    \centering
    \scalebox{0.8}
    {
\begin{tabular}{|c|c|c|}
\hline
$\cos \theta \leq 1/3 $ & $1/3 \leq \cos \theta \leq 2/3$ &  $2/3 \leq \cos \theta$ \\
$m_{t \bar t} \leq 420 \, \text{GeV}$ & $m_{t \bar t} \leq 420 \, \text{GeV}$ & $m_{t \bar t} \leq 420 \, \text{GeV}$ \\
$\bm{\sigma_{\text{bin}}/\sigma = 9.0 \%} $ & $\bm{\sigma_{\text{bin}}/\sigma = 10.0 \%} $  & $\bm{\sigma_{\text{bin}}/\sigma = 12.3 \%} $ \\
\hline
$\cos \theta \leq 1/3 $ & $1/3 \leq \cos \theta \leq 2/3$ &  $2/3 \leq \cos \theta$ \\
$420 \, \text{GeV} \leq m_{t \bar t} \leq 600 \, \text{GeV}$ & $420 \, \text{GeV} \leq m_{t \bar t} \leq 600 \, \text{GeV}$ & $420 \, \text{GeV} \leq m_{t \bar t} \leq 600 \, \text{GeV}$ \\
$\bm{\sigma_{\text{bin}}/\sigma = 9.1 \%} $ & $\bm{\sigma_{\text{bin}}/\sigma = 12.6 \%} $  & $\bm{\sigma_{\text{bin}}/\sigma = 24.6 \%} $ \\
\hline
$\cos \theta \leq 1/3 $ & $1/3 \leq \cos \theta \leq 2/3$ &  $2/3 \leq \cos \theta$ \\
$600 \, \text{GeV} \leq m_{t \bar t}$ & $600 \, \text{GeV} \leq m_{t \bar t}$ & $600 \, \text{GeV} \leq m_{t \bar t}$ \\
$\bm{\sigma_{\text{bin}}/\sigma = 2.3 \%} $ & $\bm{\sigma_{\text{bin}}/\sigma = 4.0 \%} $  & $\bm{\sigma_{\text{bin}}/\sigma = 16.1 \%} $ \\
\hline
\end{tabular}
}
    \captionsetup{width=\linewidth}
    \caption{Choice of binning of $t \bar t$ phase-space considered in this work, and fraction of the total cross-section that falls in each bin. Layout reproduces that of our differential plots.}
    \label{fig:binning}
    \end{figure}
    
    \begin{figure}[H]
    \centering
    \includegraphics[width=0.82\linewidth]{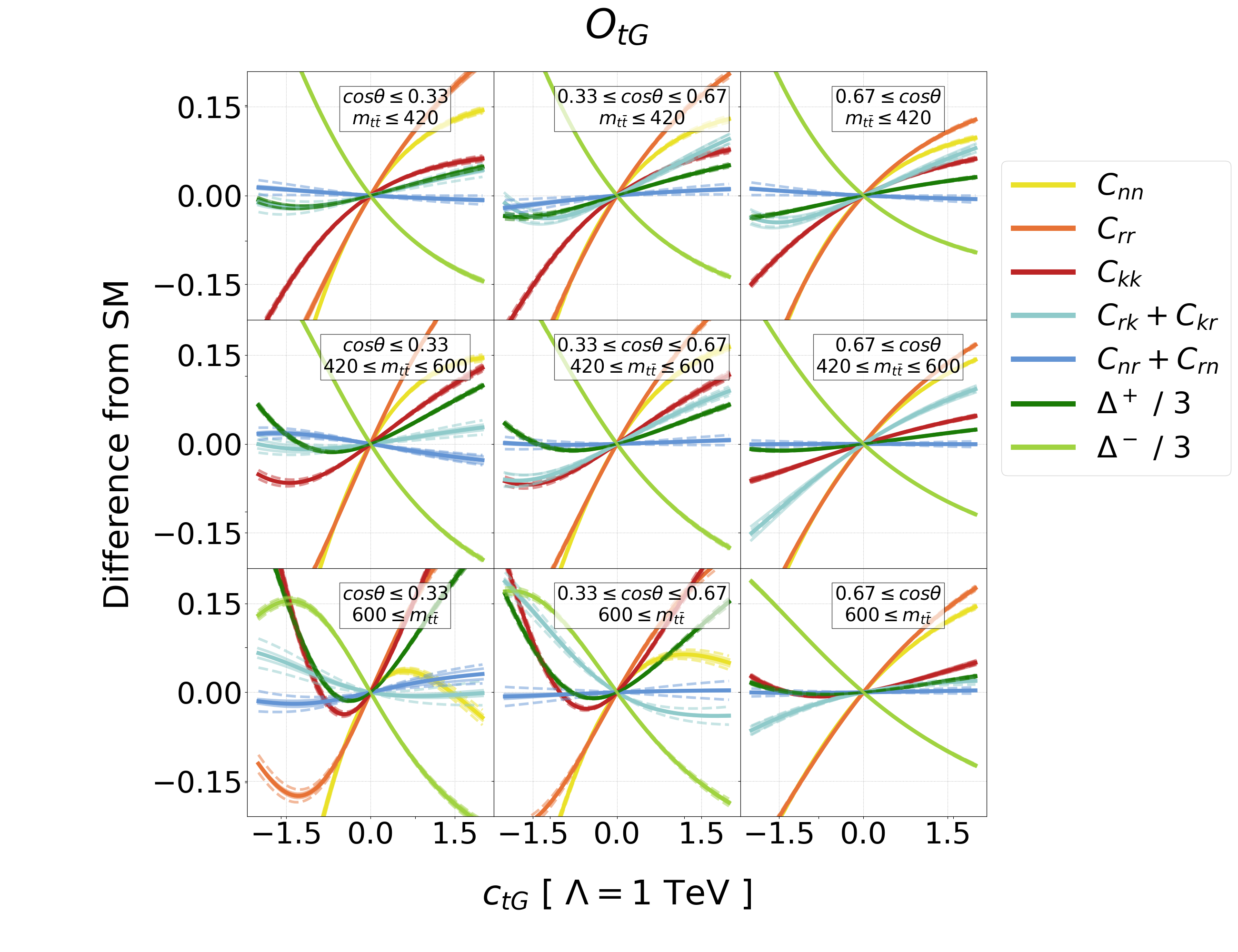}
    \vspace{-5mm}
    \captionsetup{width=\linewidth}
    \caption{Change at NLO from the SM value for spin observables for the operator $\mathcal O_{tG}$. Phase space is divided in the $t \bar t$ invariant mass along $y$, and in $\cos \theta$ along $x$, as indicated by the insets. Continuous lines around each plotted curve indicate scale variation, dashed lines indicate MC uncertainty. Only curves deviating appreciably from zero are shown.}
    \label{fig:DIM62F_24_binned}
    \end{figure}
    
    \begin{figure}[H]
    \centering
    \includegraphics[width=0.82\linewidth]{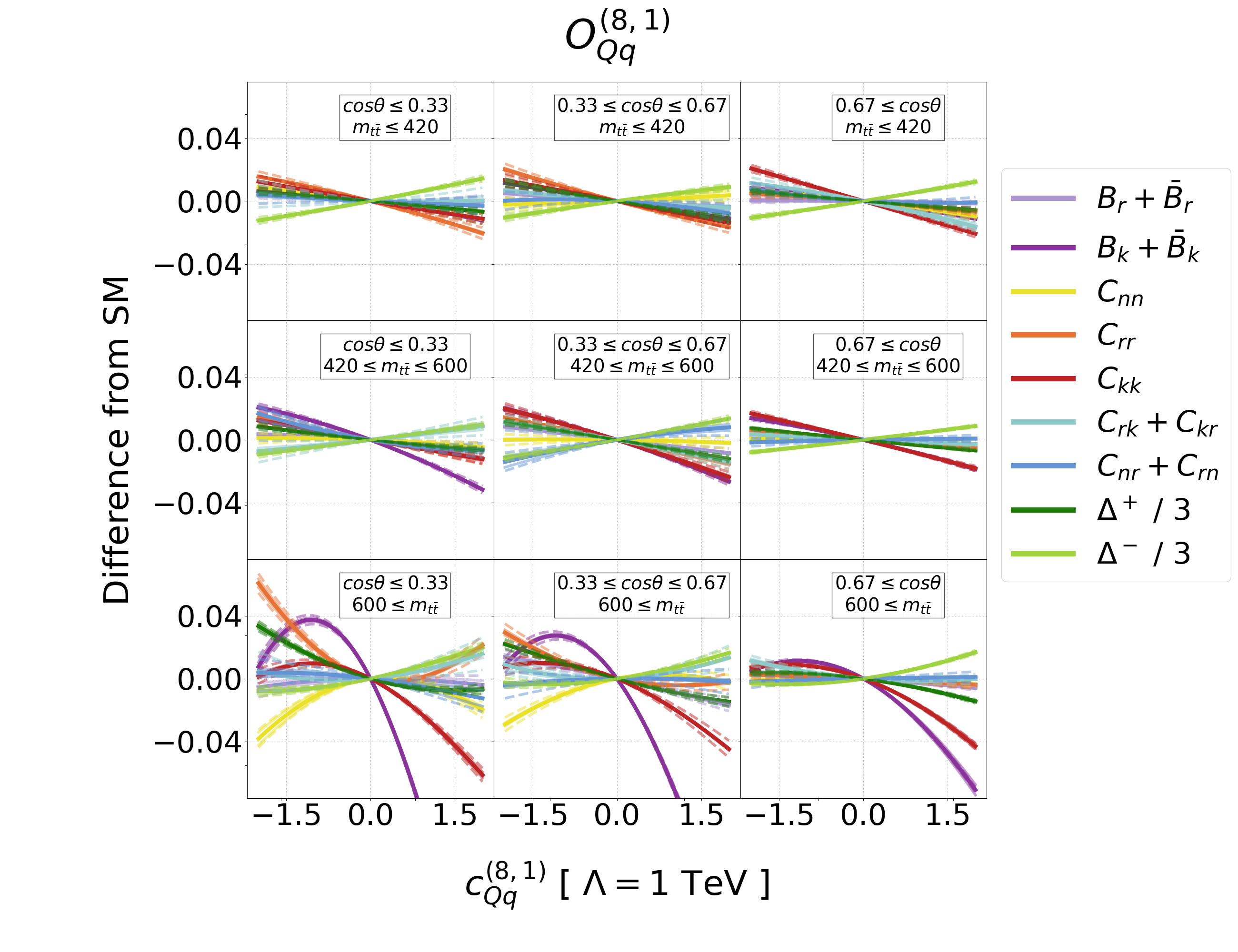}
     \vspace{-5mm}
    \captionsetup{width=\linewidth}
    \caption{Results for the operator $\mathcal O_{Qq}^{(8,1)}$, similar to Figure \ref{fig:DIM62F_24_binned}.}
    \label{fig:DIM64F_2_binned_A}
    \end{figure}
    
    \begin{figure}[H]
    \centering
    \includegraphics[width=0.82\linewidth]{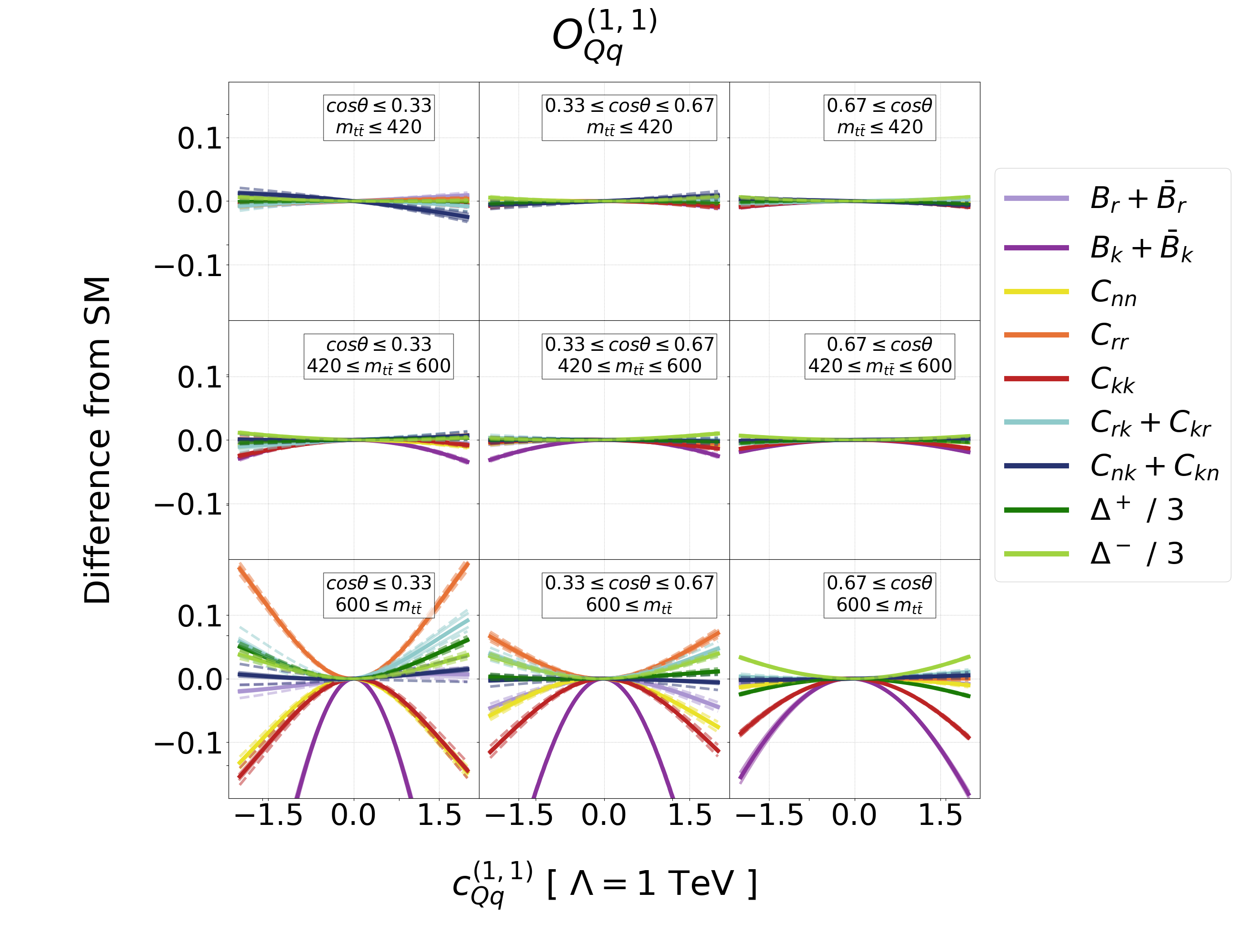}
     \vspace{-5mm}
    \captionsetup{width=\linewidth}
    \caption{Results for the operators $\mathcal O_{Qq}^{(1,1)}$, similar to Figure \ref{fig:DIM62F_24_binned}.}
    \label{fig:DIM64F_2_binned_B}
    \end{figure}
	
	This Section contains plots for selected SMEFT operators, that have been chosen as their behaviour is representative of the others. Appendix \ref{app:binned} contains plots for the remaining operators considered in this work. Contrary to the inclusive plots, the doubly-differential plots split the sources of theory uncertainty, with scale variation and Monte Carlo uncertainty shown separately. This is to highlight that, should it become the limiting factor, MC uncertainty is improvable arbitrarily. 
	
	We find that spin observables are heavily affected by the presence of new physics, with a rich, and non--trivial structure in $t \bar t$ phase space. Even a modest cut on the invariant mass, for example our lowest row $m_{t \bar t} > 600 \, \text{GeV}$, can enhance the sensitivity to new physics by a factor of $10$ over an inclusive measurement.  On top of the expected growth of deviations from the SM with energy, we find that the $t \bar t$ kinematical regions that are most sensitive to new physics are those at high-$p_T$, already investigated in the context of the observation of quantum entanglement between tops. Aggressive phase space selections in the high-$p_T$ region will yield increasing gains in sensitivity, but will realistically come with a corresponding worsening of experimental uncertainty, both statistical and systematic.

    The variation of the spin correlation observables with different phase space regions is most obvious for the four-fermion operator coefficients. In the Wilson coefficient range we consider, the deviation from the SM predictions varies from the percent level in the low invariant mass region to more than 10\% for the high-mass and high-$p_T$ region for certain operators.

    It is also interesting that in several phase-space regions the relation between Wilson coefficients and spin observables is not only departing from the linear regime, but it is also observed to be deviating from a quadratic approximation. This is evident, for instance, for the entanglement marker $\Delta^-$ in the high $m_{t \bar t}$ region of the $\mathcal O_{tG}$ plot, where two saddle points can be seen. Such a rich and non-trivial behaviour highlights the advantage of our full parametrization \eqref{ratio} over series expansions.
	
	To conclude this Section, we note that for certain operators, even after enhancing the impact on the observables by focusing on the high-energy regions, the deviation from the SM remains at the 2-3\% level in the Wilson coefficient range still allowed by other measurements. For these operators it will be challenging to obtain stringent constraints on the coefficients using spin observables, as this would require extremely precise measurements, probably beyond the reach of present colliders.

    \section{Constraints on the Wilson coefficients}	\label{sec:constraints}	
    
	The early Run 2 top spin polarization and $t \bar t$ spin correlation measurement from the CMS Collaboration \cite{CMS:2019nrx} is the only experimental result, as of 2022, where the full spin density matrix is reconstructed. The measurement covers $35.9 \, \text{fb}^{-1}$ of integrated luminosity, and is performed inclusively in phase space, which largely washes out the intricate structure of spin observables. 
	
	As we do not expect that constraints obtained from \cite{CMS:2019nrx} will be competitive with the global fit of top data from the LHC \cite{Ethier:2021bye}, we do not perform a global fit employing the state-of-the-art statistical machinery developed in \cite{Ethier:2021bye}, but only consider one SMEFT operator at a time, and a $\chi^2$ parameter estimation. Our fit considers our most accurate SMEFT results, at NLO QCD and quadratic in $c/\Lambda^2$. The most accurate SM predictions available are at NLO QCD + EW \cite{Bernreuther:2015yna} and NNLO QCD \cite{Czakon:2020qbd}. Due to the large uncertainty of the experimental measurement, the use of NNLO QCD or NLO QCD + EW SM makes a negligible difference on the final fit results. The numbers we quote in the following use \cite{Bernreuther:2015yna} as the SM central value. We consider as experimental sources of uncertainty the full statistical and systematic covariance matrices of \cite{CMS:2019nrx}; theory uncertainty includes SM scale variation, SMEFT scale variation, and the full statistical covariance matrix of the SMEFT contribution. We assume Gaussian probability density functions for the uncertainties. \smallskip
	
	The $2 \, \sigma$ allowed range for the Wilson coefficient of all SMEFT operators considered in this work are in the first column of Table \ref{tab:cmsfit}. The range we obtain for $c_{tG}$ is similar to the one already obtained by the CMS Collaboration in \cite{CMS:2019nrx} using a SMEFT theory prediction accurate at LO QCD and linear in $c/\Lambda^2$. As noted above, this is due to the nature of the top chromomagnetic moment operator $\mathcal O_{tG}$, which leads to a significant interference with LO QCD SM amplitudes. A selection of color-octet four-fermion operators has also been constrained in \cite{CMS:2019nrx}. While it is true that, after accounting for the different conventions and restricting our results to LO QCD and linear in $c/\Lambda^2$, we are able to obtain similar exclusion ranges, we find that quadratic SMEFT (and, to a lesser extent, higher order QCD) contributions amount to sizable corrections. In our fit we also constrain all seven directions in the four-fermion color-octet parameter space, compared to the six considered in \cite{CMS:2019nrx}, and constrain the four-fermion color-singlet sector for the first time. 
	
    The range obtained from the global fit of LHC data \cite{Ethier:2021bye} is also reported in Table \ref{tab:cmsfit}. The bounds we obtain from the CMS inclusive measurement \cite{CMS:2019nrx} are as expected significantly worse than the results obtained in global analyses of the top sector as these typically involve over a hundred data points constraining these operators. For most four-fermion operators the constraints we obtain are less stringent than the global fit by up to one order of magnitude, whilst for the chromomagnetic operator $\mathcal O_{tG}$ the bounds are different by a factor of a few. 
	
    To explore the potential of spin observables in constraining SMEFT operators in the future, we obtain estimates for a binned measurement using LHC Run 3 data, proceeding as follows. We assume the binning used by experimental collaborations will be the one presented above, $3 \times 3$ in $\cos \theta - m_{t \bar t}$. The experimental statistical covariance is obtained by rescaling the one in \cite{CMS:2019nrx} to an integrated luminosity of $300 \, \text{fb}^{-1}$ at $\sqrt{s} = 13.6 \, \text{TeV}$. We note that in practice this means setting the statistical uncertainty to zero, as the existing measurement is already systematics-dominated. The systematic covariance matrix is obtained by assuming the systematics of the inclusive measurement \cite{CMS:2019nrx} are replicated in each bin, neglecting bin-to-bin correlations. While it is more than likely that between the early Run 2 measurement \cite{CMS:2019nrx} and the end of Run 3 significant improvements on the systematic uncertainties will be obtained, moving from an inclusive to a differential measurement will inevitably come with a worsening of the experimental resolution. It is impossible, {\it a priori}, to assess precisely either of these two effects. Assigning the early Run 2 systematic uncertainties in each bin of our projected Run 3 measurement is a compromise we consider to be realistic for the near future and conservative in the medium term. We also consider the SM and SMEFT scale variation, as described above, as a source of theory uncertainty. Monte Carlo uncertainty is not considered, as it can be lowered as needed. The bounds we obtain in this way are in the second column of Table \ref{tab:cmsfit}. 
    
    Whilst the bounds obtained from the CMS inclusive measurement \cite{CMS:2019nrx} are typically approximately an order of magnitude broader than those obtained from the global fit, a binned measurement with LHC Run 3 statistics and present-day systematics would yield bounds competitive or better than those in the global fit. 
    Even though we do not perform a global fit, we advocate that spin observables should be included in future global fits. In particular, entanglement markers such as $\Delta^\pm$, being measurable to a significantly greater accuracy than the individual $C_{ij}$'s, should also be considered, and may in fact be the observables yielding the best sensitivity. For instance, in \cite{CMS:2019nrx}, $C_{rr}$ was measured to a relative accuracy of $46\%$, $C_{rk} + C_{kr}$ to $31\%$, while $\Delta^-$ was measured to $5\%$.
    
    The expected improvement in the global fit driven by the inclusion of spin/entanglement measurements is a result of two concurring facts. First, the effect of most operators on the cross-sections $\sigma_S$ are qualitatively different from the effect of the same operators on the total $t \bar t$ cross-section $\sigma$. In several cases, the effects on $\sigma_S$ and $\sigma$ have similar magnitude but opposite sign, so that the resulting shift on the ratio $S = \sigma_S/\sigma$ is enhanced with respect to the shift of $\sigma$ alone. In other cases, as noted in previous Sections, $\sigma_S$ is sensitive to particular combinations of Wilson coefficients, e.g. \eqref{bk_uu} and \eqref{bk_dd} for $B_k + \overline{B_k}$, that do not appear in $\sigma$ at all. Second, the sheer number of new observables, that in fact probe new, never added before, $t \bar t$ degrees of freedom, significantly improves the fit quality.
	
	It is not inconceivable that, assuming differential and doubly-differential spin measurements turn out to be experimentally feasible, in the future the bounds for several Wilson coefficients that enter $t \bar t$ physics will receive significant contributions from spin polarization, spin correlation, and spin entanglement observables. 

\begin{table}[H]
\centering
\scalebox{0.9}{
\begin{tabular}{|c|cc|c|} \hline
\multirow{2}{*}{Operator} & CMS \cite{CMS:2019nrx} & Run III Projection & \multirow{2}{*}{Current Global Fit}  \\
  & $36 \, \text{fb}^{-1}$ Inclusive  & $300 \, \text{fb}^{-1}$ Differential &   \\ \hline
  $\mathcal O_{tG}$ & $[-0.18,0.18]$ & $[-0.03,0.04]$ & $[0.00, 0.11]$ \\ \hline
 $\mathcal O_{tu}^{8}$ & $[-5.8,3.6]$ & $[-1.0, 0.7]$ & $[-0.9, 0.3]$ \\
 $\mathcal O_{td}^{8}$ & $[-7.9,5.2]$ & $[-1.3, 1.0]$ & $[-1.3, 0.6]$ \\
$\mathcal O_{tq}^{8}$ & $[-4.2,3.1]$ & $[-0.7, 0.5]$ & $[-0.5, 0.4]$ \\
$\mathcal O_{Qu}^{8}$ & $[-9.4,4.6]$ & $[-0.7, 0.6]$ & $[-1.0, 0.5]$ \\
$\mathcal O_{Qd}^{8}$ & $[-11.7,5.8]$ & $[-0.9, 0.8]$ & $[-1.6, 0.9]$ \\
$\mathcal O_{Qq}^{(1,8)}$ & $[-5.8,-4.6] \cup [-1.7,2.5]$ & $[-0.4, 0.3]$ & $[-0.4, 0.3]$ \\
$\mathcal O_{Qq}^{(3,8)}$ & $[-5.0,4.2]$ & $[-1.1, 0.8]$ & $[-0.5, 0.4]$ \\ \hline
$\mathcal O_{tu}^{1}$ & $[-2.1,2.1]$ & $[-0.5,0.5]$ & $[-0.4, 0.3]$ \\
 $\mathcal O_{td}^{1}$ & $[-2.7,2.6]$ & $[-0.6, 0.6]$ & $[-0.4, 0.4]$ \\
$\mathcal O_{tq}^{1}$ & $[-1.7,1.8]$ & $[-0.4, 0.4]$ & $[-0.2, 0.3]$ \\
$\mathcal O_{Qu}^{1}$ & $[-2.1,2.4]$ & $[-0.4, 0.5]$ & $[-0.3, 0.4]$ \\
$\mathcal O_{Qd}^{1}$ & $[-2.8, 3.0]$ & $[-0.6, 0.6]$ & $[-0.3, 0.4]$ \\
$\mathcal O_{Qq}^{(1,1)}$ & $[-1.8,1.8]$ & $[-0.4, 0.4]$ & $[-0.3, 0.2]$ \\
$\mathcal O_{Qq}^{(3,1)}$ & $[-1.8,1.8]$ & $[-0.4, 0.4]$ & $[-0.1, 0.2]$ \\ \hline
\end{tabular}
}
\captionsetup{width=\linewidth}
\caption{$2 \sigma$ bounds on Wilson coefficients of SMEFT operators considered in this work. The scale $\Lambda$ is set to 1 TeV.  First column: bounds from the inclusive early LHC Run 2 experimental measurement \cite{CMS:2019nrx}. Second column: projected bounds from one binned, LHC Run 3 measurement. Third column: current individual bounds from the global fit of top LHC data \cite{Ethier:2021bye}.}
\label{tab:cmsfit}
\end{table}

\section{Conclusions} \label{sec:conclusion}

	We have presented a numerical simulation of the effect of new physics on top spin polarization, $t \bar t$ spin correlation, and $t \bar t$ spin entanglement at the LHC, in the framework of the Standard Model Effective Field Theory. Our simulation is accurate at NLO QCD in the $t \bar t$ production, and considers both $\mathcal O(c/\Lambda^2)$ and $\mathcal O(c^2/\Lambda^4)$ effects, which is state of the art. All dimension-6 operators containing top quarks have been considered, with the exception of those highly suppressed by the proton PDF and those {\it only} entering in NLO corrections. This amounts to 15 operators in total. We have presented results for the expected deviations from the SM as a function of all Wilson coefficients for all non-CP-violating terms in the $t \bar t$ density matrix.
	
  We find that the inclusion of NP-squared $\mathcal O(c^2/\Lambda^4)$ contributions to cross-sections is almost always important for describing the effects of SMEFT operators on all spin observables. This is true not only for color-singlet operators, whose interference with the LO SM QCD amplitude vanishes, but also for the color-octets, in the range of Wilson coefficients still allowed by other measurements. The notable exception is the top chromomagnetic moment operator $\mathcal O_{tG}$, for which a LO analysis is found to be adequate. Next-to-leading order QCD corrections, evaluated here for the first time, do not radically change the patterns seen at LO, but the appearance of relatively large $K$-factors on several cross-sections often makes LO scale variation a poor estimate of missing higher orders. As expected, NLO corrections also significantly reduce the uncertainties due to missing higher orders as quantified by scale variations.
	
  We presented results both for an inclusive measurement, and for a doubly-differential measurement in the top pair phase space. Already at the inclusive level, we find that spin observables depend on Wilson coefficients in a largely different way than the total $t \bar t$ cross-section, thus providing a novel sensitivity to new physics in the top sector. Our second scenario is for a measurement binned in the top pair invariant mass and in the top scattering angle. All spin observables show an intricate structure in phase space. The kinematical region that seems most promising for the detection of NP is the one at large top $p_T$, already considered in the context of detection of quantum effects, such as entanglement between the tops' spin and possible violations of Bell Inequalities. 

    Only one experimental result is available as of this writing, an early Run 2 measurement from the CMS Collaboration \cite{CMS:2019nrx}, where the $t \bar t$ spin density matrix has been fully measured, however without any phase space cuts. No significant deviation from the SM was observed. We argue that this should not be taken as discouraging, as the effects induced by new physics are localized in phase-space, and are washed out by an inclusive measurement.
    
    Using our results and precision SM calculations, we extract bounds for Wilson coefficients from the CMS measurement \cite{CMS:2019nrx}, and produce a projection for an LHC Run 3 differential measurement. We find that just one doubly-differential measurement of the $t \bar t$ spin density matrix would yield constraints competitive with the current full global fit to LHC data.
    
    Apart from the prospects of detecting quantum entanglement at the TeV scale \cite{Afik:2020onf, Afik:2022kwm, Aguilar-Saavedra:2022uye}, and possibly observing Bell Inequalities violations \cite{Fabbrichesi:2021npl, Severi:2021cnj}, we have shown how the inclusion of spin measurements in the physics program of experimental collaborations would yield a novel and promising road for detecting beyond the SM effects, and provided a systematic and complete framework for these studies.

\section{Acknowledgements}
    We are grateful to Fabio Maltoni and Luca Mantani for comments on the manuscript, and to Rafael Aoude and Olivier Mattelaer for useful discussions. We also thank Ioannis Tsinikos for collaboration at the early stages of this work, and Rikkert Frederix and Timea Vitos for the valuable help.  The Authors are supported by the European Union’s Horizon 2020 research and innovation program under the EFT4NP project (grant agreement no. 949451) and by a Royal Society University Research Fellowship through grant URF/R1/201553.

\clearpage

\appendix

\section{Numerical results inclusive in phase space} \label{app:inclusive}

The following Tables \ref{tab:SM_DIM2F_24} - \ref{tab:DIM64F_8} collect numerical results needed to reproduce our plots in Figures \ref{fig:DIM62F_24} - \ref{fig:DIM64F_7_8}. We present all cross sections at LO and NLO appearing in the numerator and denominator of \eqref{ratio} for the spin polarization observables $B_i + \bar B_i$, and the spin correlation observables $C_{ii}$ and $C_{ij} + C_{ji}$, for all operators we considered. The first quoted uncertainty, in brackets, is Monte Carlo, in units of the least significant digit of the central value. The second one, $^{+x}_{-y}$, is scale variation\footnote{The scale variation we quote on \eqref{ratio} is {\it not} obtained by propagating the scale variation from each term, but it is extracted by evaluating the full ratio at each scale choice.}.

\begin{table}[H]
\centering
\scalebox{0.82}
{
\begin{tabular}{cc|r|rr}
  & & SM & \multicolumn{2}{c}{$\mathcal O_{tG}$} \\ \hline
Observable & Order & $\sigma^{(0)}$ [pb] & $\sigma^{(1)}$ [pb] & $\sigma^{(2)}$ [pb] \\ \hline
\multirow{ 2}{*}{ $\sigma$ }& \small{LO} & $ 29.12 \scriptstyle{(2)} \, \displaystyle{^{+9.10}_{-6.49}}  $   & $ 9.32 \scriptstyle{(7)} \, \displaystyle{^{+2.90}_{-2.08}}  $ & $ 1.63 \scriptstyle{(1)} \, \displaystyle{^{+0.52}_{-0.37}}  $  \\ 
 & \small{NLO} & $ 36.23 \scriptstyle{(2)} \, \displaystyle{^{+1.70}_{-3.14}}  $   & $ 11.50 \scriptstyle{(6)} \, \displaystyle{^{+0.51}_{-0.97}}  $ & $ 2.01 \scriptstyle{(1)} \, \displaystyle{^{+0.09}_{-0.17}}  $  \\ \hline
\multirow{ 2}{*}{ $\sigma_{\displaystyle B_{n}+\bar{B}_{n}}$ }& \small{LO} & $ -0.09 \scriptstyle{(5)} \, \displaystyle{^{+0.02}_{-0.03}}  $   & $ -0.01 \scriptstyle{(19)} \, \displaystyle{^{+0.00}_{-0.00}}  $ & $ -0.00 \scriptstyle{(4)} \, \displaystyle{^{+0.00}_{-0.00}}  $  \\ 
 & \small{NLO} & $ -0.04 \scriptstyle{(6)} \, \displaystyle{^{+0.00}_{-0.01}}  $   & $ 0.00 \scriptstyle{(24)} \, \displaystyle{^{+0.01}_{-0.00}}  $ & $ 0.01 \scriptstyle{(5)} \, \displaystyle{^{+0.00}_{-0.00}}  $  \\ \hline
\multirow{ 2}{*}{ $\sigma_{\displaystyle B_{r}+\bar{B}_{r}}$ }& \small{LO} & $ 0.04 \scriptstyle{(6)} \, \displaystyle{^{+0.01}_{-0.01}}  $   & $ -0.01 \scriptstyle{(23)} \, \displaystyle{^{+0.00}_{-0.00}}  $ & $ -0.00 \scriptstyle{(4)} \, \displaystyle{^{+0.00}_{-0.00}}  $  \\ 
 & \small{NLO} & $ 0.03 \scriptstyle{(7)} \, \displaystyle{^{+0.03}_{-0.02}}  $   & $ -0.01 \scriptstyle{(27)} \, \displaystyle{^{+0.01}_{-0.01}}  $ & $ -0.00 \scriptstyle{(5)} \, \displaystyle{^{+0.00}_{-0.00}}  $  \\
\hline
\end{tabular}
}
\captionsetup{width=\linewidth}
\caption{Numerical results for the Standard Model and for the operator $\mathcal O_{tG}$.}
\label{tab:SM_DIM2F_24}
\end{table}

\begin{table}[H]
\centering
\scalebox{0.82}
{
\begin{tabular}{cc|rr|rr}
  & & \multicolumn{2}{c}{$\mathcal O_{Qq}^{(8,3)}$} & \multicolumn{2}{c}{$\mathcal O_{Qq}^{(1,3)}$} \\ \hline
Observable & Order & $\sigma^{(1)}$ [pb] & $\sigma^{(2)}$ [pb] & $\sigma^{(1)}$ [pb] &  $\sigma^{(2)}$ [pb] \\ \hline
\multirow{ 2}{*}{ $\sigma$ }& \small{LO} & $ 0.073 \scriptstyle{(2)} \, \displaystyle{^{+0.008}_{-0.007}}  $ & $ 0.075 \scriptstyle{(0)} \, \displaystyle{^{+0.004}_{-0.004}}  $  &   $ 0.006 \scriptstyle{(4)} \, \displaystyle{^{+0.002}_{-0.001}}  $ & $ 0.334 \scriptstyle{(1)} \, \displaystyle{^{+0.019}_{-0.017}}  $  \\ 
 & \small{NLO} & $ 0.059 \scriptstyle{(3)} \, \displaystyle{^{+0.002}_{-0.005}}  $ & $ 0.072 \scriptstyle{(1)} \, \displaystyle{^{+0.002}_{-0.001}}  $  &   $ -0.002 \scriptstyle{(3)} \, \displaystyle{^{+0.001}_{-0.000}}  $ & $ 0.476 \scriptstyle{(1)} \, \displaystyle{^{+0.025}_{-0.022}}  $  \\ \hline
\multirow{ 2}{*}{ $\sigma_{\displaystyle B_{n}+\bar{B}_{n}}$ }& \small{LO} & $ 0.002 \scriptstyle{(7)} \, \displaystyle{^{+0.000}_{-0.000}}  $ & $ -0.003 \scriptstyle{(2)} \, \displaystyle{^{+0.001}_{-0.001}}  $  &   $ -0.002 \scriptstyle{(12)} \, \displaystyle{^{+0.002}_{-0.001}}  $ & $ -0.006 \scriptstyle{(3)} \, \displaystyle{^{+0.001}_{-0.001}}  $  \\ 
 & \small{NLO} & $ -0.009 \scriptstyle{(10)} \, \displaystyle{^{+0.004}_{-0.007}}  $ & $ 0.000 \scriptstyle{(2)} \, \displaystyle{^{+0.001}_{-0.000}}  $  &   $ 0.022 \scriptstyle{(13)} \, \displaystyle{^{+0.010}_{-0.006}}  $ & $ 0.003 \scriptstyle{(3)} \, \displaystyle{^{+0.001}_{-0.001}}  $  \\ \hline
\multirow{ 2}{*}{ $\sigma_{\displaystyle B_{r}+\bar{B}_{r}}$ }& \small{LO} & $ -0.008 \scriptstyle{(7)} \, \displaystyle{^{+0.001}_{-0.001}}  $ & $ -0.011 \scriptstyle{(2)} \, \displaystyle{^{+0.001}_{-0.000}}  $  &   $ -0.002 \scriptstyle{(15)} \, \displaystyle{^{+0.003}_{-0.005}}  $ & $ -0.050 \scriptstyle{(3)} \, \displaystyle{^{+0.003}_{-0.003}}  $  \\
\hline
\end{tabular}
}
\captionsetup{width=\linewidth}
\caption{Numerical results for the operators $\mathcal O_{Qq}^{(8,3)}$ (left) and $\mathcal O_{Qq}^{(1,3)}$ (right).}
\label{tab:DIM64F_1}
\end{table}

\begin{table}[H]
\centering
\scalebox{0.82}
{
\begin{tabular}{cc|rr|rr}
  & & \multicolumn{2}{c}{$\mathcal O_{Qq}^{(8,1)}$} & \multicolumn{2}{c}{$\mathcal O_{Qq}^{(1,1)}$} \\ \hline
Observable & Order & $\sigma^{(1)}$ [pb] & $\sigma^{(2)}$ [pb] & $\sigma^{(1)}$ [pb] &  $\sigma^{(2)}$ [pb] \\ \hline
\multirow{ 2}{*}{ $\sigma$ }& \small{LO} & $ 0.385 \scriptstyle{(2)} \, \displaystyle{^{+0.046}_{-0.039}}  $ & $ 0.075 \scriptstyle{(0)} \, \displaystyle{^{+0.004}_{-0.004}}  $  &   $ -0.002 \scriptstyle{(2)} \, \displaystyle{^{+0.000}_{-0.000}}  $ & $ 0.335 \scriptstyle{(1)} \, \displaystyle{^{+0.019}_{-0.017}}  $  \\ 
 & \small{NLO} & $ 0.345 \scriptstyle{(3)} \, \displaystyle{^{+0.003}_{-0.013}}  $ & $ 0.078 \scriptstyle{(1)} \, \displaystyle{^{+0.002}_{-0.001}}  $  &   $ -0.019 \scriptstyle{(3)} \, \displaystyle{^{+0.004}_{-0.006}}  $ & $ 0.478 \scriptstyle{(1)} \, \displaystyle{^{+0.026}_{-0.022}}  $  \\ \hline
\multirow{ 2}{*}{ $\sigma_{\displaystyle B_{n}+\bar{B}_{n}}$ }& \small{LO} & $ -0.013 \scriptstyle{(7)} \, \displaystyle{^{+0.002}_{-0.003}}  $ & $ 0.001 \scriptstyle{(2)} \, \displaystyle{^{+0.001}_{-0.000}}  $  &   $ -0.006 \scriptstyle{(10)} \, \displaystyle{^{+0.001}_{-0.001}}  $ & $ 0.001 \scriptstyle{(2)} \, \displaystyle{^{+0.000}_{-0.000}}  $  \\ 
 & \small{NLO} & $ 0.005 \scriptstyle{(10)} \, \displaystyle{^{+0.003}_{-0.001}}  $ & $ 0.001 \scriptstyle{(2)} \, \displaystyle{^{+0.001}_{-0.001}}  $  &   $ -0.005 \scriptstyle{(12)} \, \displaystyle{^{+0.003}_{-0.005}}  $ & $ -0.002 \scriptstyle{(3)} \, \displaystyle{^{+0.001}_{-0.002}}  $  \\ \hline
\multirow{ 2}{*}{ $\sigma_{\displaystyle B_{r}+\bar{B}_{r}}$ }& \small{LO} & $ -0.062 \scriptstyle{(9)} \, \displaystyle{^{+0.004}_{-0.005}}  $ & $ -0.012 \scriptstyle{(2)} \, \displaystyle{^{+0.001}_{-0.001}}  $  &   $ 0.009 \scriptstyle{(11)} \, \displaystyle{^{+0.004}_{-0.002}}  $ & $ -0.054 \scriptstyle{(3)} \, \displaystyle{^{+0.002}_{-0.002}}  $  \\
\hline
\end{tabular}
}
\captionsetup{width=\linewidth}
\caption{Numerical results for the operators $\mathcal O_{Qq}^{(8,1)}$ (left) and $\mathcal O_{Qq}^{(1,1)}$ (right).}
\label{tab:DIM64F_2}
\end{table}

\begin{table}[H]
\centering
\scalebox{0.82}
{
\begin{tabular}{cc|rr|rr}
  & & \multicolumn{2}{c}{$\mathcal O_{Qu}^{8}$} & \multicolumn{2}{c}{$\mathcal O_{Qu}^{1}$} \\ \hline
Observable & Order & $\sigma^{(1)}$ [pb] & $\sigma^{(2)}$ [pb] & $\sigma^{(1)}$ [pb] &  $\sigma^{(2)}$ [pb] \\ \hline
\multirow{ 2}{*}{ $\sigma$ }& \small{LO} & $ 0.230 \scriptstyle{(2)} \, \displaystyle{^{+0.027}_{-0.023}}  $ & $ 0.046 \scriptstyle{(0)} \, \displaystyle{^{+0.003}_{-0.002}}  $  &   $ 0.001 \scriptstyle{(3)} \, \displaystyle{^{+0.000}_{-0.000}}  $ & $ 0.207 \scriptstyle{(1)} \, \displaystyle{^{+0.012}_{-0.011}}  $  \\ 
 & \small{NLO} & $ 0.193 \scriptstyle{(2)} \, \displaystyle{^{+0.005}_{-0.012}}  $ & $ 0.034 \scriptstyle{(1)} \, \displaystyle{^{+0.001}_{-0.002}}  $  &   $ -0.020 \scriptstyle{(3)} \, \displaystyle{^{+0.004}_{-0.006}}  $ & $ 0.296 \scriptstyle{(1)} \, \displaystyle{^{+0.016}_{-0.013}}  $  \\ \hline
\multirow{ 2}{*}{ $\sigma_{\displaystyle B_{n}+\bar{B}_{n}}$ }& \small{LO} & $ -0.003 \scriptstyle{(8)} \, \displaystyle{^{+0.001}_{-0.002}}  $ & $ 0.003 \scriptstyle{(2)} \, \displaystyle{^{+0.001}_{-0.001}}  $  &   $ 0.006 \scriptstyle{(10)} \, \displaystyle{^{+0.001}_{-0.001}}  $ & $ -0.001 \scriptstyle{(3)} \, \displaystyle{^{+0.000}_{-0.000}}  $  \\ 
 & \small{NLO} & $ 0.003 \scriptstyle{(10)} \, \displaystyle{^{+0.004}_{-0.005}}  $ & $ -0.002 \scriptstyle{(3)} \, \displaystyle{^{+0.001}_{-0.002}}  $  &   $ -0.004 \scriptstyle{(10)} \, \displaystyle{^{+0.003}_{-0.004}}  $ & $ -0.003 \scriptstyle{(3)} \, \displaystyle{^{+0.001}_{-0.001}}  $  \\ \hline
\multirow{ 2}{*}{ $\sigma_{\displaystyle B_{r}+\bar{B}_{r}}$ }& \small{LO} & $ -0.033 \scriptstyle{(10)} \, \displaystyle{^{+0.002}_{-0.003}}  $ & $ -0.008 \scriptstyle{(2)} \, \displaystyle{^{+0.001}_{-0.001}}  $  &   $ -0.000 \scriptstyle{(11)} \, \displaystyle{^{+0.001}_{-0.001}}  $ & $ -0.030 \scriptstyle{(3)} \, \displaystyle{^{+0.001}_{-0.001}}  $  \\
\hline
\end{tabular}
}
\captionsetup{width=\linewidth}
\caption{Numerical results for the operators $\mathcal O_{Qu}^{8}$ (left) and $\mathcal O_{Qu}^1$ (right).}
\label{tab:DIM64F_3}
\end{table}

\begin{table}[H]
\centering
\scalebox{0.82}
{
\begin{tabular}{cc|rr|rr}
  & & \multicolumn{2}{c}{$\mathcal O_{tq}^{8}$} & \multicolumn{2}{c}{$\mathcal O_{tq}^{1}$} \\ \hline
Observable & Order & $\sigma^{(1)}$ [pb] & $\sigma^{(2)}$ [pb] & $\sigma^{(1)}$ [pb] &  $\sigma^{(2)}$ [pb] \\ \hline
\multirow{ 2}{*}{ $\sigma$ }& \small{LO} & $ 0.380 \scriptstyle{(1)} \, \displaystyle{^{+0.045}_{-0.038}}  $ & $ 0.074 \scriptstyle{(0)} \, \displaystyle{^{+0.004}_{-0.004}}  $  &   $ -0.001 \scriptstyle{(3)} \, \displaystyle{^{+0.000}_{-0.000}}  $ & $ 0.335 \scriptstyle{(1)} \, \displaystyle{^{+0.019}_{-0.017}}  $  \\ 
 & \small{NLO} & $ 0.340 \scriptstyle{(3)} \, \displaystyle{^{+0.005}_{-0.015}}  $ & $ 0.056 \scriptstyle{(1)} \, \displaystyle{^{+0.002}_{-0.003}}  $  &   $ -0.030 \scriptstyle{(3)} \, \displaystyle{^{+0.005}_{-0.008}}  $ & $ 0.475 \scriptstyle{(1)} \, \displaystyle{^{+0.025}_{-0.022}}  $  \\ \hline
\multirow{ 2}{*}{ $\sigma_{\displaystyle B_{n}+\bar{B}_{n}}$ }& \small{LO} & $ -0.002 \scriptstyle{(9)} \, \displaystyle{^{+0.000}_{-0.001}}  $ & $ -0.007 \scriptstyle{(2)} \, \displaystyle{^{+0.001}_{-0.002}}  $  &   $ -0.013 \scriptstyle{(11)} \, \displaystyle{^{+0.003}_{-0.005}}  $ & $ 0.000 \scriptstyle{(2)} \, \displaystyle{^{+0.001}_{-0.001}}  $  \\ 
 & \small{NLO} & $ 0.011 \scriptstyle{(11)} \, \displaystyle{^{+0.007}_{-0.004}}  $ & $ 0.001 \scriptstyle{(3)} \, \displaystyle{^{+0.001}_{-0.001}}  $  &   $ 0.009 \scriptstyle{(13)} \, \displaystyle{^{+0.002}_{-0.001}}  $ & $ 0.002 \scriptstyle{(3)} \, \displaystyle{^{+0.001}_{-0.001}}  $  \\ \hline
\multirow{ 2}{*}{ $\sigma_{\displaystyle B_{r}+\bar{B}_{r}}$ }& \small{LO} & $ 0.049 \scriptstyle{(9)} \, \displaystyle{^{+0.003}_{-0.002}}  $ & $ 0.008 \scriptstyle{(2)} \, \displaystyle{^{+0.000}_{-0.000}}  $  &   $ -0.006 \scriptstyle{(12)} \, \displaystyle{^{+0.001}_{-0.001}}  $ & $ 0.050 \scriptstyle{(3)} \, \displaystyle{^{+0.002}_{-0.002}}  $  \\
\hline
\end{tabular}
}
\captionsetup{width=\linewidth}
\caption{Numerical results for the operators $\mathcal O_{tq}^{8}$ (left) and $\mathcal O_{tq}^1$ (right).}
\label{tab:DIM64F_4}
\end{table}

\begin{table}[H]
\centering
\scalebox{0.82}
{
\begin{tabular}{cc|rr|rr}
  & & \multicolumn{2}{c}{$\mathcal O_{Qd}^{8}$} & \multicolumn{2}{c}{$\mathcal O_{Qd}^{1}$} \\ \hline
Observable & Order & $\sigma^{(1)}$ [pb] & $\sigma^{(2)}$ [pb] & $\sigma^{(1)}$ [pb] &  $\sigma^{(2)}$ [pb] \\ \hline
\multirow{ 2}{*}{ $\sigma$ }& \small{LO} & $ 0.158 \scriptstyle{(1)} \, \displaystyle{^{+0.018}_{-0.016}}  $ & $ 0.028 \scriptstyle{(0)} \, \displaystyle{^{+0.002}_{-0.002}}  $  &   $ 0.001 \scriptstyle{(3)} \, \displaystyle{^{+0.000}_{-0.000}}  $ & $ 0.128 \scriptstyle{(1)} \, \displaystyle{^{+0.007}_{-0.007}}  $  \\ 
 & \small{NLO} & $ 0.148 \scriptstyle{(3)} \, \displaystyle{^{+0.001}_{-0.004}}  $ & $ 0.023 \scriptstyle{(1)} \, \displaystyle{^{+0.001}_{-0.000}}  $  &   $ 0.002 \scriptstyle{(3)} \, \displaystyle{^{+0.001}_{-0.001}}  $ & $ 0.182 \scriptstyle{(1)} \, \displaystyle{^{+0.009}_{-0.008}}  $  \\ \hline
\multirow{ 2}{*}{ $\sigma_{\displaystyle B_{n}+\bar{B}_{n}}$ }& \small{LO} & $ -0.007 \scriptstyle{(9)} \, \displaystyle{^{+0.002}_{-0.002}}  $ & $ -0.003 \scriptstyle{(2)} \, \displaystyle{^{+0.001}_{-0.001}}  $  &   $ -0.000 \scriptstyle{(9)} \, \displaystyle{^{+0.001}_{-0.001}}  $ & $ -0.001 \scriptstyle{(2)} \, \displaystyle{^{+0.000}_{-0.000}}  $  \\ 
 & \small{NLO} & $ 0.005 \scriptstyle{(10)} \, \displaystyle{^{+0.006}_{-0.003}}  $ & $ -0.001 \scriptstyle{(2)} \, \displaystyle{^{+0.001}_{-0.001}}  $  &   $ -0.003 \scriptstyle{(10)} \, \displaystyle{^{+0.004}_{-0.007}}  $ & $ 0.003 \scriptstyle{(3)} \, \displaystyle{^{+0.001}_{-0.001}}  $  \\ \hline
\multirow{ 2}{*}{ $\sigma_{\displaystyle B_{r}+\bar{B}_{r}}$ }& \small{LO} & $ -0.020 \scriptstyle{(9)} \, \displaystyle{^{+0.001}_{-0.002}}  $ & $ -0.005 \scriptstyle{(2)} \, \displaystyle{^{+0.001}_{-0.001}}  $  &   $ 0.011 \scriptstyle{(9)} \, \displaystyle{^{+0.003}_{-0.002}}  $ & $ -0.020 \scriptstyle{(2)} \, \displaystyle{^{+0.001}_{-0.001}}  $  \\
\hline
\end{tabular}
}
\captionsetup{width=\linewidth}
\caption{Numerical results for the operators $\mathcal O_{Qd}^{8}$ (left) and $\mathcal O_{Qd}^1$ (right).}
\label{tab:DIM64F_6}
\end{table}

\begin{table}[H]
\centering
\scalebox{0.82}
{
\begin{tabular}{cc|rr|rr}
  & & \multicolumn{2}{c}{$\mathcal O_{tu}^{8}$} & \multicolumn{2}{c}{$\mathcal O_{tu}^{1}$} \\ \hline
Observable & Order & $\sigma^{(1)}$ [pb] & $\sigma^{(2)}$ [pb] & $\sigma^{(1)}$ [pb] &  $\sigma^{(2)}$ [pb] \\ \hline
\multirow{ 2}{*}{ $\sigma$ }& \small{LO} & $ 0.220 \scriptstyle{(2)} \, \displaystyle{^{+0.025}_{-0.021}}  $ & $ 0.045 \scriptstyle{(0)} \, \displaystyle{^{+0.003}_{-0.002}}  $  &   $ 0.002 \scriptstyle{(2)} \, \displaystyle{^{+0.000}_{-0.000}}  $ & $ 0.206 \scriptstyle{(0)} \, \displaystyle{^{+0.012}_{-0.011}}  $  \\ 
 & \small{NLO} & $ 0.205 \scriptstyle{(3)} \, \displaystyle{^{+0.002}_{-0.008}}  $ & $ 0.047 \scriptstyle{(1)} \, \displaystyle{^{+0.001}_{-0.001}}  $  &   $ -0.000 \scriptstyle{(3)} \, \displaystyle{^{+0.001}_{-0.001}}  $ & $ 0.296 \scriptstyle{(1)} \, \displaystyle{^{+0.015}_{-0.013}}  $  \\ \hline
\multirow{ 2}{*}{ $\sigma_{\displaystyle B_{n}+\bar{B}_{n}}$ }& \small{LO} & $ 0.007 \scriptstyle{(7)} \, \displaystyle{^{+0.002}_{-0.001}}  $ & $ -0.002 \scriptstyle{(2)} \, \displaystyle{^{+0.000}_{-0.001}}  $  &   $ -0.005 \scriptstyle{(9)} \, \displaystyle{^{+0.000}_{-0.001}}  $ & $ -0.002 \scriptstyle{(2)} \, \displaystyle{^{+0.000}_{-0.000}}  $  \\ 
 & \small{NLO} & $ 0.000 \scriptstyle{(10)} \, \displaystyle{^{+0.003}_{-0.002}}  $ & $ -0.002 \scriptstyle{(3)} \, \displaystyle{^{+0.001}_{-0.001}}  $  &   $ 0.012 \scriptstyle{(10)} \, \displaystyle{^{+0.005}_{-0.003}}  $ & $ -0.004 \scriptstyle{(2)} \, \displaystyle{^{+0.001}_{-0.002}}  $  \\ \hline
\multirow{ 2}{*}{ $\sigma_{\displaystyle B_{r}+\bar{B}_{r}}$ }& \small{LO} & $ 0.031 \scriptstyle{(10)} \, \displaystyle{^{+0.001}_{-0.001}}  $ & $ 0.009 \scriptstyle{(2)} \, \displaystyle{^{+0.001}_{-0.001}}  $  &   $ -0.006 \scriptstyle{(9)} \, \displaystyle{^{+0.002}_{-0.003}}  $ & $ 0.032 \scriptstyle{(2)} \, \displaystyle{^{+0.001}_{-0.001}}  $  \\
\hline
\end{tabular}
}
\captionsetup{width=\linewidth}
\caption{Numerical results for the operators $\mathcal O_{tu}^{8}$ (left) and $\mathcal O_{tu}^1$ (right).}
\label{tab:DIM64F_7}
\end{table}

\begin{table}[H]
\centering
\scalebox{0.82}
{
\begin{tabular}{cc|rr|rr}
  & & \multicolumn{2}{c}{$\mathcal O_{td}^{8}$} & \multicolumn{2}{c}{$\mathcal O_{td}^{1}$} \\ \hline
Observable & Order & $\sigma^{(1)}$ [pb] & $\sigma^{(2)}$ [pb] & $\sigma^{(1)}$ [pb] &  $\sigma^{(2)}$ [pb] \\ \hline
\multirow{ 2}{*}{ $\sigma$ }& \small{LO} & $ 0.161 \scriptstyle{(2)} \, \displaystyle{^{+0.020}_{-0.017}}  $ & $ 0.030 \scriptstyle{(0)} \, \displaystyle{^{+0.002}_{-0.002}}  $  &   $ 0.001 \scriptstyle{(3)} \, \displaystyle{^{+0.000}_{-0.000}}  $ & $ 0.129 \scriptstyle{(1)} \, \displaystyle{^{+0.007}_{-0.007}}  $  \\ 
 & \small{NLO} & $ 0.139 \scriptstyle{(3)} \, \displaystyle{^{+0.003}_{-0.009}}  $ & $ 0.030 \scriptstyle{(1)} \, \displaystyle{^{+0.002}_{-0.001}}  $  &   $ -0.018 \scriptstyle{(3)} \, \displaystyle{^{+0.004}_{-0.005}}  $ & $ 0.183 \scriptstyle{(1)} \, \displaystyle{^{+0.010}_{-0.008}}  $  \\ \hline
\multirow{ 2}{*}{ $\sigma_{\displaystyle B_{n}+\bar{B}_{n}}$ }& \small{LO} & $ -0.005 \scriptstyle{(8)} \, \displaystyle{^{+0.001}_{-0.001}}  $ & $ -0.002 \scriptstyle{(2)} \, \displaystyle{^{+0.000}_{-0.001}}  $  &   $ 0.005 \scriptstyle{(8)} \, \displaystyle{^{+0.002}_{-0.002}}  $ & $ -0.002 \scriptstyle{(2)} \, \displaystyle{^{+0.001}_{-0.001}}  $  \\ 
 & \small{NLO} & $ 0.006 \scriptstyle{(10)} \, \displaystyle{^{+0.001}_{-0.001}}  $ & $ -0.001 \scriptstyle{(2)} \, \displaystyle{^{+0.000}_{-0.000}}  $  &   $ -0.011 \scriptstyle{(10)} \, \displaystyle{^{+0.005}_{-0.007}}  $ & $ -0.002 \scriptstyle{(2)} \, \displaystyle{^{+0.000}_{-0.000}}  $  \\ \hline
\multirow{ 2}{*}{ $\sigma_{\displaystyle B_{r}+\bar{B}_{r}}$ }& \small{LO} & $ 0.026 \scriptstyle{(8)} \, \displaystyle{^{+0.002}_{-0.002}}  $ & $ 0.005 \scriptstyle{(2)} \, \displaystyle{^{+0.000}_{-0.000}}  $  &   $ 0.006 \scriptstyle{(9)} \, \displaystyle{^{+0.002}_{-0.002}}  $ & $ 0.021 \scriptstyle{(3)} \, \displaystyle{^{+0.001}_{-0.001}}  $  \\
\hline
\end{tabular}
}
\captionsetup{width=\linewidth}
\caption{Numerical results for the operators $\mathcal O_{td}^{8}$ (left) and $\mathcal O_{td}^1$ (right).}
\label{tab:DIM64F_8}
\end{table}

\clearpage

\section{Additional differential plots} \label{app:binned}

This Appendix completes Section \ref{sec:differential}, showing results differential in $t \bar t$ phase space for all other dimension-6 operators considered in this work.

\begin{figure}[H]
\centering
\includegraphics[width=0.82\linewidth]{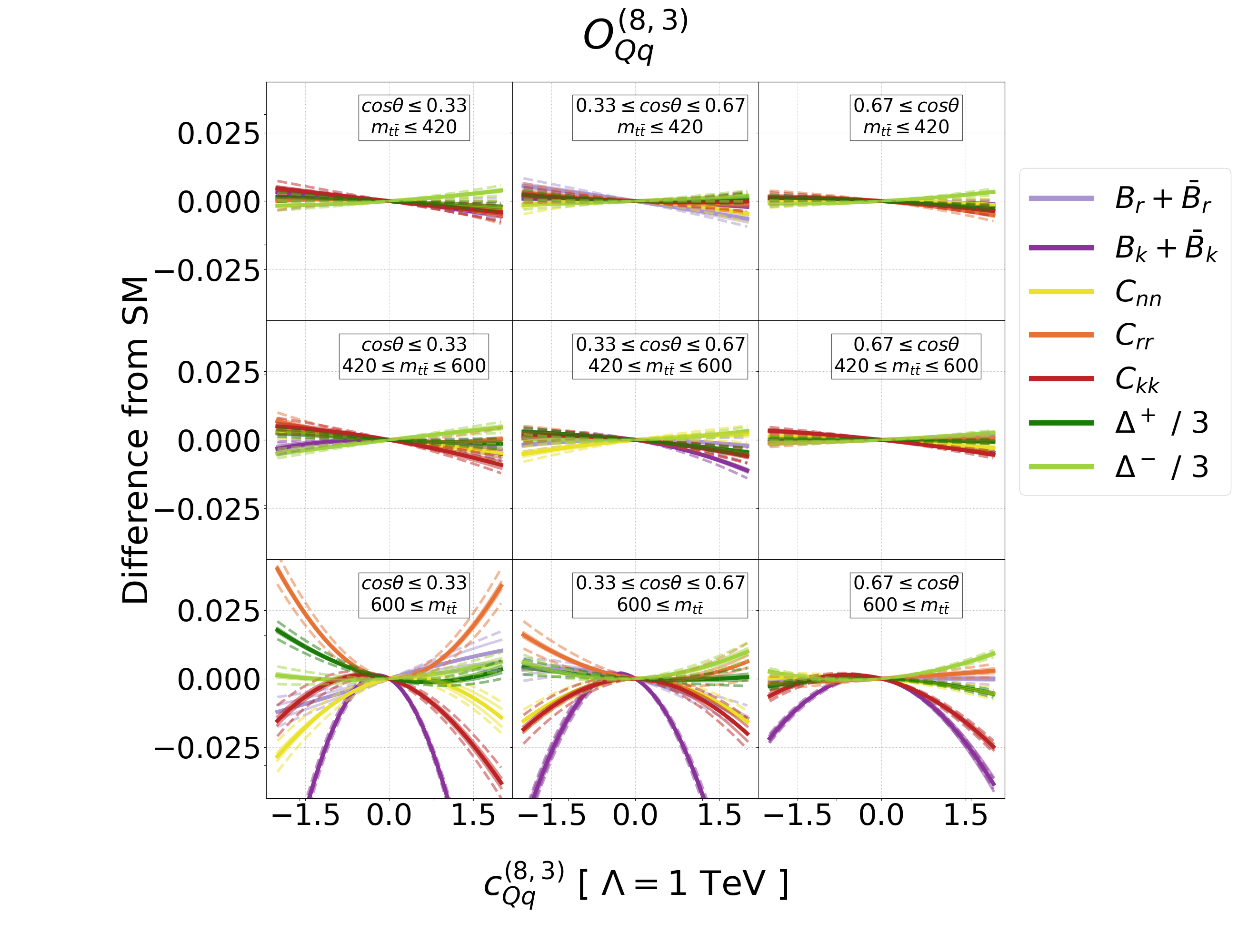}
\includegraphics[width=0.82\linewidth]{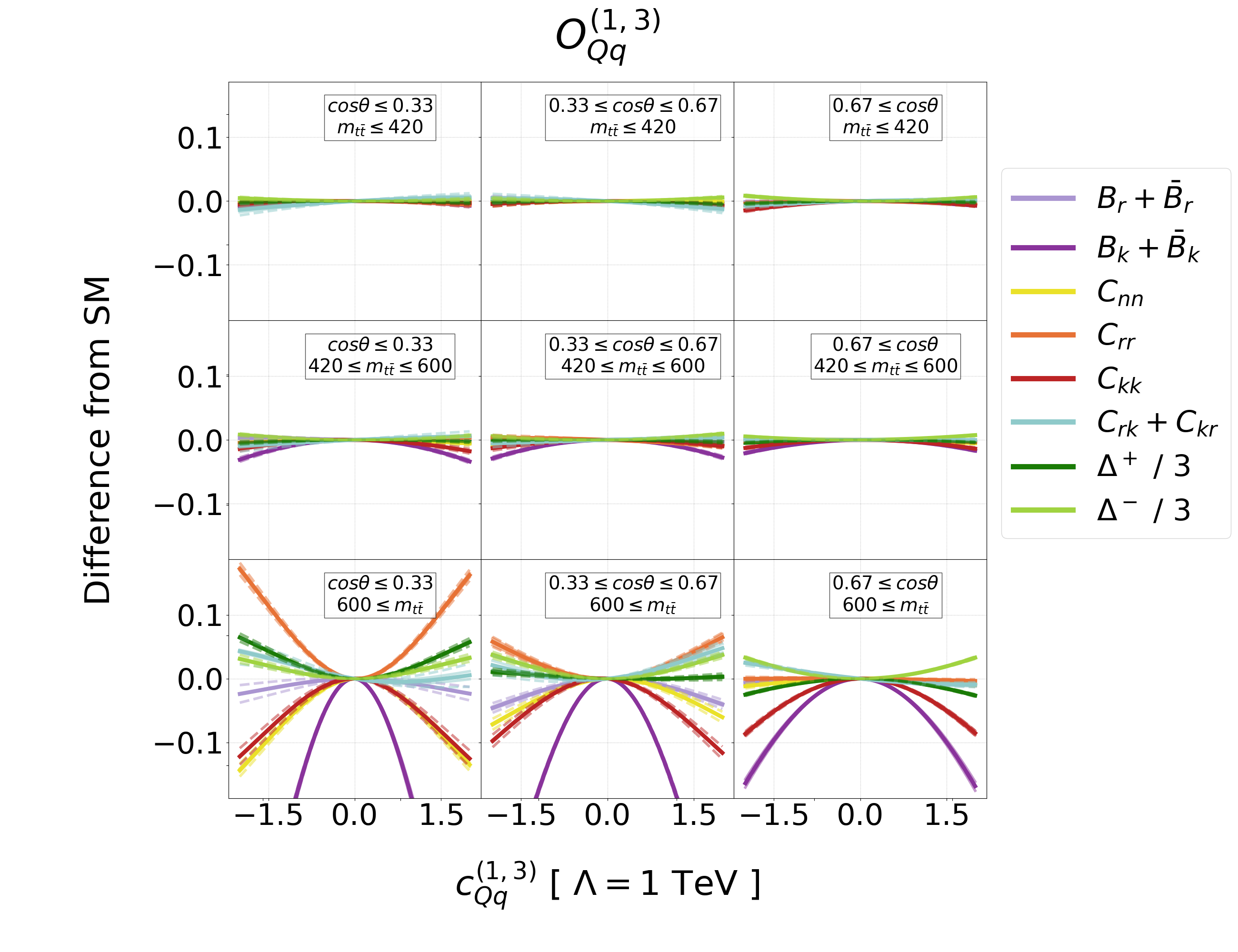}
 \vspace{-5mm}
\captionsetup{width=\linewidth}
\caption{Results for the operators $\mathcal O_{Qq}^{(8,3)}$ (top) and $\mathcal O_{Qq}^{(1,3)}$ (bottom), similar to Figure \ref{fig:DIM62F_24_binned}.}
\label{fig:DIM64F_1_binned}
\end{figure}

\begin{figure}[H]
\centering
\includegraphics[width=0.82\linewidth]{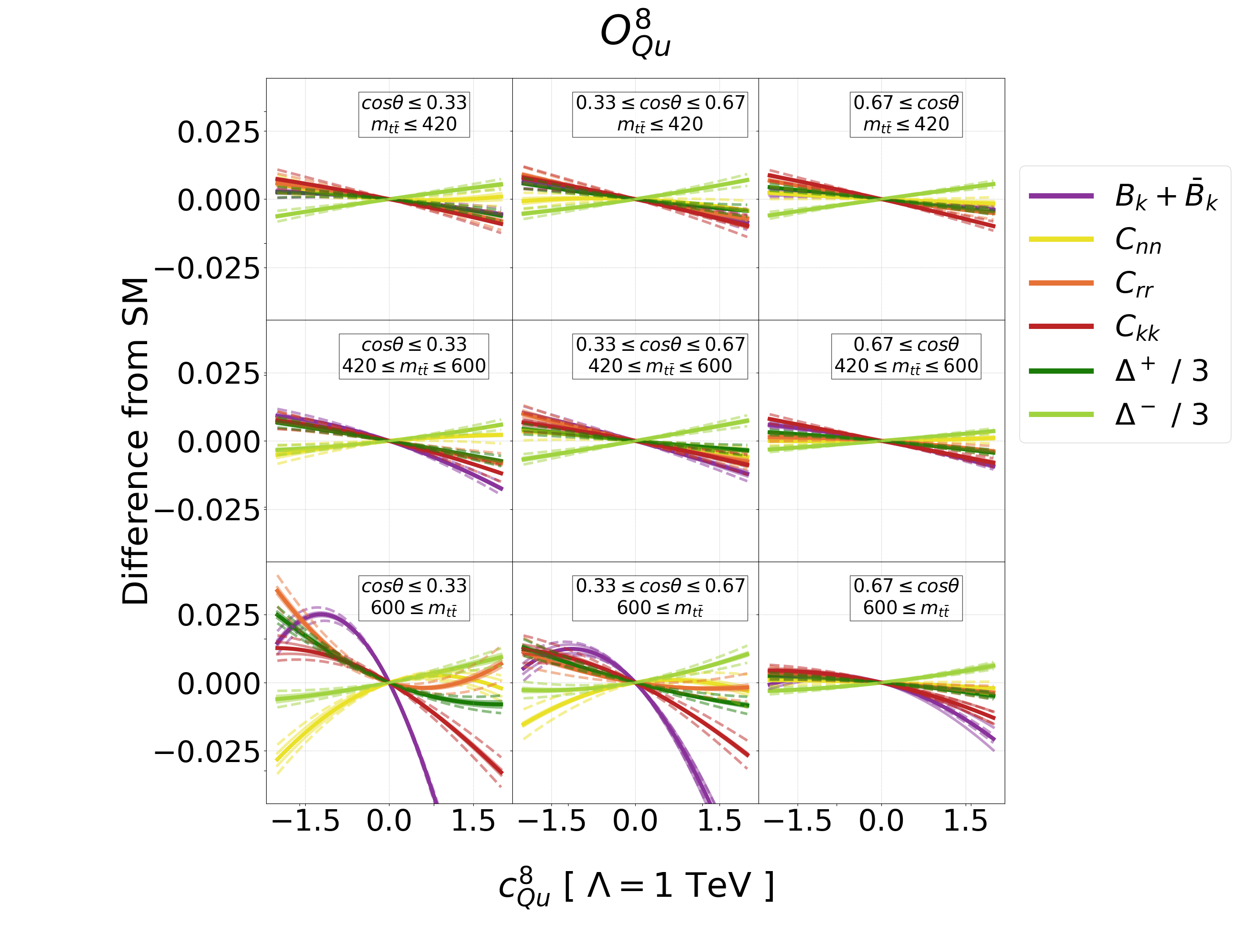}
\includegraphics[width=0.82\linewidth]{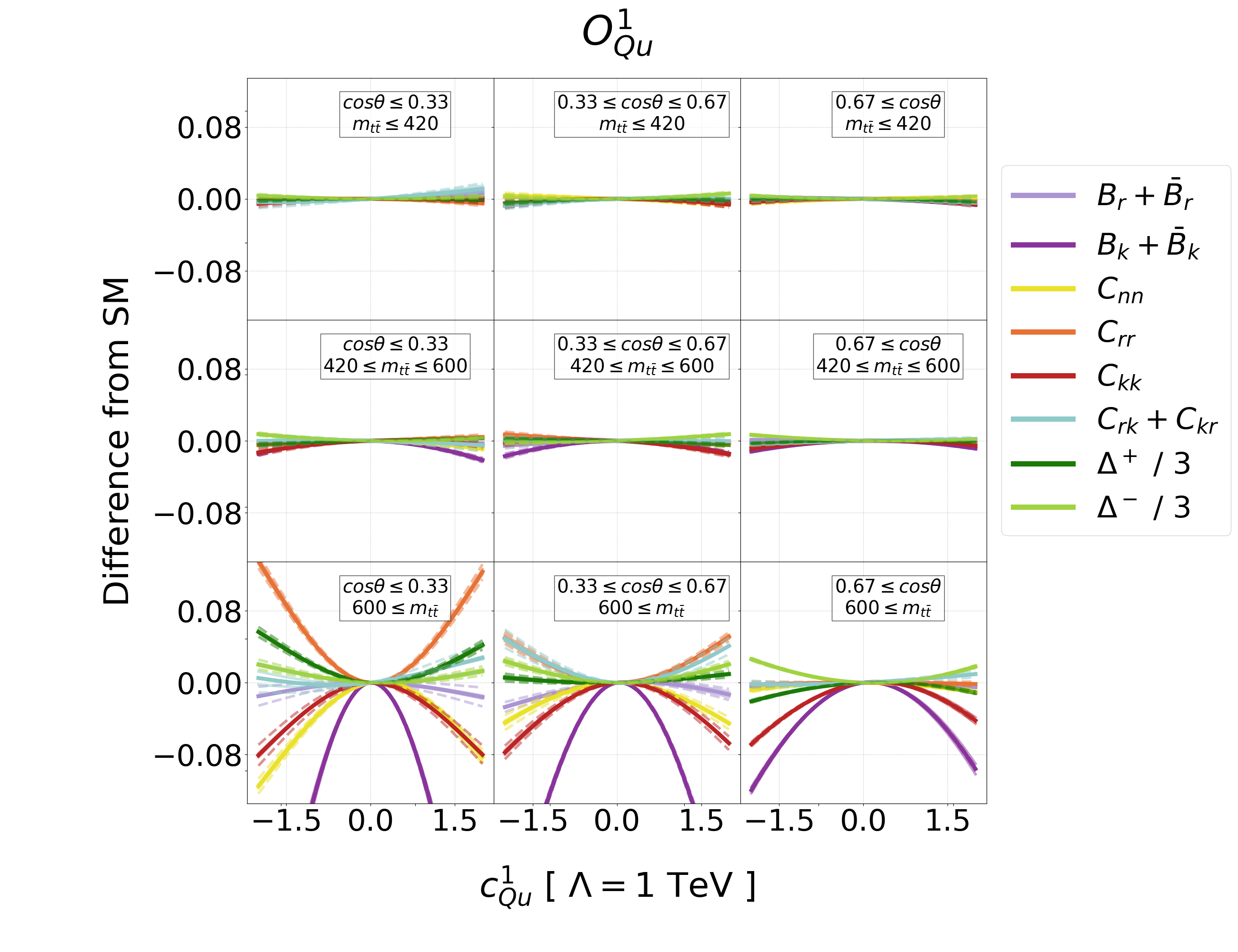}
 \vspace{-5mm}
\captionsetup{width=\linewidth}
\caption{Results for the operators $\mathcal O_{Qu}^{8}$ (top) and $\mathcal O_{Qu}^1$ (bottom), similar to Figure \ref{fig:DIM62F_24_binned}.}
\label{fig:DIM64F_3_binned}
\end{figure}

\begin{figure}[H]
\centering
\includegraphics[width=0.82\linewidth]{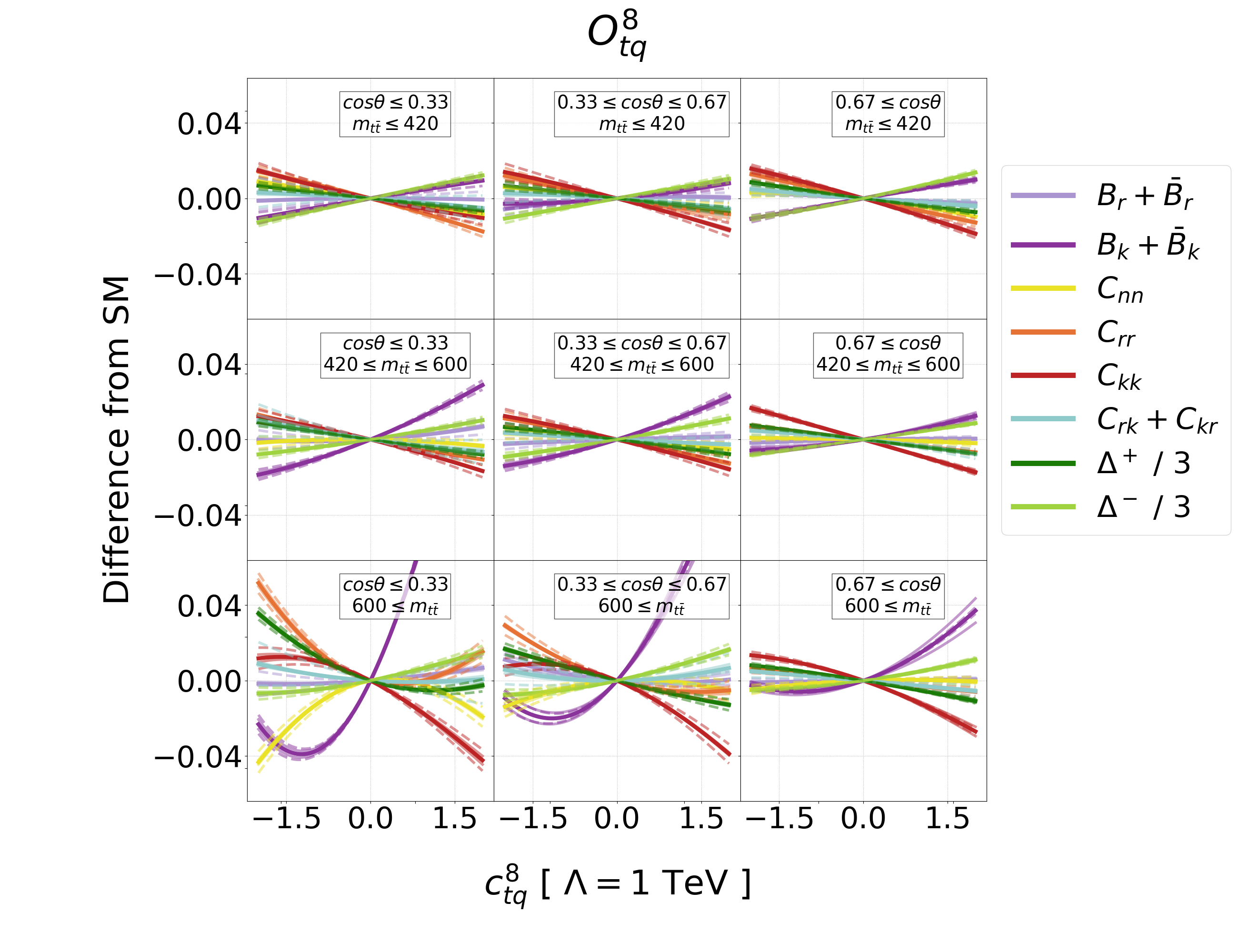}
\includegraphics[width=0.82\linewidth]{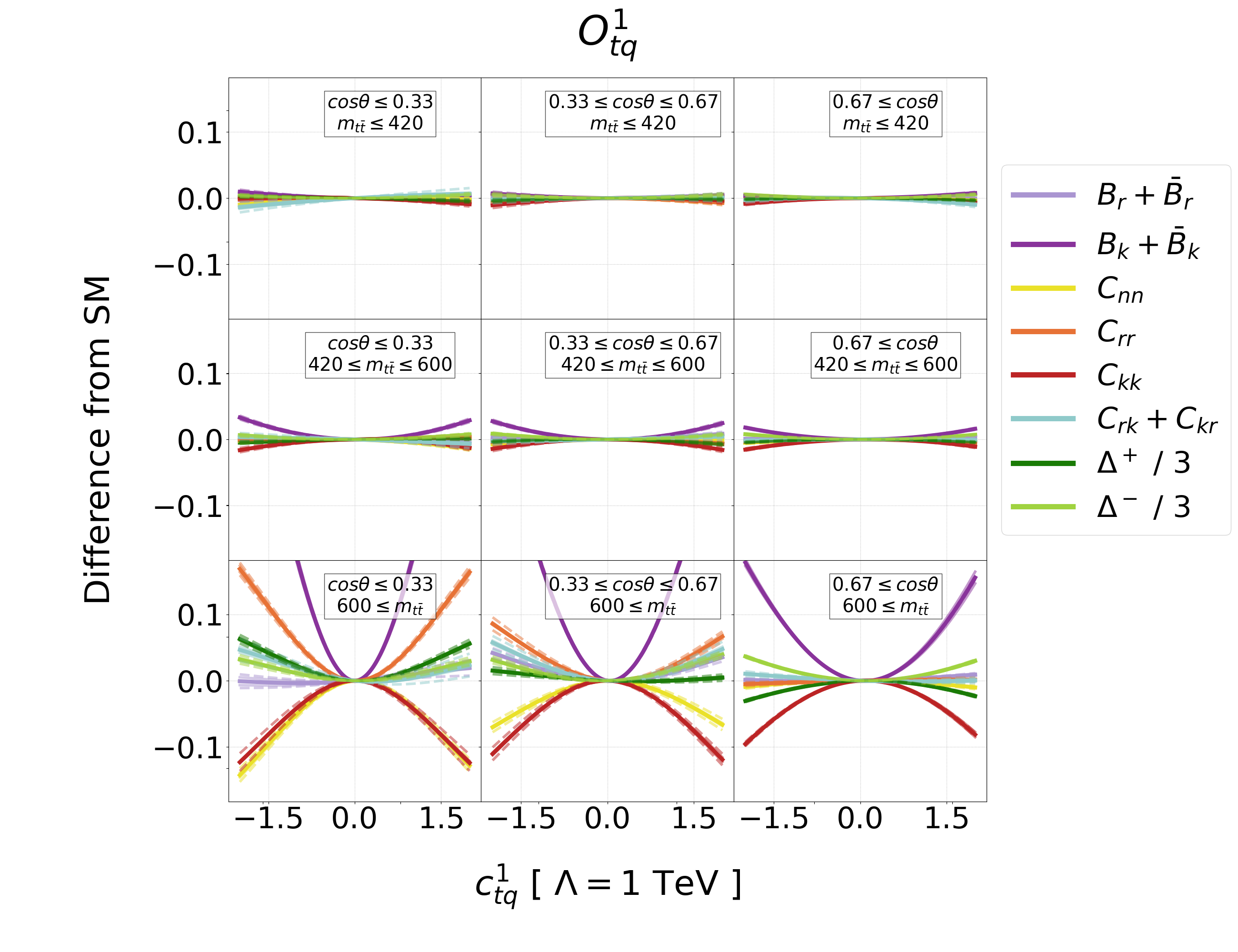}
 \vspace{-5mm}
\captionsetup{width=\linewidth}
\caption{Results for the operators $\mathcal O_{tq}^{8}$ (top) and $\mathcal O_{tq}^1$ (bottom), similar to Figure \ref{fig:DIM62F_24_binned}.}
\label{fig:DIM64F_4_binned}
\end{figure}

\begin{figure}[H]
\centering
\includegraphics[width=0.82\linewidth]{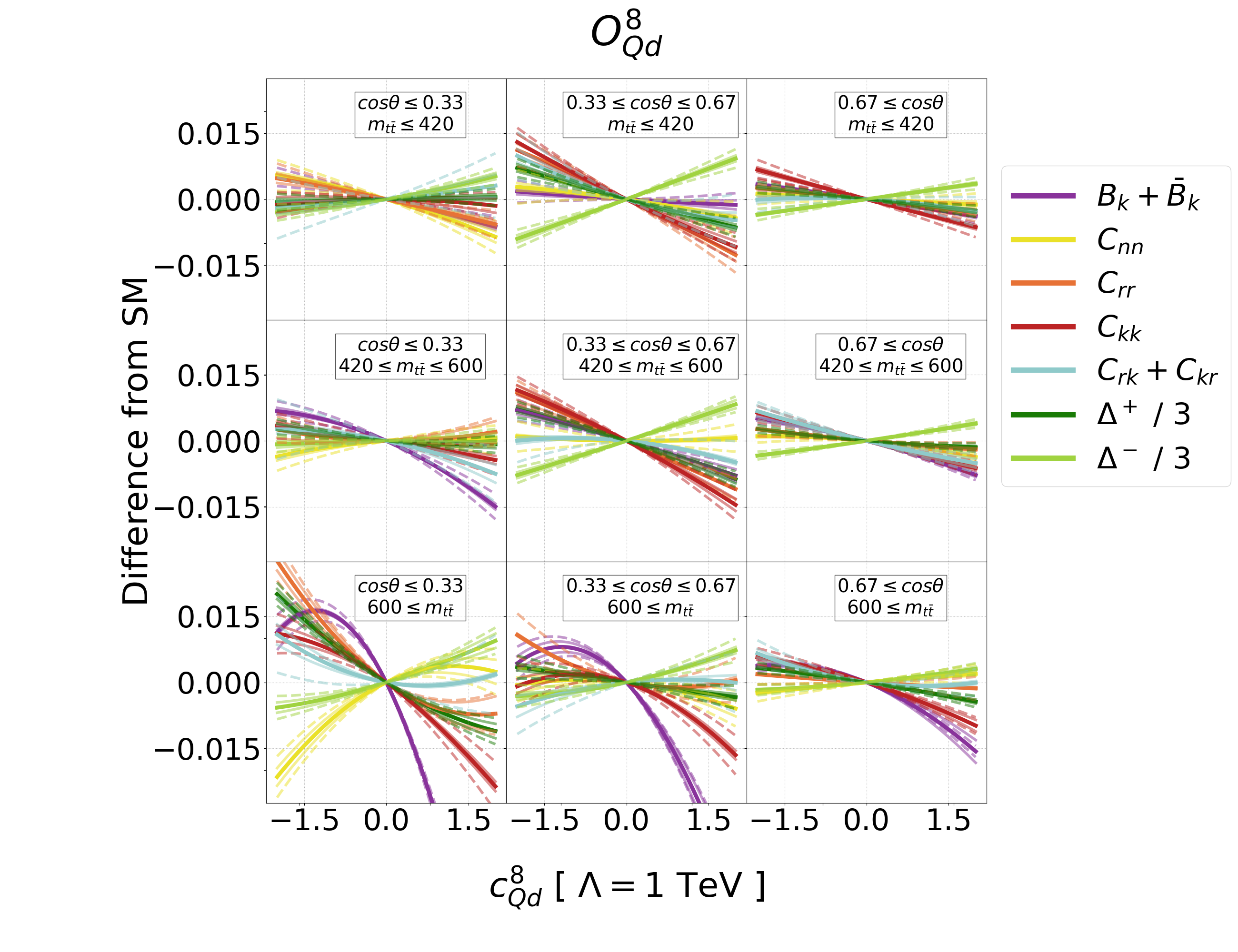}
\includegraphics[width=0.82\linewidth]{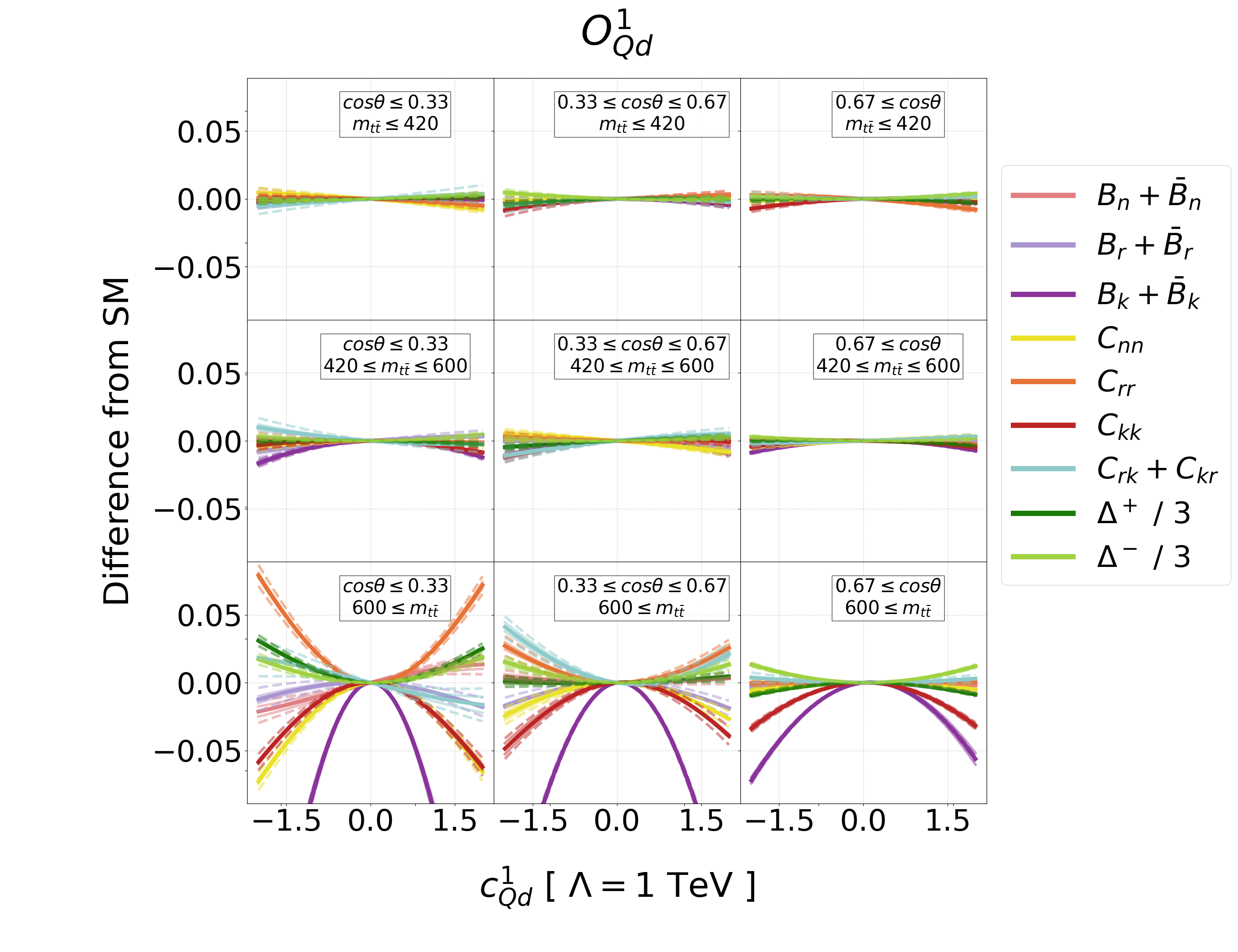}
 \vspace{-5mm}
\captionsetup{width=\linewidth}
\caption{Results for the operators $\mathcal O_{Qd}^{8}$ (top) and $\mathcal O_{Qd}^1$ (bottom), similar to Figure \ref{fig:DIM62F_24_binned}.}
\label{fig:DIM64F_6_binned}
\end{figure}

\begin{figure}[H]
\centering
\includegraphics[width=0.82\linewidth]{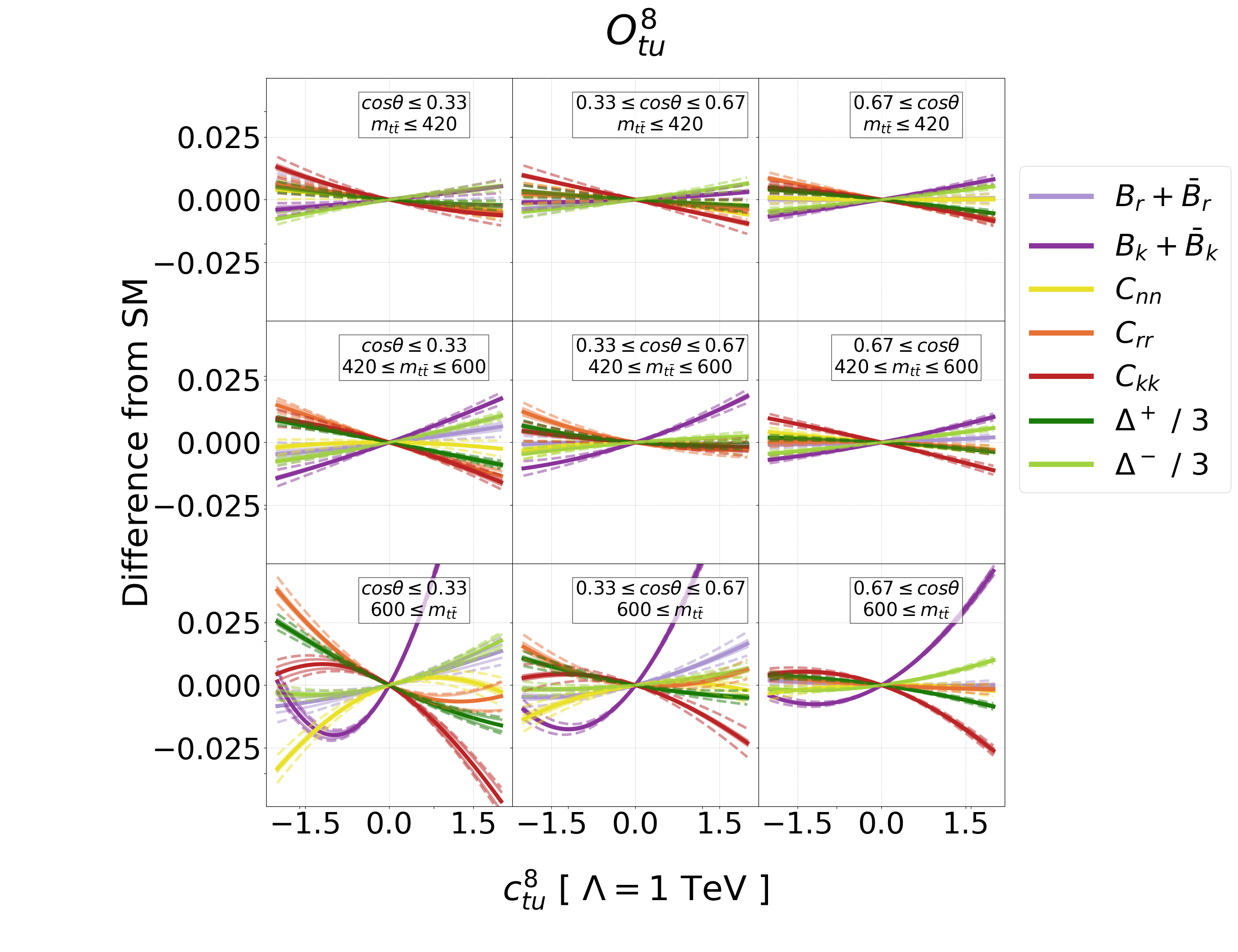}
\includegraphics[width=0.82\linewidth]{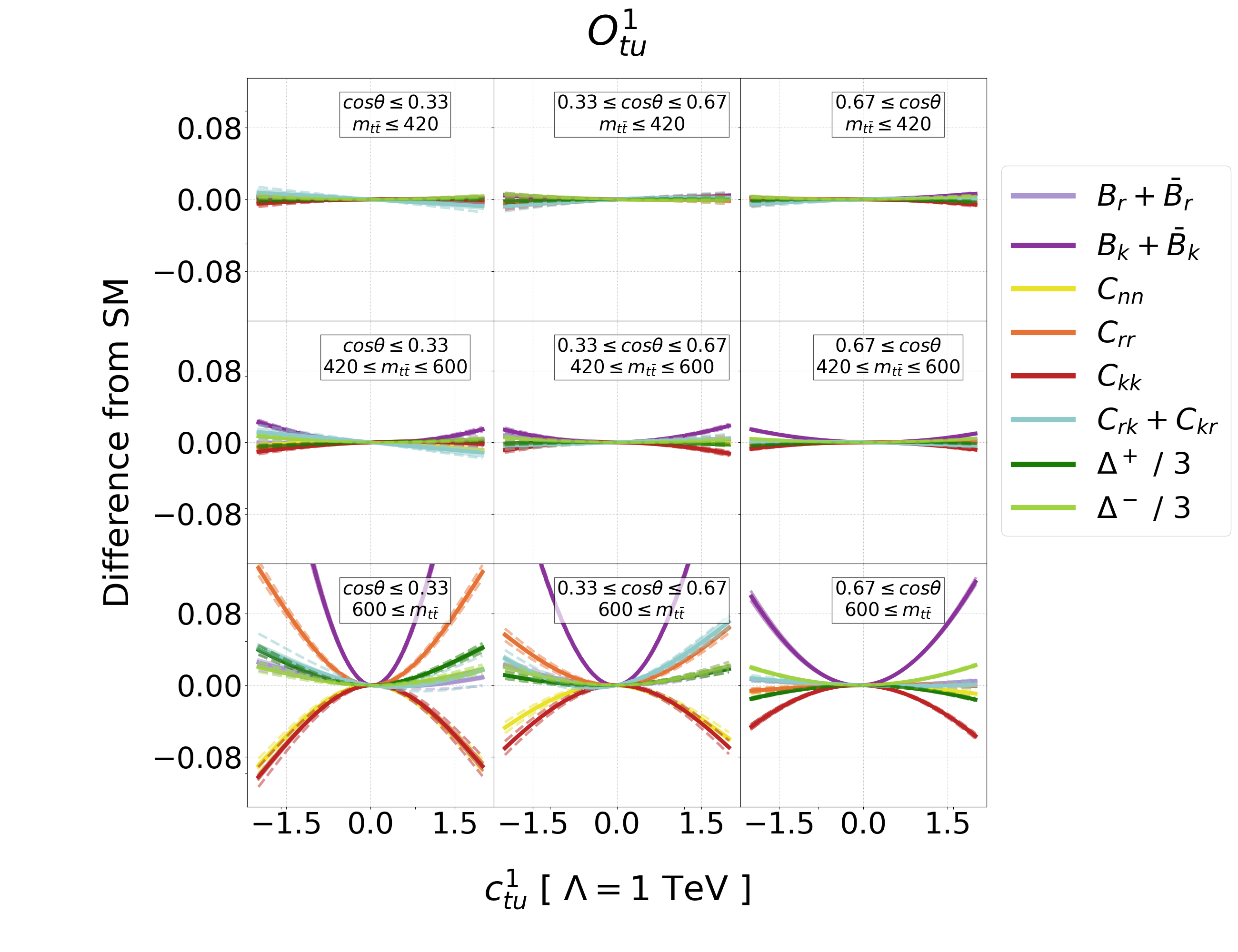}
 \vspace{-5mm}
\captionsetup{width=\linewidth}
\caption{Results for the operators $\mathcal O_{tu}^{8}$ (top) and $\mathcal O_{tu}^1$ (bottom), similar to Figure \ref{fig:DIM62F_24_binned}.}
\label{fig:DIM64F_7_binned}
\end{figure}

\begin{figure}[H]
\centering
\includegraphics[width=0.82\linewidth]{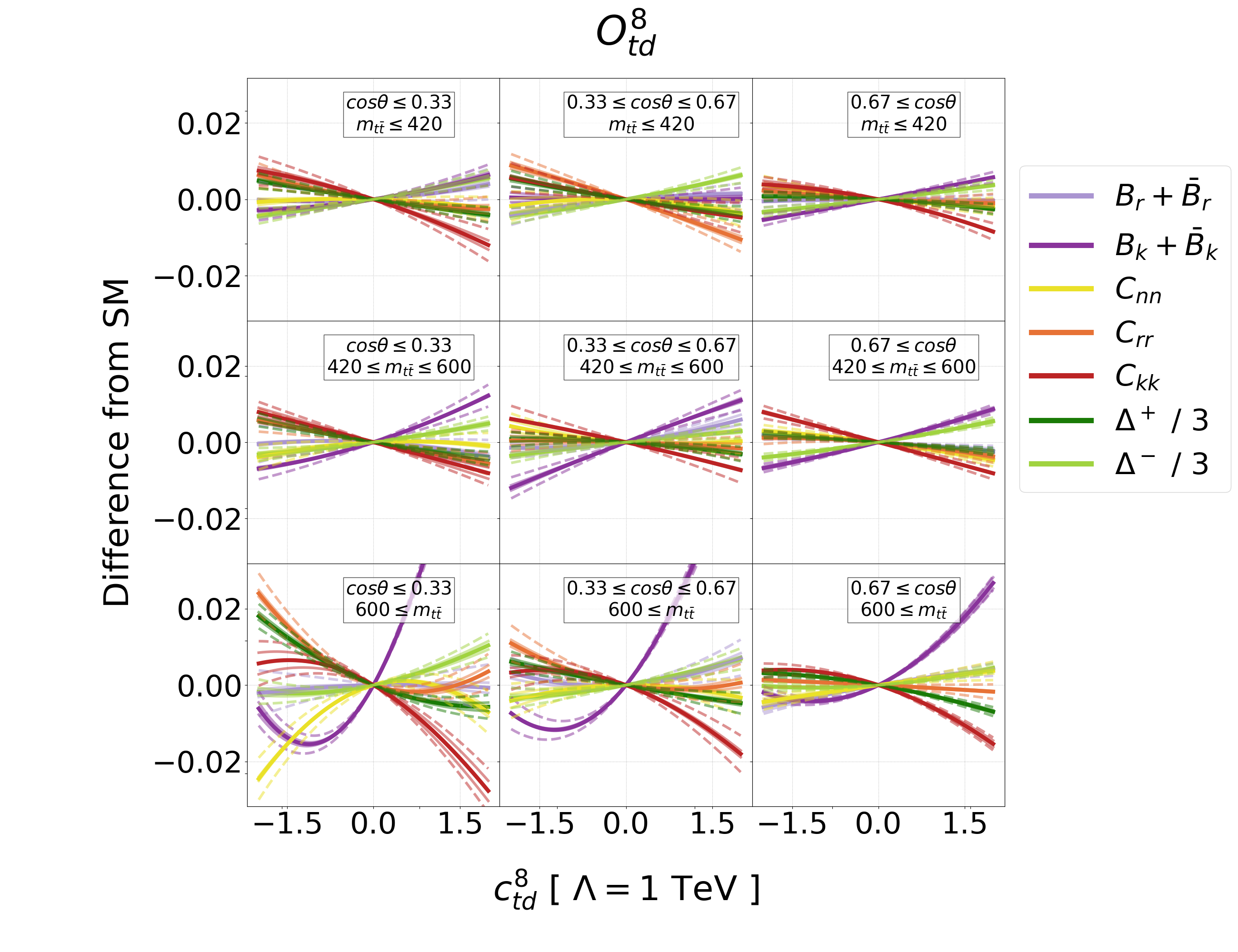}
\includegraphics[width=0.82\linewidth]{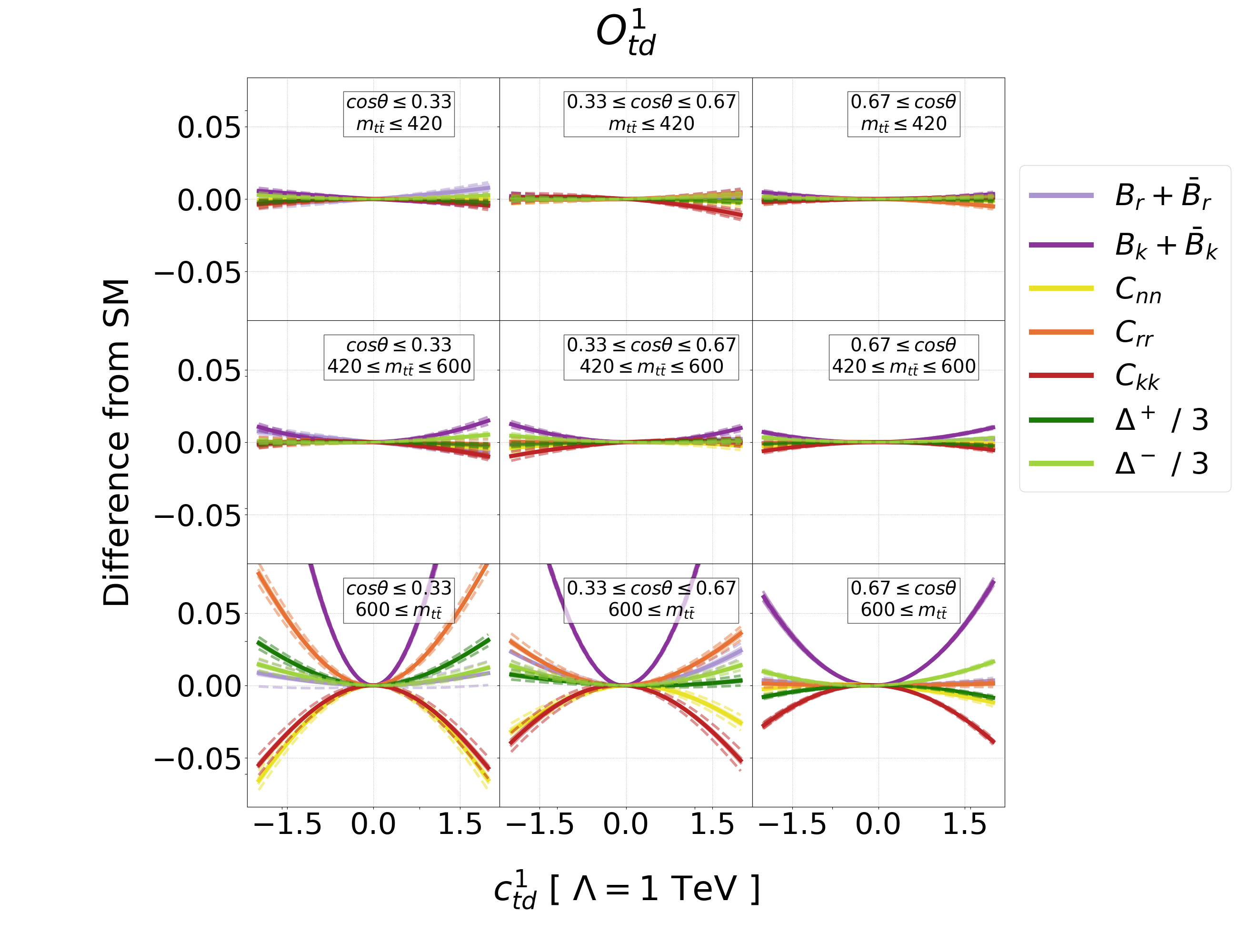}
 \vspace{-5mm}
\captionsetup{width=\linewidth}
\caption{Results for the operators $\mathcal O_{td}^{8}$ (top) and $\mathcal O_{td}^1$ (bottom), similar to Figure \ref{fig:DIM62F_24_binned}.}
\label{fig:DIM64F_8_binned}
\end{figure}

\clearpage

\bibliographystyle{JHEP}
\bibliography{biblio}

\end{document}